\documentclass[acmlarge, anonymous=false,authorversion]{acmart}

\acmJournal{IMWUT}
\acmVolume{3}
\acmNumber{1}
\acmArticle{10}
\acmYear{2019}
\acmMonth{3}
\acmArticleSeq{10}
\acmPrice{15.00}

\setcopyright{acmlicensed}

\acmDOI{10.1145/3314397}

\received{August 2018}
\received[revised]{November 2018}
\received[accepted]{January 2019}

\usepackage{booktabs} 
\usepackage[acronym, nowarn]{glossaries}
\makeglossaries

\newacronym{ap}{AP}{access point}
\newacronym{ble}{BLE}{Bluetooth Low Energy}
\newacronym{iot}{\textsc{IoT}}{Internet of Things}
\newacronym{sdp}{\textsc{SDP}}{secure device paring}
\newacronym{zip}{\textsc{ZIP}}{zero-interaction pairing}
\newacronym{zia}{\textsc{ZIA}}{zero-interaction authentication}
\newacronym{zis}{\textsc{ZIS}}{zero-interaction security}
\newacronym{dc}{\textsc{DC}}{device context}
\newacronym{ac}{\textsc{AC}}{application context}
\newacronym{far}{FAR}{False Accept Rate}
\newacronym{frr}{FRR}{False Reject Rate}
\newacronym{eer}{EER}{Equal Error Rate}
\newacronym{pki}{PKI}{Public Key Infrastructure}
\newacronym{fft}{FFT}{fast Fourier transform}
\newacronym{ntp}{NTP}{Network Time Protocol}
\newacronym{cv}{CV}{cross-validation}
\newacronym{auc}{AUC}{Area Under the Curve}

\usepackage{amsmath,amssymb,amsfonts}
\usepackage{graphicx}
\usepackage{textcomp}
\usepackage{caption}
\usepackage{subcaption}

\usepackage{tabularx}
\usepackage[flushleft]{threeparttable}
\usepackage{array,booktabs,ragged2e}
\usepackage{makecell}

\usepackage{wasysym}
\usepackage{gensymb}
\usepackage{amssymb}
\usepackage{pifont}

\newcommand{\cmark}{\ding{51}}

\makeatletter
\def\@makefnmark{%
  \leavevmode
  \raise.9ex\hbox{\fontsize\sf@size\z@\normalfont\tiny\@thefnmark}}
\makeatother

\usepackage{xcolor,colortbl}  

\definecolor{Gray}{gray}{0.8}
\definecolor{OliveGreen}{rgb}{0,0.6,0}

\usepackage{array,multirow}

\usepackage{siunitx}

\DeclareSIUnit\foot{ft}

\usepackage{tablefootnote}

\usepackage{hyperref}
\usepackage{cleveref}

\begin{document}
\title[Perils of Zero Interaction Security in the Internet of Things]{Perils of Zero-Interaction Security in the Internet of Things}

\author{Mikhail Fomichev}
\authornote{Both authors contributed equally to this work.}
\orcid{0000-0001-9697-0359}
\affiliation{%
 \department{Secure Mobile Networking Lab}
 \institution{TU Darmstadt}
 \country{Germany}}
\email{mfomichev@seemoo.tu-darmstadt.de}

\author{Max Maass}
\orcid{0000-0001-9346-8486}
\authornotemark[1]
\affiliation{%
 \department{Secure Mobile Networking Lab}
 \institution{TU Darmstadt}
 \country{Germany}}
\email{mmaass@seemoo.tu-darmstadt.de}

\author{Lars Almon}
\orcid{0000-0003-1296-2920}
\affiliation{%
 \department{Secure Mobile Networking Lab}
 \institution{TU Darmstadt}
 \country{Germany}}
\email{lalmon@seemoo.tu-darmstadt.de}

\author{Alejandro Molina}
\orcid{0000-0003-4509-9174}
\affiliation{%
 \department{Machine Learning Group}
 \institution{TU Darmstadt}
 \country{Germany}}
\email{molina@cs.tu-darmstadt.de}

\author{Matthias Hollick}
\orcid{0000-0002-9163-5989}
\affiliation{%
 \department{Secure Mobile Networking Lab}
 \institution{TU Darmstadt}
 \country{Germany}}
\email{mhollick@seemoo.tu-darmstadt.de}

\renewcommand{\shortauthors}{M. Fomichev et al.}

\begin{abstract}

The \gls{iot} demands authentication systems which can provide both security and usability. 
Recent research utilizes the rich sensing capabilities of smart devices to build security schemes operating without human interaction, such as \gls{zip} and \gls{zia}. 
Prior work proposed a number of \gls{zip} and \gls{zia} schemes and reported promising results. 
However, those schemes were often evaluated under conditions which do not reflect realistic \gls{iot} scenarios. 
In addition, drawing any comparison among the existing schemes is impossible due to the lack of a common public dataset and unavailability of scheme implementations.

In this paper, we address these challenges by conducting the first large-scale comparative study of \gls{zip} and \gls{zia} schemes, carried out under realistic conditions. 
We collect and release the most comprehensive dataset in the domain to date, containing over 4250 hours of audio recordings and 1 billion sensor readings from three different scenarios, and evaluate five state-of-the-art schemes based on these data.
Our study reveals that the effectiveness of the existing proposals is highly dependent on the scenario they are used in. In particular, we show that these schemes are subject to error rates between 0.6\% and 52.8\%.
\end{abstract}

%
%


%
%



\maketitle


\section{Introduction}
\label{sec:intro}

\glsresetall

The \gls{iot} has become a reality, with many smart devices communicating autonomously to provide users with seamless services like climate control at home, security systems in the workspace, and traffic management in public areas. 
To provide those services, \gls{iot} devices use multiple sensors to ``perceive'' their environment and act on it to make appropriate decisions. 

Numerous security incidents \cite{mirai:2017} have shown that establishing and maintaining secure communications between \gls{iot} devices is challenging. 
First, centralized approaches such as \glspl{pki} are impractical due to their high complexity and limited scalability \cite{Elkhodr:2016}. 
Second, many \gls{iot} devices do not feature user interfaces, making it impossible to perform security establishment using traditional mechanisms like passwords \cite{Fomichev:2018}.
To address this problem, recent research proposed using device sensor readings of the ambient environment, often called \textit{context information} \cite{Perera:2014}.
This information is used to build context-based security schemes operating without user interaction such as \gls{zip} \cite{Schurmann:2013, Miettinen:2014, Xi:2016} and \gls{zia} \cite{Truong:2014, Shrestha:2014, Karapanos:2015}.
We further refer to both as \gls{zis} schemes. 

The security of \gls{zis} schemes is based on the assumption that context information has high entropy, changes frequently, and is unpredictable from outside the specified environment \cite{Sigg:2011}.
Context information, obtained from the ambient environment of an \gls{iot} device, is used to derive a shared secret key between colocated devices in \gls{zip} or to serve as a proof of physical proximity between devices in \gls{zia}. 
For example, similarity in ambient audio sensed by two colocated devices was successfully used in both \gls{zip} \cite{Schurmann:2013} and \gls{zia} \cite{Karapanos:2015}, with the latter scheme becoming part of a commercial product \cite{Futurae:2017}. 
Other research explored the applicability of different context information in \gls{zis} schemes: temperature, humidity, pressure, and luminosity \cite{Shrestha:2014, Miettinen:2014}, magnetic fields, acceleration and rotation rates \cite{Shrestha:2016, Schurmann:2017}, as well as observed WiFi and Bluetooth beacons \cite{Truong:2014}. 

\gls{zis} schemes have three main advantages compared to traditional approaches. 
First, they offer high usability by minimizing user involvement in pairing and authentication procedures.
Second, \gls{zis} schemes can scale to a large number of devices, including those that do not share a common sensing modality \cite{Han:2018}.
Third, \gls{zis} schemes can be built on top of devices' sensing capabilities, reducing modification overhead and facilitating interoperability.

Despite the great potential of \gls{zis} schemes to enable a more secure and usable \gls{iot}, prior work raised questions about their practical applicability \citep{Shepherd:2017} and security soundness \citep{Shrestha:2015, Shrestha:2016spf}. 
The evaluation of the proposed schemes in realistic \gls{iot} scenarios is crucial yet mostly missing. 
In our work, we fill this gap by conducting the first large-scale comparative study of existing \gls{zis} schemes.
We reproduce five state-of-the-art \gls{zis} schemes \cite{Schurmann:2013, Truong:2014, Miettinen:2014, Shrestha:2014, Karapanos:2015} and evaluate their ability to distinguish authorized and unauthorized devices on comprehensive datasets of context information collected in three realistic \gls{iot} scenarios: a \textit{connected car}, \textit{smart office} and \textit{smart office with mobile heterogeneous devices}.
Our evaluation reveals trade-offs between different kinds of context information and context features, and gives insights into pitfalls of the reproduced schemes in practice. 
  
In our scenarios, we collect seven different kinds of context information, given in \autoref{tab:contextinfo}: audio, WiFi and \gls{ble} beacons, barometric pressure, humidity, luminosity, and temperature, from which we compute 16 distinct context features.
We implement \gls{iot} scenarios by distributing sensing devices among various spots, each reflecting a potential \gls{iot} functionality like a smart light, with people following their daily routine in these scenarios. 
When evaluating mobility, we additionally supply users with different sensing devices.
 
From our car, office and mobile scenarios, we collect context information datasets of 1.7, 214 and 23.2 GB, respectively, and annotate them with the ground truth.
To our knowledge, these are the largest datasets of annotated context information collected in the \gls{zis} domain so far. 
Our analysis reveals that many of the reproduced schemes are challenged by our scenarios and often cannot maintain the classification accuracy found by their original authors, reaching error rates between 0.6\% and 52.8\%.
We also observe that many schemes have limited adaptability to difficult circumstances and frequently are not robust, with parameters optimal for one scenario leading to notably lower classification performance in the other.

To facilitate future research, we publicly release the collected context information in an anonymized form, ground truth information, the computed context features, machine learning datasets and results for all reproduced schemes, as well as the full source code used in data collection and evaluation procedures, including metadata for reproducibility (cf. \autoref{sec:design:dataset}) \cite{dataset}.
We further enhance reproducibility by releasing raw audio recordings from the mobile scenario \cite{audio_dataset}, making our dataset the first of its kind in the domain of \gls{zis}.

In summary, we make the following contributions:
\begin{itemize}
	\item We reproduce five state-of-the-art \gls{zis} schemes 
	\cite{Schurmann:2013, Truong:2014, Miettinen:2014, Shrestha:2014, Karapanos:2015}
	and design three realistic \gls{iot} scenarios from which we collect comprehensive datasets of diverse context information.
	
	\item We evaluate the scheme performance and robustness for use in different scenarios. We also provide insights into pitfalls of the reproduced schemes. 

	\item We release the first open-source toolkit, containing datasets of diverse context information, including audio, together with the source code used to collect these data and implementations of the five \gls{zis} schemes.
\end{itemize}

\begin{table}[!htb]
	\caption{Context information use by the reproduced schemes}
	\label{tab:contextinfo}
	\begin{minipage}{\columnwidth}
	\begin{center}
	\begin{tabular}{lcccccccc}
		\toprule
		\gls{zis} scheme & Audio & \gls{ble} & WiFi & Press. & Hum. & Lum. & Temp. & Details \\
		\midrule
		Karapanos \textit{et al.} \cite{Karapanos:2015} & \cmark & \raisebox{-0.75ex}{\APLminus} &\raisebox{-0.75ex}{\APLminus} & \raisebox{-0.75ex}{\APLminus} & \raisebox{-0.75ex}{\APLminus} & \raisebox{-0.75ex}{\APLminus} & \raisebox{-0.75ex}{\APLminus} & \textsection \ref{subsec:appx1-karapanos} \\
		Sch{\"u}rmann and Sigg \cite{Schurmann:2013} & \cmark & \raisebox{-0.75ex}{\APLminus} & \raisebox{-0.75ex}{\APLminus} & \raisebox{-0.75ex}{\APLminus} & \raisebox{-0.75ex}{\APLminus} & \raisebox{-0.75ex}{\APLminus} & \raisebox{-0.75ex}{\APLminus} & \textsection \ref{subsec:appx1-schurmann} \\
		Miettinen \textit{et al.} \cite{Miettinen:2014} & \cmark & \raisebox{-0.75ex}{\APLminus} & \raisebox{-0.75ex}{\APLminus} & \raisebox{-0.75ex}{\APLminus} & \raisebox{-0.75ex}{\APLminus} & \cmark & \raisebox{-0.75ex}{\APLminus} & \textsection \ref{subsec:appx1-miettinen} \\
		Truong \textit{et al.} \cite{Truong:2014} & \cmark & \cmark & \cmark & \raisebox{-0.75ex}{\APLminus} & \raisebox{-0.75ex}{\APLminus} & \raisebox{-0.75ex}{\APLminus} & \raisebox{-0.75ex}{\APLminus} & \textsection \ref{subsec:appx1-truong} \\
		Shrestha \textit{et al.} \cite{Shrestha:2014} & \raisebox{-0.75ex}{\APLminus} & \raisebox{-0.75ex}{\APLminus} & \raisebox{-0.75ex}{\APLminus} & \cmark & \cmark & \raisebox{-0.75ex}{\APLminus} &  \cmark & \textsection \ref{subsec:appx1-shrestha} \\
		\bottomrule
	\end{tabular}
	\end{center}
	\smallskip\centering
	\center{\cmark{} = used; \raisebox{-0.75ex}{\APLminus{}} = not used}
	\end{minipage}
\end{table}

\section{Background}
\label{sec:bkgrd}
In this section we provide the terminology used in this paper, present our system and threat model, and describe the \gls{zis} schemes that we reproduced and evaluated.  

\subsection{Terminology}
\label{sec:bkgrd:terminology}
We start by clarifying the relevant terms in the domain of \gls{zis}.

\textit{Context information}. We define context information as the data collected from device sensors (e.g., microphones, light sensors, etc.), augmented with metadata like timestamps \cite{Perera:2014}.

\textit{Context}. We refer to a set of context information collected by a device from its ambient environment over time as the context of the device.

\textit{Colocation}. We define colocation as a set of devices residing in the same physical space. 
In our scenarios, the spaces are different cars and offices, thus devices within the same car or office are colocated, otherwise non-colocated.
The term colocation highly depends on the use case of the \gls{zis} scheme.
In the case of wearables, colocated devices are on the same body \cite{Brusch:2018}, whereas for a smart home, colocated devices are inside a house \cite{Han:2018}.

\textit{Context feature}. We define context feature as a concise context property computed from context information.
Context features are based on a snapshot of context information \cite{Schurmann:2013, Truong:2014, Shrestha:2014, Karapanos:2015} or on relative changes of context information over time \cite{Miettinen:2014}. 
They calculate a distance or similarity metric between two samples of context information \cite{Truong:2014, Shrestha:2014, Karapanos:2015}, or derive a binary fingerprint vector from a sample of context information \cite{Schurmann:2013, Miettinen:2014}.

\subsection{System and Threat Model}
\label{sec:bkgrd:model}
We assume an \gls{iot} scenario containing a number of devices that are colocated and equipped with a set of sensors to collect context information. 
The goal of \gls{zis} schemes is to have two colocated devices establish a secure connection (\gls{zip}) or a proof of proximity (\gls{zia}) without user interaction, utilizing context features to secure the process.
We assume no established infrastructure and, in the case of \gls{zip}, no prior trust between devices. 

Our adversary is based on the models used in the reproduced \gls{zis} schemes. 
The adversary is an \gls{iot} device located in an adjacent car or office. 
This device can be benign, accidentally trying to pair or authenticate with proximate devices in its wireless range (e.g., \gls{iot} device in a neighbor's car), or it can be malicious, intentionally trying to pair or authenticate with non-colocated devices. 
The adversary is non-colocated with benign devices, thus it can neither observe their context information, nor compromise benign devices to circumvent a \gls{zis} scheme. 
However, the adversary is physically close to benign devices (i.e., adjacent car or office), equipped with the same sensing hardware to collect context information, and can communicate with them over a wireless link.  

The goal of the adversary is to obtain similar enough context information to fool benign devices into believing that it is co-located with them.
Compared to threat models of the reproduced schemes our adversary is more powerful as it possesses two extra capabilities. 
First, it remains permanently in close proximity to benign devices, including times of low context activity such as during the night.
Second, due to symmetric deployment of devices in our scenarios, the adversary has much better chances of following the same trends in context information (e.g., lighting conditions) as benign devices.

We purposely make our adversary powerful to evaluate the scheme performance in challenging scenarios.
This allows us to establish the worst-case \textit{baseline adversary}, facilitating comparison of the reproduced \gls{zis} schemes (discussed in \autoref{sec:disc}), as well as gain first insights into possible attack vectors for an active adversary \cite{Shrestha:2015}.

\subsection{Reproduced ZIS schemes}
\label{sec:bkgrd:repro}
To select \gls{zis} schemes for our study, we surveyed frequently cited schemes published at top security venues in the last five years.
We selected schemes that targeted \gls{iot} scenarios and utilized different context information or context features.  
We excluded schemes based on behavioral biometrics, e.g., gait \cite{Brusch:2018}, gesture \cite{Shrestha:2016theft} or keystroke dynamics \cite{Mare:2014}, as these schemes designed for wearable \gls{iot} scenarios.
In the end, we arrived at five schemes, which we reproduced from the ground up, relying on the help of the original authors to ensure the correctness of our implementations.
We briefly introduce each scheme in its respective result subsection (cf. \autoref{sec:res}) and refer to the Appendix for detailed descriptions.

\section{Study Design}
\label{sec:sdesign}
We designed our study to cover the majority of relevant context information used in current \gls{zis} schemes.
We selected three realistic IoT scenarios: in the first two, we used identical sensing devices to collect context information, minimizing the effects of hardware variations on our results.
In the third scenario, we used different sensing devices to evaluate the impact of device heterogeneity. 
This section describes the design and conduct of our experiments, as well as ethical concerns when dealing with sensitive personal data collected in our study.

\subsection{Data Collection}
\label{sec:design:datacoll}
The goal of our experiment was to collect a comprehensive real-world dataset of context information that can serve as a baseline for comparing current and future \gls{zis} schemes.
In the first two scenarios, we collected data using a \textit{Texas Instruments SensorTag CC2650} and a \textit{Raspberry Pi 3}.
Audio data was collected using a \textit{Samson Go} USB microphone, which recorded a mono audio stream with a 16 kHz sampling rate, and encoded it using the lossless \textit{FLAC} format.
The Raspberry Pi also collected all visible wireless \glspl{ap} and \gls{ble} devices, including their signal strength, every ten seconds.
The remaining context information (accelerometer, barometer, gyroscope, humidity, light intensity, magnetometer, and temperature) was recorded using the SensorTag, connected to the Raspberry Pi using \gls{ble}. \footnote{While accelerometer, gyroscope and magnetometer are not used by any scheme, we collected their data for use in future schemes.}
Sensor data was recorded with a sampling rate of 10 Hz. 

In the third scenario, we additionally used \textit{Samsung Galaxy S6} smartphones and \textit{Samsung Gear S3} smartwatches to collect the same context information.  
Since those devices are not equipped with temperature and humidity sensors we combined them with a \textit{RuuviTag+}. 
We tried to obtain the same sampling rate on all our devices, however, the Galaxy S6 limits barometric pressure and luminosity readings to 5 Hz. 
The summary of used sensing devices and sampling rates is given in \autoref{tab:sensing-devices} in the Appendix.
All events that could influence the context information (e.g., windows/doors being opened or closed, people entering or leaving the recording area, traces of mobile devices, etc.) were documented automatically or by hand in a ground truth sheet.

\subsection{Scenario 1: Car}
\label{sec:design:car}
\begin{figure}
\centering
	\begin{minipage}[b]{.5\textwidth}
  		\centering
  		\includegraphics[width=.9\linewidth]{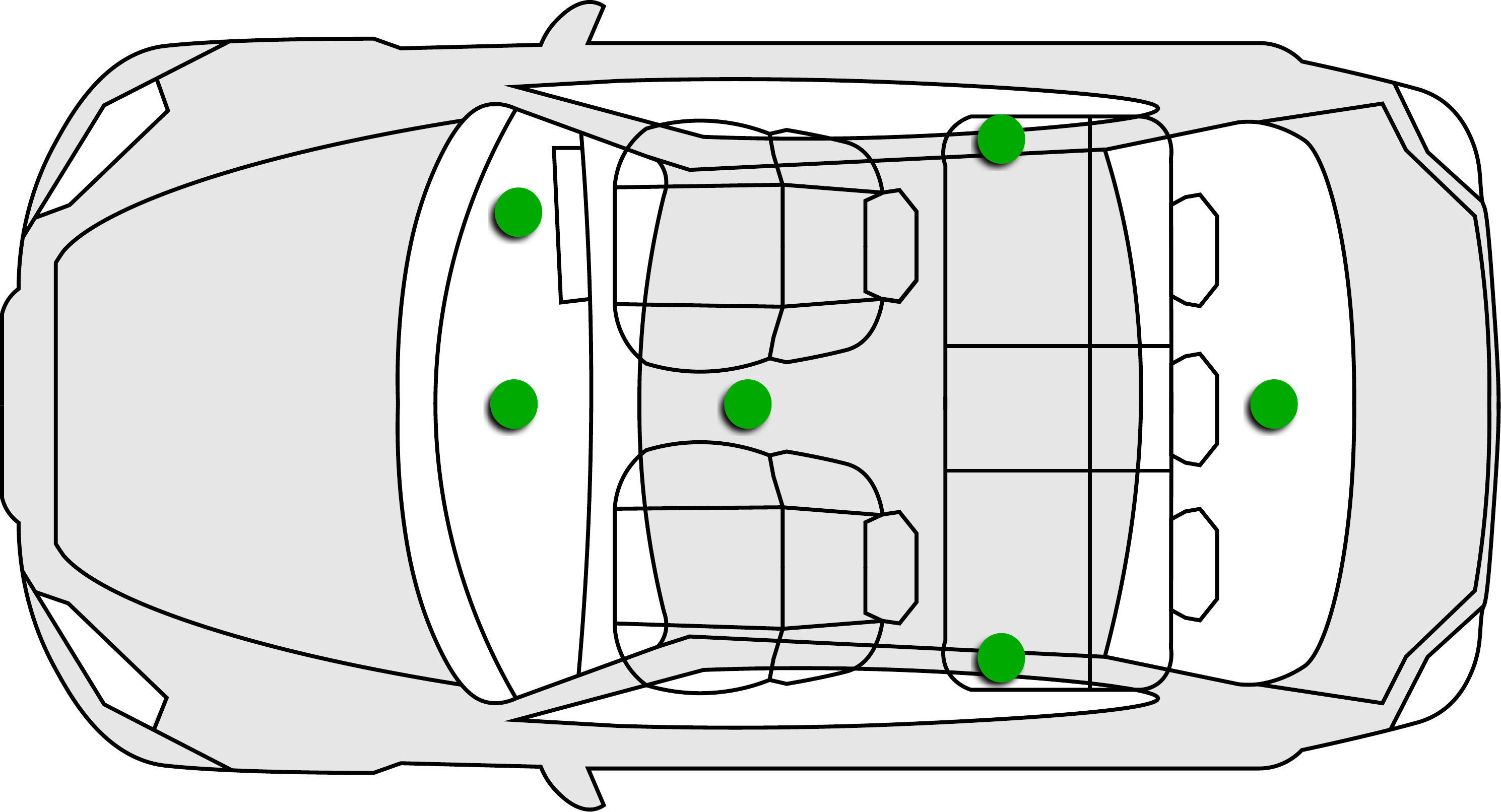}
  		\captionof{figure}{Device location map in the car scenario}
  		\label{fig:car}
	\end{minipage}%
	\begin{minipage}[b]{.5\textwidth}
  		\centering
  		\includegraphics[width=.9\linewidth]{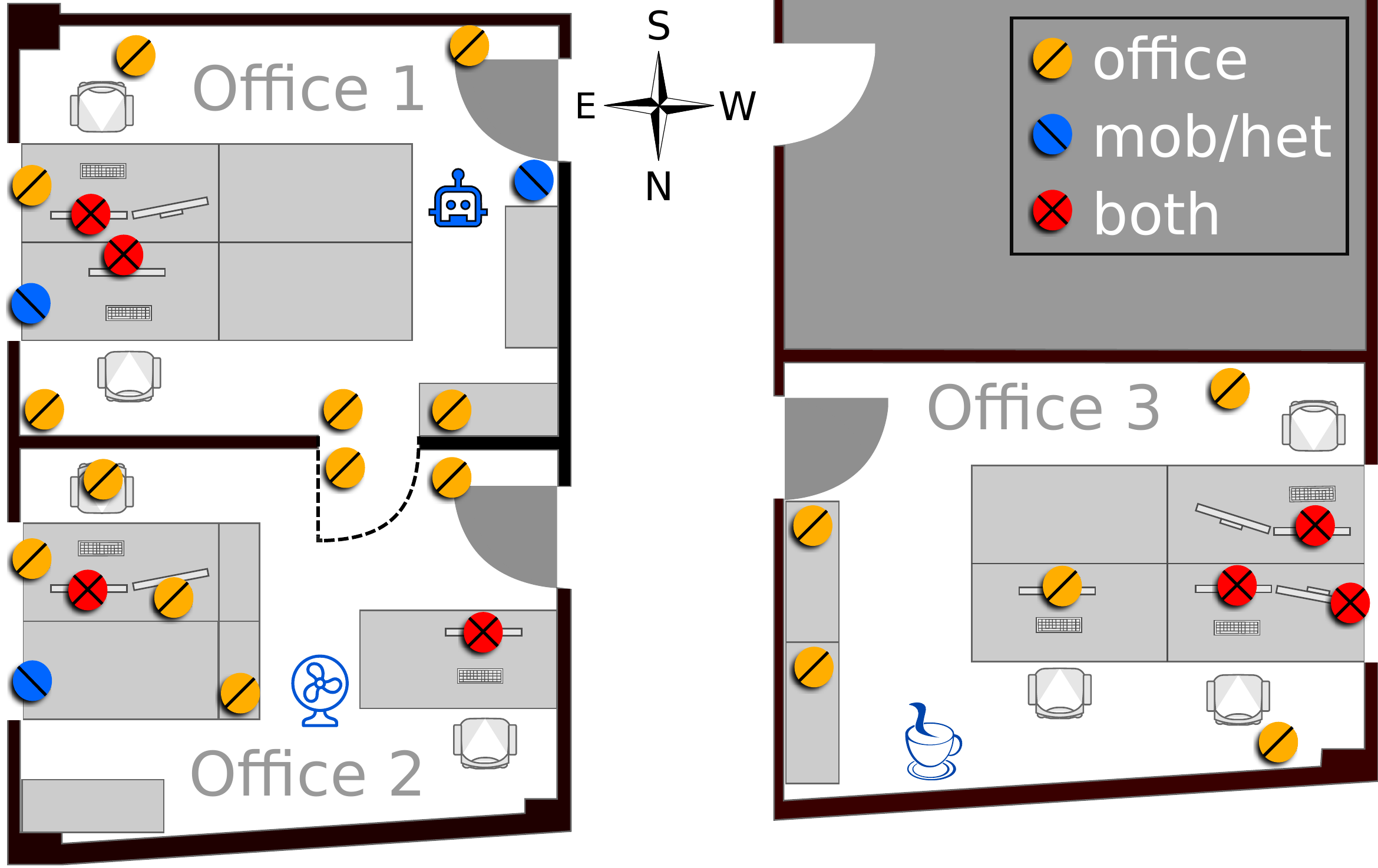}
  		\captionof{figure}{Device location map in the office and mob/het scenarios}
  		\label{fig:office}
	\end{minipage}
\end{figure}

In the first scenario, we used two cars from different manufacturers. 
Each car was equipped with six sensing devices distributed inside the vehicle as shown in Figure \ref{fig:car}.
The devices occupied similar spots in both cars: 
one device was placed on top of the dashboard facing the windshield, inside the glove compartment, in between the front seats facing upwards, attached to each handhold above the two rear doors, and put in the middle of the trunk.
This placement covers all prominent spots one might expect a sensor or a personal device inside a car (cf. \autoref{tab:car-mapping} in the Appendix).

After setting up the cars, we drove a predefined route of three hours and \SI{120}{\kilo\meter} (\SI{74}{\nauticalmile}) on the afternoon of an autumn day.
The time was chosen to ensure that the collection began while the sun was visible and ended after sunset, to collect a variety of lighting conditions.
The route included city traffic, country roads, and highway (cf. \autoref{fig:car-route} in the Appendix for a map).
We drove both cars close to each other within a distance of \SI{20}{\metre} (\SI{65}{\foot}), which we varied from time to time.
In addition, we took a short break, with the cars parked side by side. 

The challenge for the \gls{zis} schemes is to identify colocated devices in a single car, while excluding devices from different cars that might be nearby or just listening to the same radio station.

\subsection{Scenario 2: Office}
\label{sec:design:office}
A typical application for IoT devices is the deployment in a smart home or office.
To collect realistic context information in this scenario, we deployed eight sensing devices in three office rooms as shown in Figure \ref{fig:office}.
We put the devices in similar places, representing typical \gls{iot} spots:
one device was attached to the main screen of a workplace (smart workstation, several spots), above the windows (smart shades), near the ceiling lights (smart lights), in a closed cupboard (smart device), near the door at around two meters height (smart room/motion sensor) and in a corner at around 2.5 meters (environmental sensor).
The summary of device locations in the office scenario is given in \autoref{tab:office-mapping} in the Appendix. 

We collected context information for one full week, resulting in five work days with people present and two days of the weekend, when offices 1 and 2 were empty, and one person was working in office 3. 
Offices 1 and 2 were adjacent and connected with a door, which was closed most of the time.
Office 3 was on the opposite side of the floor.
All three rooms had a similar setup in terms of size and position of furniture but a different number of participants working in them (one in office 2, two in office 1 and three in office 3).

The collected dataset is intended for testing \gls{zis} schemes designed for smart homes and offices.
The challenge here is to distinguish between the three different rooms.
Ideally, a scheme identifies all colocated devices in one room but excludes all others.

\subsection{Scenario 3: Office with Mobile Heterogeneous Devices (Mob/het)}
\label{sec:design:mobile}
We extended the office scenario by including both static devices permanently residing inside offices, and mobile devices carried by users (cf. Figure \ref{fig:office}).  
We added a number of appliances (i.e., vacuum robot and its station in office 1, fan in office 2, coffee machine in office 3), facilitating device mobility when users move to use them. 
Each office was equipped with four static devices (SensorTag), covering similar spots and the appliances: one device was attached to the main screen of a workplace (smart workstation, several spots), near a power plug (smart plug), on top of a vacuum robot station (smart robot station) and coffee machine (smart coffee maker), near a fan (smart fan).
We equipped four participants with three mobile devices each: a laptop (with attached smartphone to collect context information), smartphone and smartwatch.
We also placed a smartphone on top of the robot vacuum cleaner.
Device locations are summarized in \autoref{tab:mob-het} in the Appendix. 

We collected context information for eight hours from 9 am till 5 pm, representing a typical working day.
Over the course of the day participants moved freely between the offices to get a cup of coffee, have a meal or attend a meeting, each time carrying a set of their mobile devices. 
We also moved the vacuum robot between the offices, letting it autonomously run a complete cleaning cycle. 

Similarly to the office scenario, the challenge for \gls{zis} schemes is to distinguish devices present in the same office, while excluding devices in others.

\subsection{Reproducibility and Re-usability}
\label{sec:design:dataset}
In total, our dataset contains 239 GB of context information, including more than 4250 hours of audio recordings, over 1 billion sensor readings, and over 12 million WiFi and \gls{ble} beacons.
Computing the context features of the reproduced schemes took over 300\,000 CPU hours.
The audio-based features were computed using Matlab on a high-performance cluster.
The remaining features were implemented in Python on a high-performance server.
After compression, they utilize almost 1 TB of disk space.
This includes the computed features, aggregated statistics, and metadata for reproduction and validation following the recommendations by Benureau \textit{et al.} \cite{Benureau:2017}.

To facilitate future reuse, we release the source code of the entire data collection and evaluation stack, as well as the collected context information in an anonymized form, all intermediate and final data files (including machine learning models) and the code used to generate the visualizations.
Privacy concerns prevent us from releasing the audio data recorded in the Car and Office scenarios, but we are able to provide researchers with the audio recordings from the Mobile scenario upon request \cite{audio_dataset}.
See \cite{dataset} for an index of all released data and code.

\subsection{Ethical Considerations}	
\label{sec:design:ethics}
The study was approved by our institutional ethical review board, data protection officer, and workers' council.
Participants gave informed consent for the collection, use, and release of the data.
During collection, the audio data was encrypted with keys controlled by the affected participants,
requiring their explicit consent and cooperation to decrypt the data for processing.
In the mob/het scenario, we gave participants the chance to inspect the recordings before obtaining informed consent for their release.

\section{Results}
\label{sec:res}
\begin{table}
	\caption{Overview of the reproduced schemes and evaluation result}
	\begin{tabular}{lcl|ccc}
		\toprule
		&&& \multicolumn{3}{c}{Best EER}\\
		Scheme & Sect. & Conclusion & Car & Office & Mob/het \\
		\toprule
		\makecell{Karapanos \\\textit{et al.} \cite{Karapanos:2015}} & \ref{sec:res:karapanos} & \makecell{Best EER for car scenario among the reproduced schemes.\\ Limited robustness on intervals from 5 to 15 seconds.\\Breaks down in heterogeneous setting.} & 0.006 & 0.098 & 0.157 \\
		\rowcolor{gray!12}
		\makecell{Sch{\"u}rmann, \\Sigg \cite{Schurmann:2013}} & \ref{sec:res:schurmann} & {\setlength{\fboxsep}{0pt}\colorbox{gray!12}{\makecell{Generates fingerprints with good randomness properties,\\ but shows varying performance on subscenarios, and \\provides only limited robustness.}}} & 0.154 & 0.241 & 0.140 \\
		\makecell{Miettinen \\\textit{et al.} \cite{Miettinen:2014}} & \ref{sec:res:miettinen} & \makecell{Insufficient fingerprint randomness leads to some error rates\\exceeding 0.5. Low robustness. Audio-based fingerprints \\perform better than luminosity-based fingerprints.} & 0.226 & 0.120 & --- \\
		\rowcolor{gray!12}
		{\setlength{\fboxsep}{0pt}\colorbox{gray!12}{\makecell{Truong \\\textit{et al.} \cite{Truong:2014}}}} & \ref{sec:res:truong} & {\setlength{\fboxsep}{0pt}\colorbox{gray!12}{\makecell{Achieves the best error rates in office and mob/het scenario, \\but shows low robustness and high reliance on audio feature,\\and struggles with heterogeneous settings.}}} & 0.104 & 0.069 & 0.123 \\
		\makecell{Shrestha \\\textit{et al.} \cite{Shrestha:2014}} & \ref{sec:res:shrestha} & \makecell{Promising performance in the car scenario, but lower \\performance in the office and mob/het scenario, low \\robustness, and high redundancy and ambiguity in features.} & 0.115 & 0.247 & 0.141 \\
		\bottomrule
	\end{tabular}
	\label{tab:contents}
\end{table}

In this section, we report on the performance in distinguishing colocated and non-colocated devices for the five reproduced schemes (cf. \autoref{tab:contents}).
The performance evaluation of each scheme is structured as follows.
We first provide a concise overview of the scheme by explaining the context features used to distinguish colocated and non-colocated devices. 
Then, we explain the methodology of the original scheme and provide details of our evaluation. 
Next, we present and interpret the performance results of the scheme for each scenario.
To quantify the performance we compute the \gls{eer}, which is the point of equal \gls{far} and \gls{frr}.
In addition, we assess how much usability the schemes can deliver if a specific security level is required by setting a number of target \glspl{far} (between 0.1\% and 5\%) and analyzing the resulting \glspl{frr}.

We evaluate the scheme robustness by analyzing an increase in error rates (either \gls{far} or \gls{frr}) from the original \gls{eer} when applying parameters found to be optimal in one scenario to another. 
This simulates a scheme being used in a scenario it was not trained on, like an IoT device optimized for office use being deployed in a car. 
We further summarize each studied scheme by comparing our results with the original findings and providing key takeaways from our evaluation. 
This facilitates a direct comparison of the different schemes in our scenarios. 

We introduce subscenarios to investigate the impact of changes in the environment (e.g., time of day, moving vs. parked cars) on the scheme performance. 
A subscenario represents a subset of context information collected at a specific stage in the scenario. 
For the car scenario, we distinguish three subscenarios:
the \textit{city} and \textit{highway} subscenarios contain context information of the cars driving inside city limits or on the highway, respectively, and the \textit{parked} subscenario includes context information from the time the cars were parked. 
Similarly, we construct three subscenarios for the office scenario:
the \textit{weekday} subscenario contains context information collected from Monday to Friday from 8 am to 9 pm, the \textit{night} includes context information for all seven days from 9 pm to 8 am, and the \textit{weekend} consists of context information from Saturday and Sunday in the timeframe from 8 am to 9 pm.
We omit the subscenario evaluation in the mob/het scenario as there were no specific stages in this scenario.

We assess the performance of all schemes except \cite{Truong:2014} and \cite{Shrestha:2014} on time intervals of 5, 10, 15, 30, 60 and 120 seconds with the length denoted $t$.
The interval represents a timeframe over which context information is aggregated to compute a  context feature, e.g., a 5 second audio snippet or a 30 second WiFi capture.
\cite{Truong:2014} is evaluated on time intervals of 10 and 30 seconds, as the scheme is less well-suited to an arbitrary interval length due to the used features, while \cite{Shrestha:2014} does not use any intervals.

\subsection{Karapanos \textit{et al.}}
\label{sec:res:karapanos}
Karapanos \textit{et al.} \cite{Karapanos:2015} proposed using maximum cross-correlation between snippets of ambient audio from two devices to decide if they are colocated.
The cross-correlation is computed on a set of one-third octave bands \citep{ANSI:2004} and averaged to a similarity score.
One-third octave bands split the audible spectrum (20 Hz to 20 kHz) into 32 frequency ranges of different sizes.
To prevent erroneous authentication when audio activity is low, a power threshold is applied to discard audio snippets with insufficient average power. 
The similarity score is checked against a fixed similarity threshold to decide if two devices are colocated, and can thus be authenticated. 
Tuning the similarity threshold allows trading usability for security and vice versa. 
The authors evaluated their scheme in several scenarios such as a quiet office, lecture hall, and caf\'{e}. 
The scheme details are given in Appendix \ref{subsec:appx1-karapanos}. 

\subsubsection{Methodology}
To investigate the scheme performance we compute similarity scores between colocated and non-colocated devices on different interval lengths. 
We increase the minimum length of audio snippet and maximum correlation lag to achieve a comparable level of synchronization to the original implementation. 
These changes have a negligible impact on the similarity score computation, as stated in Appendix \ref{subsec:appx1-karapanos}. 
To understand factors affecting the performance, we analyze the behavior of similarity scores on different octave bands.

\subsubsection{Car}
\label{sec:res:spf:car}

\begin{table}
\centering
	\caption{\gls{eer} summary for Karapanos \textit{et al.}}
	\label{tab:res:spf:sum}
	\begin{tabular}{c|cccc|cccc|c}
		\toprule
		&
		\multicolumn{4}{c|}{Car} &
		\multicolumn{4}{c|}{Office} & 
		\multicolumn{1}{c}{Mob/het}\\
		$t$ & Full & City & Highway & Parked & Full & Night & Weekday & Weekend & Full \\
		\midrule
		5 & 0.050 & 0.071 & 0.009 & 0.124 & 0.141 & 0.140 & 0.135 & 0.143 & 0.157 \\
		10 & 0.032 & 0.049 & 0.003 & 0.071 & 0.133 & 0.132 & 0.128 & 0.136 & 0.168 \\ 
		15 & 0.026  & 0.043 & 0.002 & 0.060 & 0.128 & 0.126 & 0.123 & 0.129 & 0.170$^*$\tablefootnote{In cases where \gls{far} and \gls{frr} do not match to three digits after the decimal, we average them and denote the result as \gls{eer}$^*$.} \\ 
		30 & 0.017 & 0.031 & 0.001 & 0.022 & 0.118 & 0.115 & 0.116 & 0.115 & 0.172 \\ 
		60 & 0.008 & 0.014 & 0.002 & 0.007 & 0.107 & 0.102 & 0.109 & 0.099 & 0.179$^*$ \\ 
		120 & 0.006 & 0.010 & 0.000 & 0.037 & 0.098 & 0.090 & 0.103 & 0.081 & 0.183$^*$ \\ 
		\bottomrule
	\end{tabular}
\end{table}

\begin{figure}
\centering
\begin{subfigure}[b]{0.33\textwidth}
	\centering
   	\includegraphics[width=0.99\textwidth]{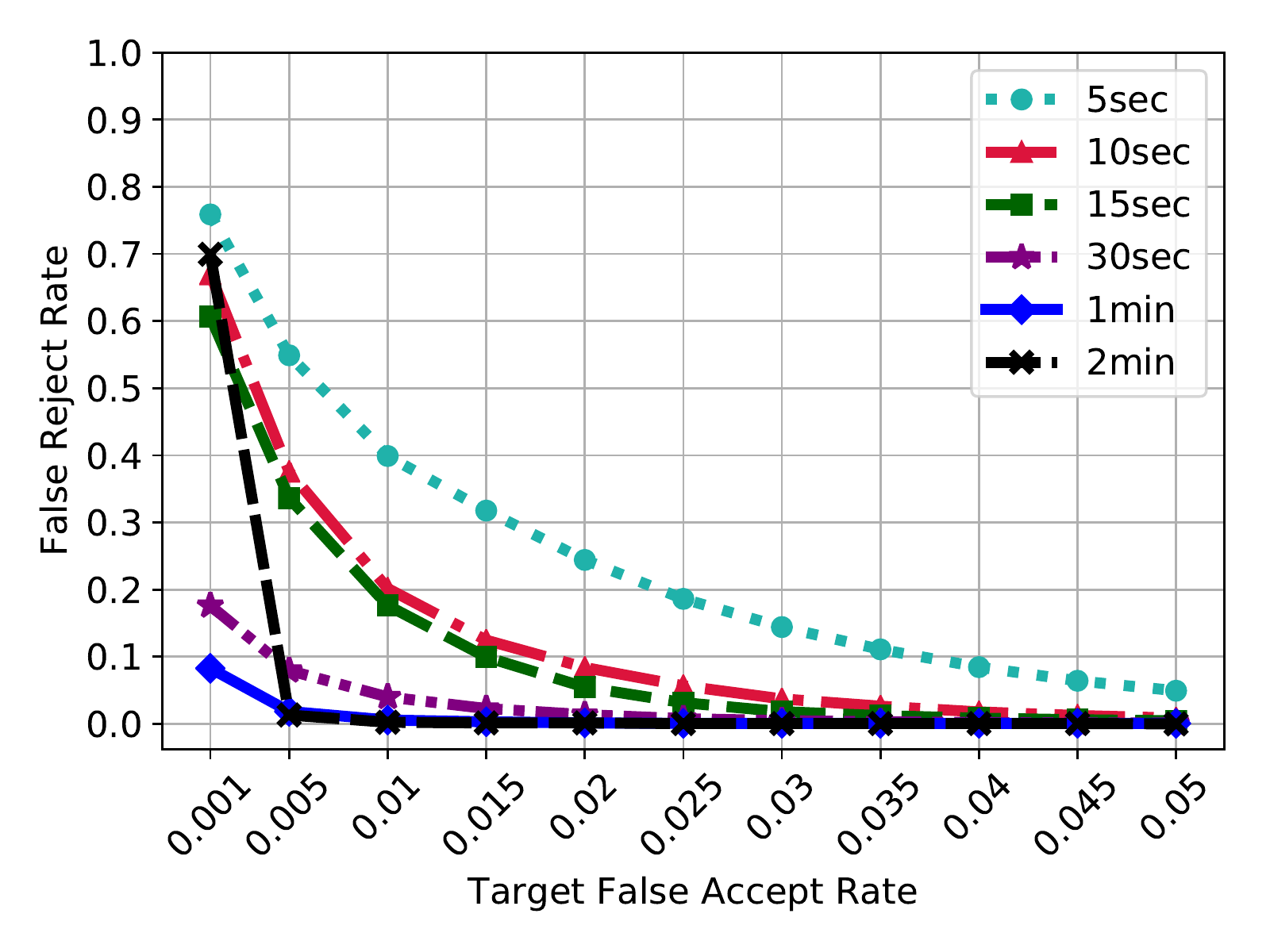}
   	\caption{Car}
   	\label{sf:spf-car-all}
\end{subfigure}
\begin{subfigure}[b]{0.33\textwidth}
	\centering
   	\includegraphics[width=0.99\textwidth]{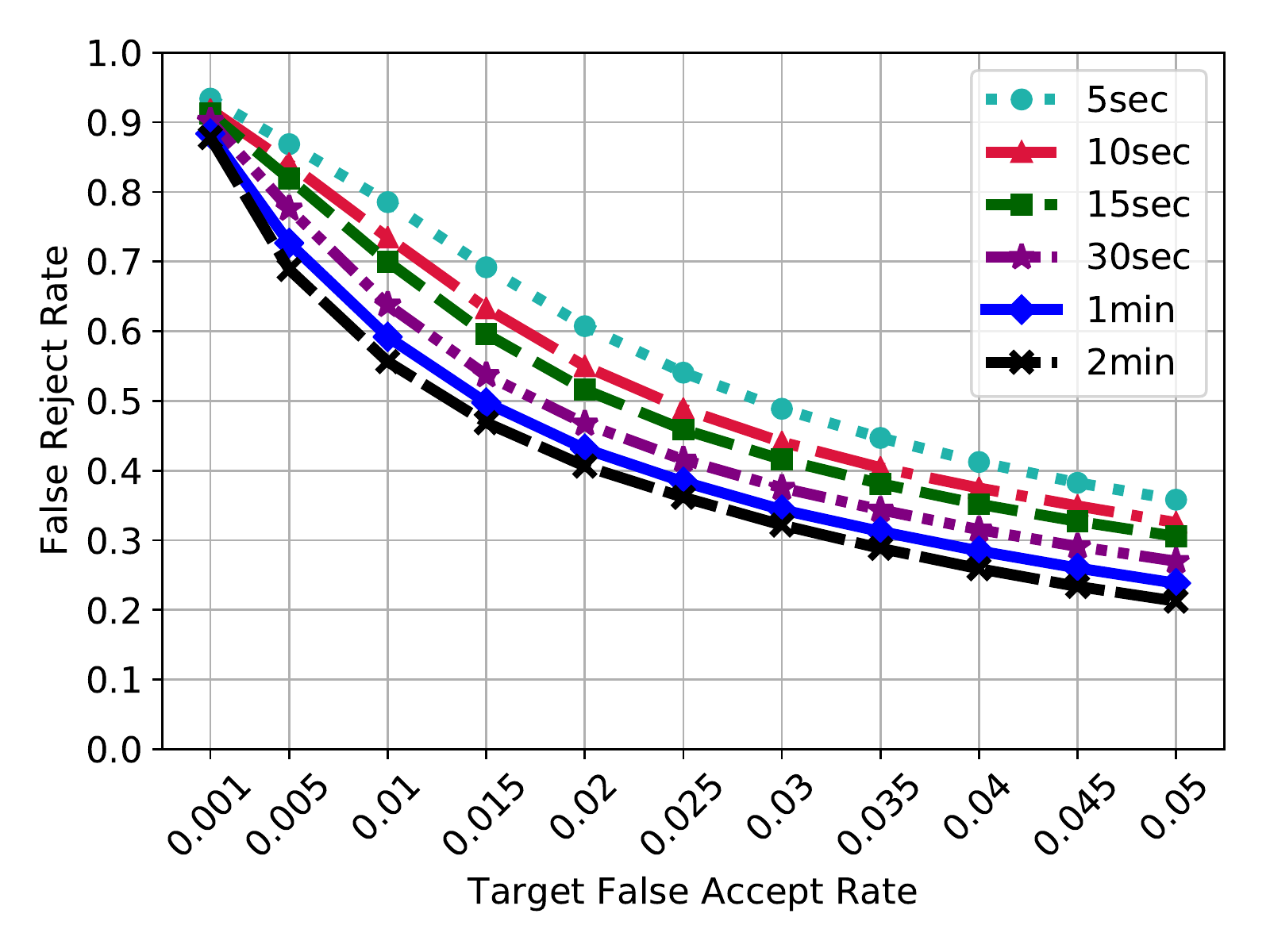}
   	\caption{Office}
    \label{sf:spf-office-all}
\end{subfigure}
\begin{subfigure}[b]{0.33\textwidth}
	\centering
   	\includegraphics[width=0.99\textwidth]{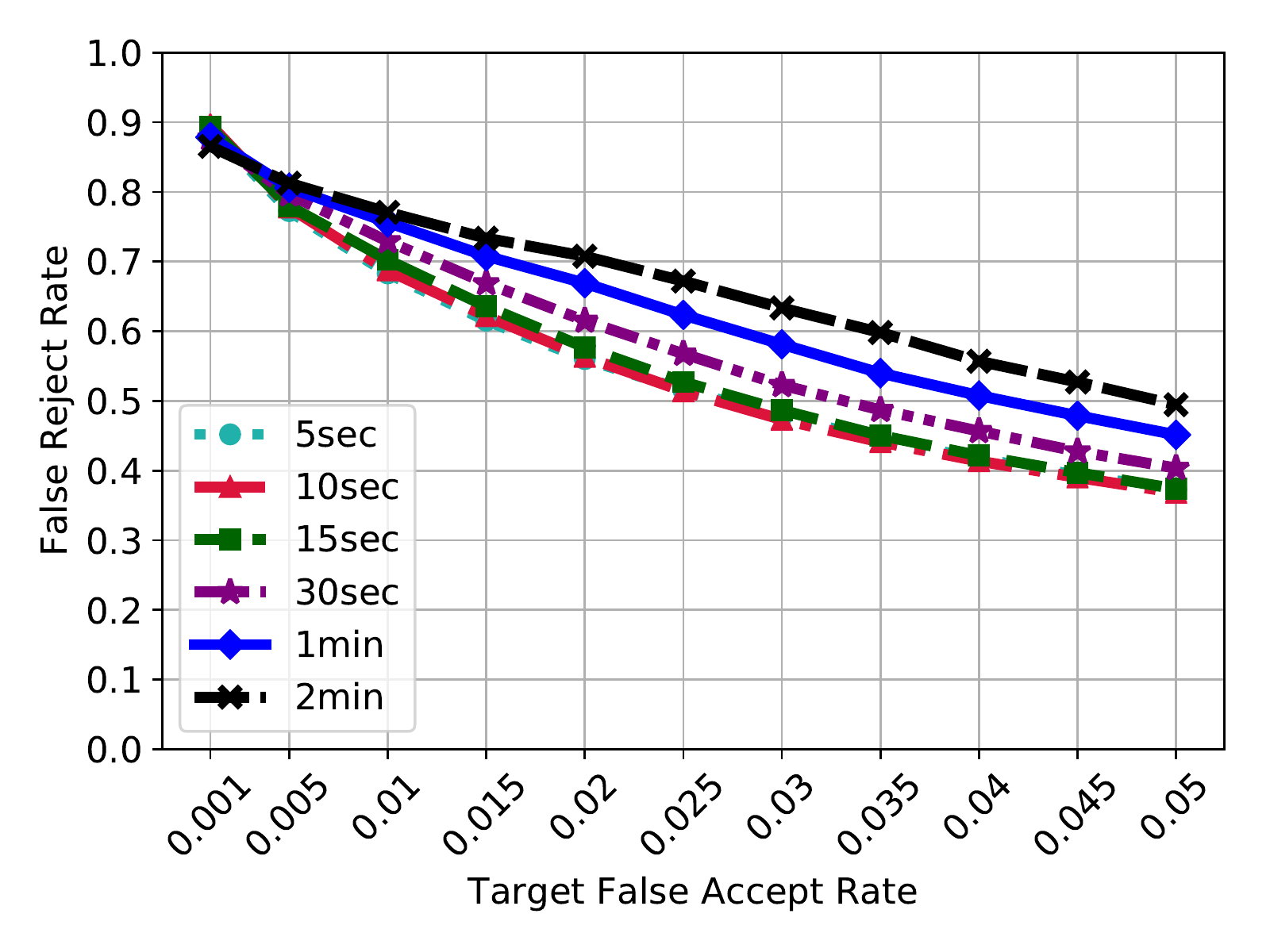}
   	\caption{Mob/het}
   	\label{sf:spf-mobile-all}
\end{subfigure}
\caption{\glspl{frr} with target \glspl{far} for Karapanos \textit{et al.} in the full car, office and mob/het scenarios}
\end{figure}

We observe \glspl{eer} between 0.006 and 0.050, decreasing with rising interval length (cf. \autoref{tab:res:spf:sum}).
To understand this behavior we compute the distributions of colocated and non-colocated similarity scores for each interval. 
Overlaps of these distributions explain the corresponding error rates: in the car scenario, the overlaps range from 1.1\% to 8.5\%.
We observe a clearer separation between colocated and non-colocated similarity scores at longer intervals, caused by a sharper drop of non-colocated similarity scores. 
When targeting low \glspl{far}, the resulting \glspl{frr} are below 0.2 on the intervals above $t = 15$, dropping rapidly with a growing \gls{far} (cf. \autoref{sf:spf-car-all}). 


Our octave band analysis shows the profound influence of lower frequencies (below 315 Hz) caused by a running car on the overall similarity score.
This explains the lowest \glspl{eer} reaching 0.0 in the uniform sound environment of a highway (cf. \autoref{tab:res:spf:sum}). 
The more diverse sound environment of a city shows a severalfold increase in \glspl{eer} compared to the highway subscenario. 
Surprisingly, in a low-activity environment of parked cars, the \glspl{eer} are only a few percentage points above the city subscenario. 
Investigating this phenomenon revealed that the power threshold discards up to 90\% of similarity scores in the parked subscenario, retaining only those scores that resulted from intense audio activity. 

Applying office and mob/het \gls{eer} thresholds to the car dataset leads to a marginal increase in error rates below 1 percentage point on the intervals $t = 5$ to $15$ for the office, and on $t = 10, t = 15$ for the mob/het, with other intervals showing severalfold growths in error rates. 
Among subscenarios, we see limited robustness between quiet (parked) and active environments (city and highway) at $t = 120$, as well as when applying city thresholds to the highway dataset for $t  = 60, t = 120$.

\subsubsection{Office}
\label{sec:res:spf:office}
In the office scenario, we observe \glspl{eer} between 0.098 and 0.141, decreasing with growing interval length (cf. \autoref{tab:res:spf:sum}).
We attribute these \glspl{eer} to larger overlaps between colocated and non-colocated classes, ranging from 19\% to 28\%. 
We see a clear trend of higher similarity scores between non-colocated devices in adjacent offices (offices 1 and 2 in \autoref{fig:office}). 
Our octave band analysis reveals close resemblance between these scores on lower frequencies below 250 Hz and on higher frequencies above 1250 Hz. 
Thus, both low frequencies penetrating adjacent offices and high frequency sounds like a police siren can increase non-colocated similarity scores. 
When targeting low \glspl{far}, the resulting \glspl{frr} start around 0.9 and never drop below 0.2 (cf. \autoref{sf:spf-office-all}).

We observe that higher audio activity of weekdays results in lower \glspl{eer} on the intervals below $t = 30$. 
However, on longer intervals the \glspl{eer} of low-activity environments (i.e., night and weekend) become lower compared to the weekday. 
Investigating this phenomenon in more detail reveals two reasons for such a behavior. 
First, the power threshold retains a few similarity scores originated from intense audio activity in the night and weekend subscenarios.
Second, in low-activity environments sounds are infrequent, localized and short-term, making them easier to capture on longer intervals by colocated devices and less prone to be leaked to non-colocated devices. 

Applying car and mob/het \gls{eer} thresholds to the office dataset results in a minor increase in error rates below 2 percentage points on the intervals $t = 10, t = 15$ for the car, and on $t = 10$ for the mob/het, with other error rates rising a few extra percentage points.
In subscenarios, we observe robustness only between low-activity environments of night and weekend, showing an increase in error rates below 2 percentage points on all intervals.  

\subsubsection{Mob/het}
\label{sec:res:spf:mobile}
Investigating similarity scores in the mob/het scenario revealed that 75\% to 100\% of the scores generated by smartphones and watches were discarded by the default power threshold (40 dB) in the absence of intense audio activity (e.g., running vacuum robot). 
We adjusted the power thresholds for smartphones and watches to 38 and 35 dB respectively, significantly increasing 
the scheme availability in the cases of medium audio activity (e.g., low-voiced conversation), while still discarding the similarity scores from quiet environments. 

With the new power thresholds, we observe increased \glspl{eer} between 0.157 and 0.183, rising with interval length, reversing the trend seen in the car and office scenarios (cf. \autoref{tab:res:spf:sum}).
Once again, higher \glspl{eer} are explained by larger overlaps between colocated and non-colocated classes, ranging from 33\% to 36\%.
When targeting low \glspl{far}, the resulting \glspl{frr} vary almost linearly between 0.9 and 0.37 (cf. \autoref{sf:spf-mobile-all}). 

We found that microphone diversity and device mobility are likely reasons for the reversed \gls{eer} trend. 
The similarity scores among heterogeneous devices are generally lower, decreasing significantly towards longer intervals.
Our analysis suggests that the main reason for these lowered scores is diverse sensitivity and frequency response of heterogeneous microphones \cite{Kardous2016}. 
We empirically observed that smartwatch microphones are optimized for human voice but rather insensitive to low frequencies, while on smartphones low frequencies cause a lot of noise in recordings, and the USB microphones show the best signal quality on a wide frequency range.  
On longer intervals, device mobility further increases signal variation: the probability of capturing a unique signal (e.g., a keystroke by smartwatch) or wide-band scratching noises (e.g., smartphone rubbing against a pocket) increases.

Applying car and office \gls{eer} thresholds to the mob/het dataset leads to a minor increase in error rates up to 1.5 percentage points on the intervals $t = 10, t = 15$ for the car, and $t = 10$ for the office, with other intervals showing several percentage points extra growths in error rates. 

\subsubsection{Conclusion}
Our results show that the scheme by Karapanos \textit{et al.} can reliably distinguish colocated and non-colocated devices in the car scenario, but degrades in performance in the office and mob/het. 
We generally achieve higher \glspl{eer} compared to the authors, who observe an \gls{eer} of 0.002.  
Possible reasons for that are the increased distance between colocated devices and sustained closeness of non-colocated devices in our scenarios. 

When the scheme is used among homogeneous devices (car and office scenarios) we observe better performance with increasing interval length and more intense audio activity. 
The difference between car and office \glspl{eer} is due to a smaller distance between colocated devices in the car, and more intense audio activity, especially on lower frequencies (highway).
We see that highway \glspl{eer} decrease marginally towards longer intervals, suggesting the use of short- to medium-sized intervals in active environments, reducing the run-time overhead of the scheme.

With heterogeneous devices (mob/het scenario) using longer intervals decreases the scheme performance, and intense audio activity is only beneficial if heterogeneous microphones can similarly record it (e.g., human voice), otherwise the performance will further decrease, especially on longer intervals.  
Considering that built-in microphones in mobile devices are user-interaction oriented, the scheme can benefit from shorter intervals and audio activity in the frequency range of human voice in heterogeneous settings. 

The power threshold allows the scheme to cope with quiet environments, sometimes at the price of excluding a significant portion of the dataset (e.g., parked car), trading off availability for security.
However, as we have seen in the mob/het scenario, the power threshold proposed by the authors severely decreases scheme availability already in the cases of medium audio activity, urging the need to carefully select this parameter, depending on the characteristics of the microphones. 

The scheme consistently shows robustness on medium-sized intervals ($t = 10, t = 15$) among our scenarios, suggesting that it can potentially adapt to new environments on these intervals.

\subsection{Sch{\"u}rmann and Sigg}
\label{sec:res:schurmann}
Sch{\"u}rmann and Sigg \cite{Schurmann:2013} propose encoding a snippet of ambient audio into a binary fingerprint to pair two devices.
The generated fingerprint consists of 16 individual shorter fingerprints that reflect the energy changes of successive frequency bands in the audio snippet over shorter timeframes.
The similarity between the fingerprints derived by two devices informs a pairing decision.
These fingerprints need to exhibit good randomness in order to secure a key establishment procedure between devices via fuzzy commitments.
The authors evaluated their scheme in a series of deployments, ranging from staged lab measurements to recordings in a busy canteen and near a road.
A detailed description of the scheme can be found in Appendix \ref{subsec:appx1-schurmann}.

\begin{table}
\centering
	\caption{\gls{eer}$^*$ summary for Sch{\"u}rmann and Sigg}
	\label{tab:res-schuermann}
	\begin{tabular}{c|cccc|cccc|c}
		\toprule
		&
		\multicolumn{4}{c|}{Car} &
		\multicolumn{4}{c|}{Office} &
		Mob/het \\
		$t$ & Full & City & Highway & Parked & Full & Night & Weekday & Weekend & Full \\
		\midrule
		5 & 0.271 & 0.228 & 0.247 & 0.362 & 0.419 & 0.423 & 0.406 & 0.440 & 0.363 \\   
		10 & 0.226 & 0.175 & 0.199 & 0.359 & 0.351 & 0.365 & 0.319 & 0.380 & 0.257 \\  
		15 & 0.211  & 0.157 & 0.170 & 0.361 & 0.317 & 0.340 & 0.267 & 0.347 & 0.215 \\ 
		30 & 0.179 & 0.121 & 0.126 & 0.361 & 0.277 & 0.308 & 0.215 & 0.309 & 0.175 \\  
		60 & 0.160 & 0.100 & 0.106 & 0.359 & 0.256 & 0.287 & 0.194 & 0.280 & 0.154 \\  
		120 & 0.154 & 0.096 & 0.112 & 0.328 & 0.241 & 0.275 & 0.178 & 0.253 & 0.140 \\ 
		\bottomrule
	\end{tabular}
\end{table}

\subsubsection{Methodology}
We evaluate the performance of the scheme by generating fingerprints using different intervals $t$.
Due to hardware constraints, we use a lower audio sampling rate, which reduces the length of the fingerprint from 512 to 496 bits.
This change introduces a marginal deviation from the original implementation as detailed in Appendix \ref{subsec:appx1-schurmann}.
To evaluate the similarity of the generated fingerprints of two devices, we calculate the similarity percentage as $1-(hamming\_dist / length)$.

The scheme uses a fixed similarity threshold that distinguishes colocated from non-colocated devices. 
In addition, we investigate the randomness of the fingerprints by interpreting them as random walks, with 1- and 0-bits representing steps in the positive and negative direction \cite{Brusch:2018}.
The outcomes will follow a binomial distribution if the fingerprints are uniformly random.
We also investigate bit transition probabilities by interpreting each bit of the fingerprint as a state in a Markov chain.

\begin{figure}
\centering
\begin{subfigure}[b]{0.33\textwidth}
	\centering
   	\includegraphics[width=0.99\textwidth]{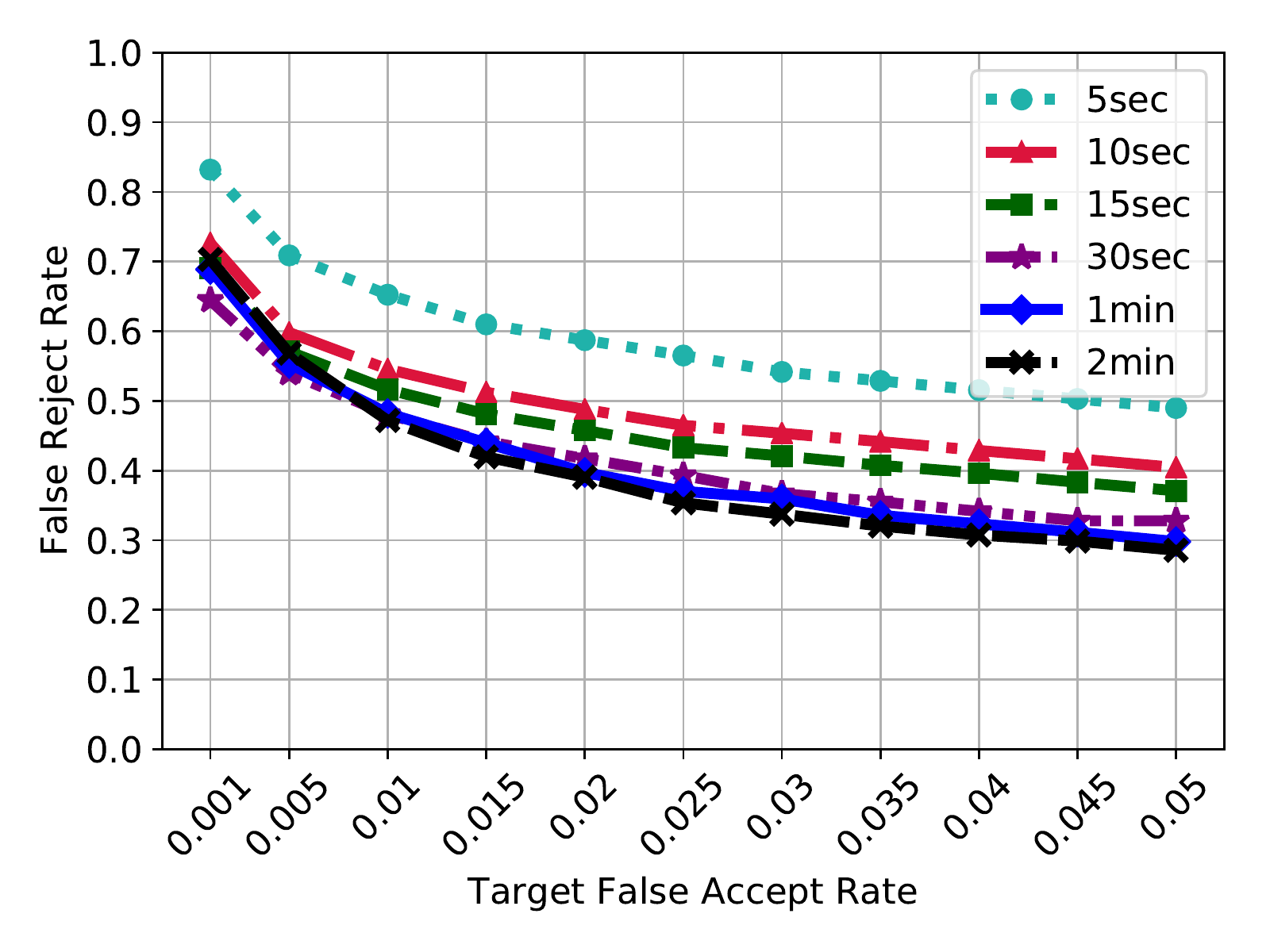}
   	\caption{Car}
   	\label{sf:afp-car-all}
\end{subfigure}
\begin{subfigure}[b]{0.33\textwidth}
	\centering
   	\includegraphics[width=0.99\textwidth]{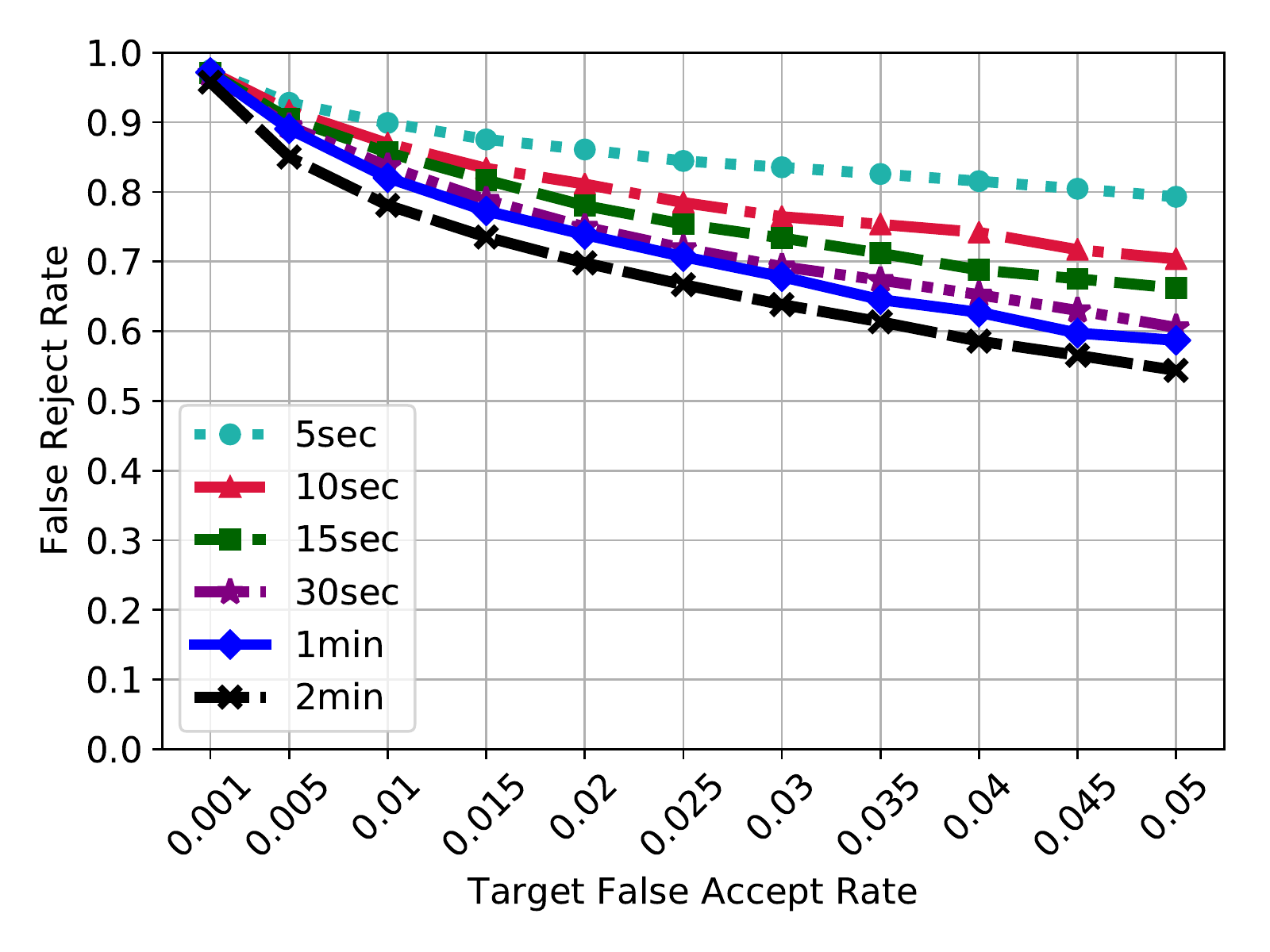}
   	\caption{Office}
    \label{sf:afp-office-all}
\end{subfigure}
\begin{subfigure}[b]{0.33\textwidth}
	\centering
   	\includegraphics[width=0.99\textwidth]{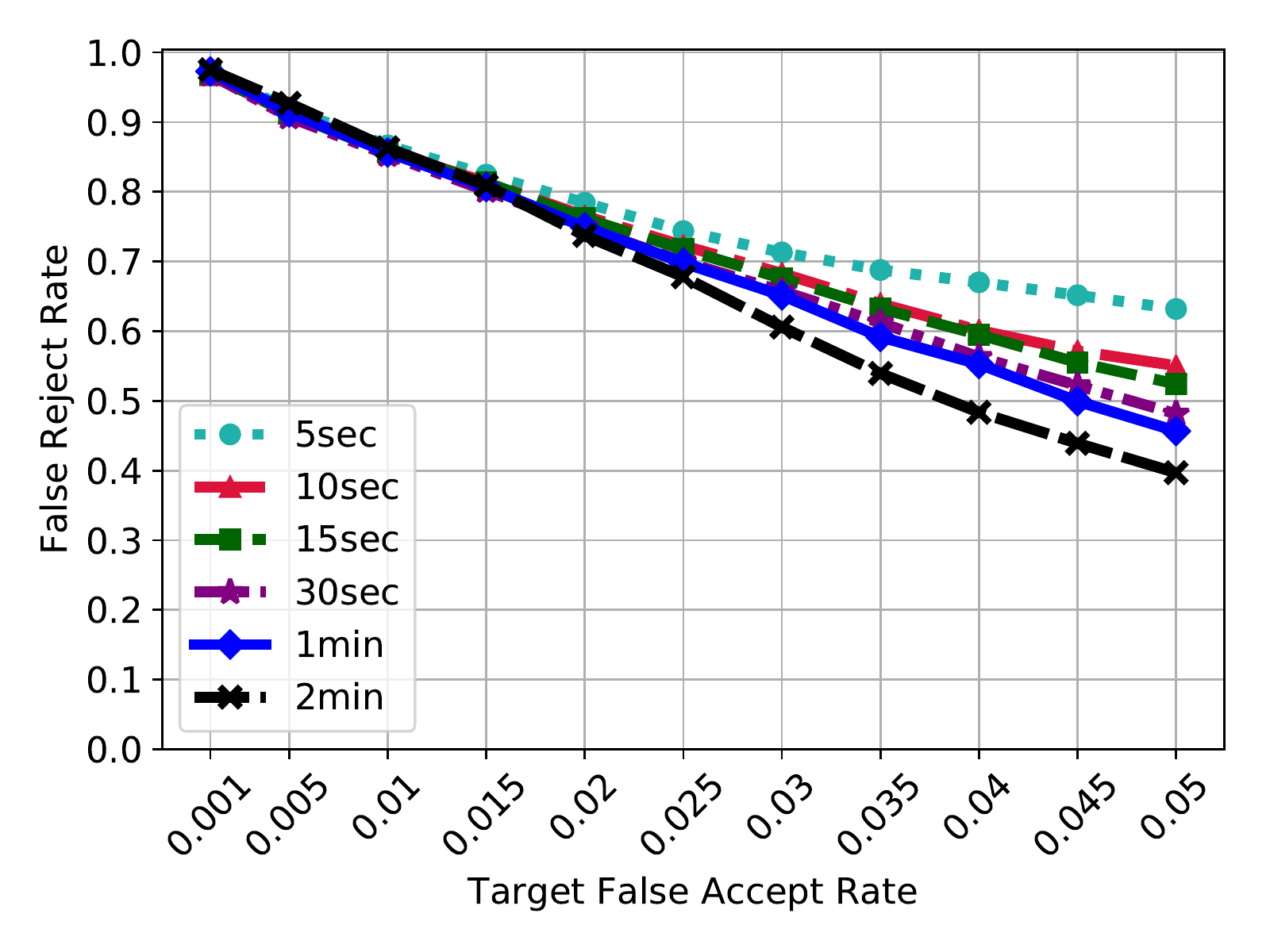}
   	\caption{Mob/het}
   	\label{sf:afp-mobile-all}
\end{subfigure}
\caption{\glspl{frr} with target \glspl{far} for Sch{\"u}rmann and Sigg in the full car, office and mob/het scenarios}
\end{figure}

\subsubsection{Car}
We observe \glspl{eer} between 0.154 and 0.271, decreasing with increasing interval length $t$ (cf. \autoref{tab:res-schuermann}).
These error rates correspond to the observed overlaps in similarity between colocated and non-colocated devices, which ranges between 30\% and 51\%.
When optimizing for a low \gls{far}, the resulting \glspl{frr} exceed 0.8 for certain parameters and never drop significantly below 0.3 (cf. \autoref{sf:afp-car-all}).
The system performs best in scenarios with diverse sound environments, like driving within city limits, showing consistently lower \glspl{eer} in the city subscenario (cf. \autoref{tab:res-schuermann}), dropping as low as 0.096.
Environments with a uniform sound environment, like driving on the highway, show slightly increased error rates, but still remain consistently below the error rates for the full dataset.
In low-activity environments like parked cars, the scheme shows significantly increased error rates of up to 0.362---an increase of 0.134 over the city environment with the same parameters.

\begin{figure}[t]
	\centering
	\begin{minipage}[b]{.47\textwidth}
		\centering
		\subcaptionbox{Full fingerprint, Device A\label{sf:afp-s6-full}}{%
			\includegraphics[width=0.49\textwidth]{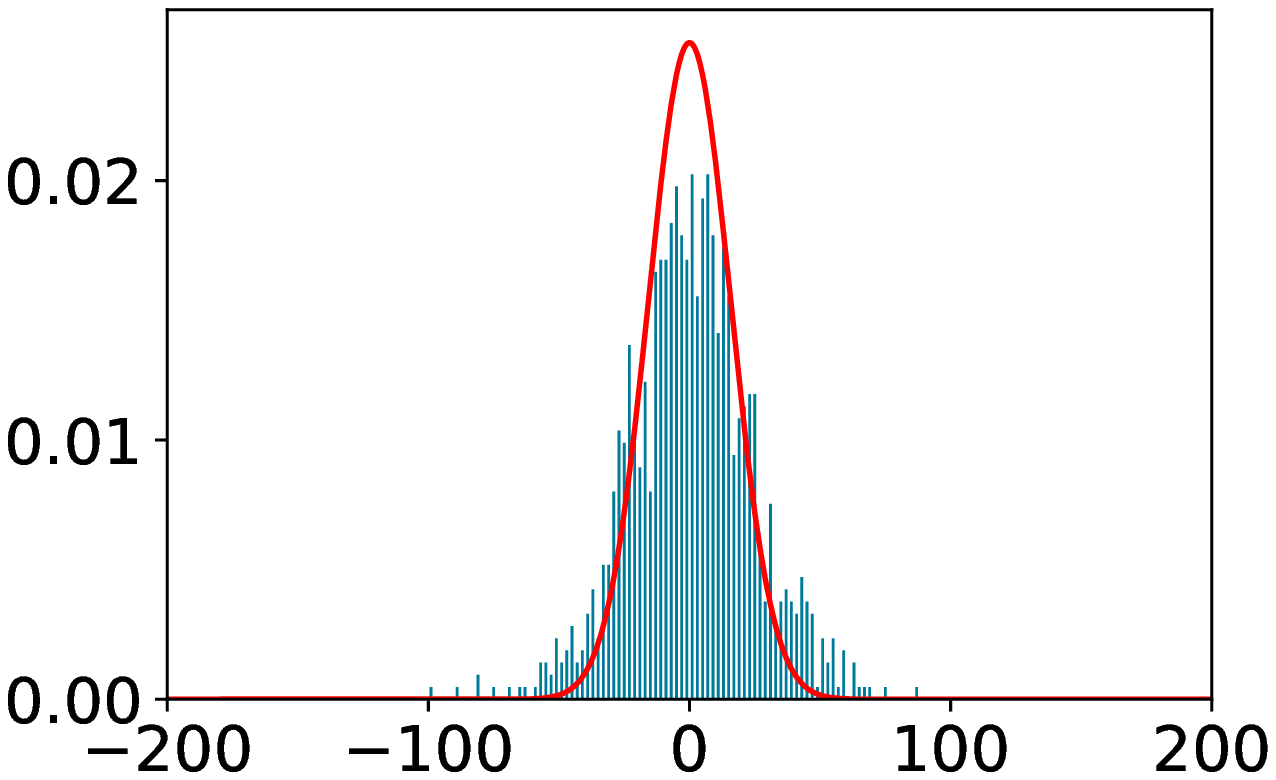}%
		}
		\subcaptionbox{Full fingerprint, Device B\label{sf:afp-s11-full}}{%
			\includegraphics[width=0.49\textwidth]{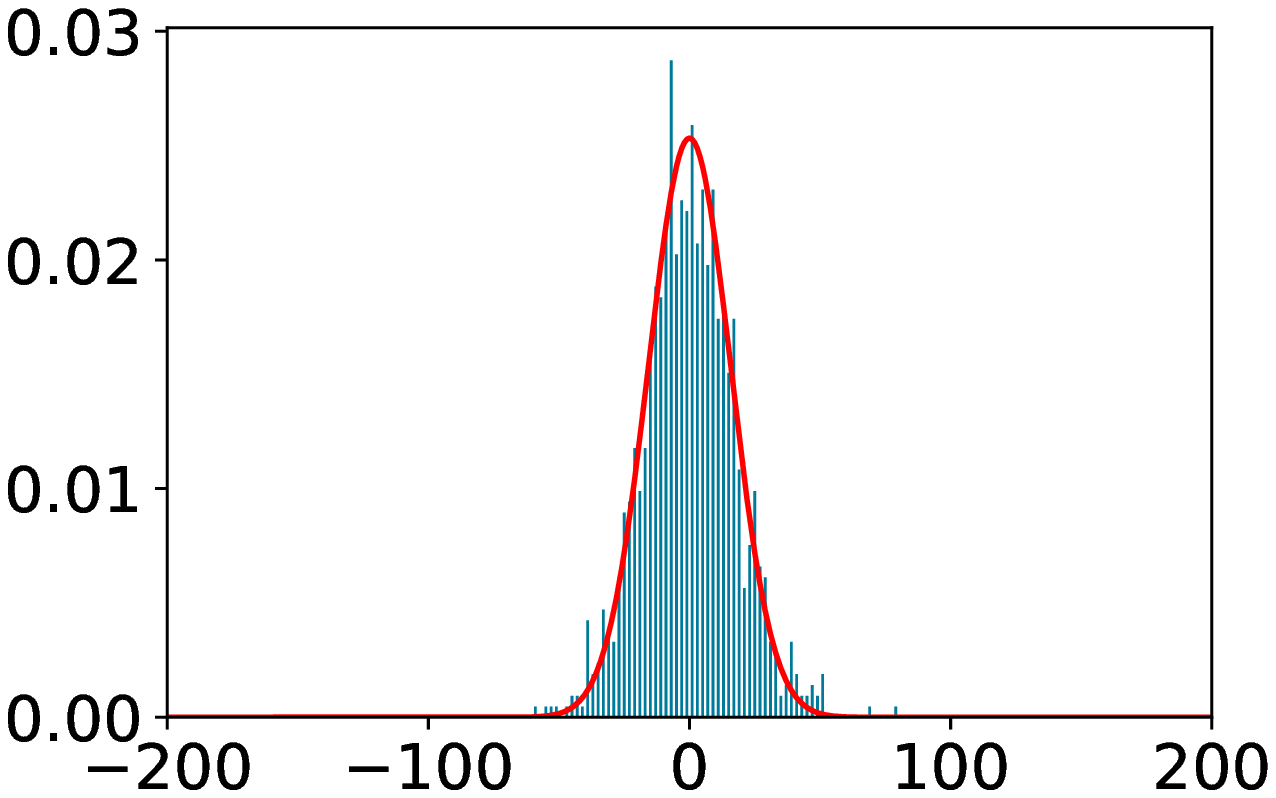}%
		}
	
		\subcaptionbox{Sub-fingerprint, Device A\label{sf:afp-s6-sub}}{%
			\includegraphics[width=0.49\textwidth]{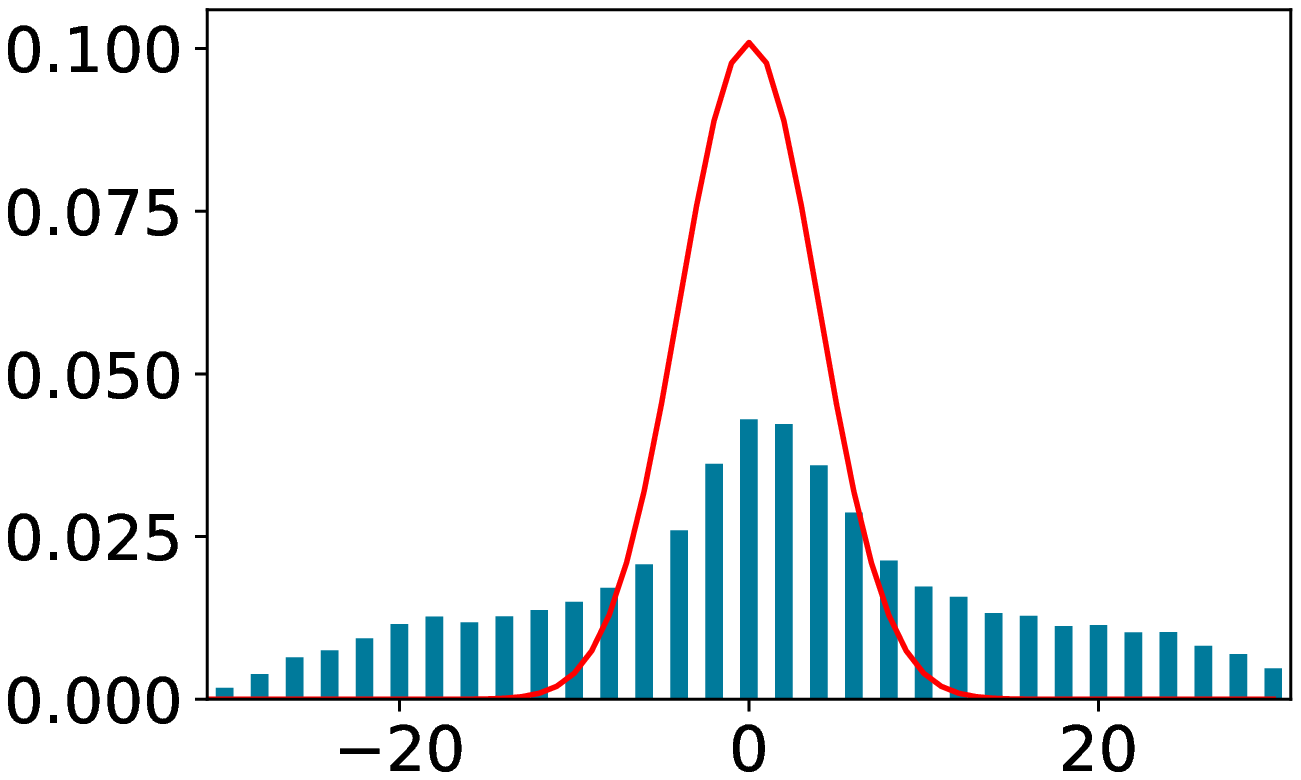}%
		}
		\subcaptionbox{Sub-fingerprint, Device B\label{sf:afp-s11-sub}}{%
			\includegraphics[width=0.49\textwidth]{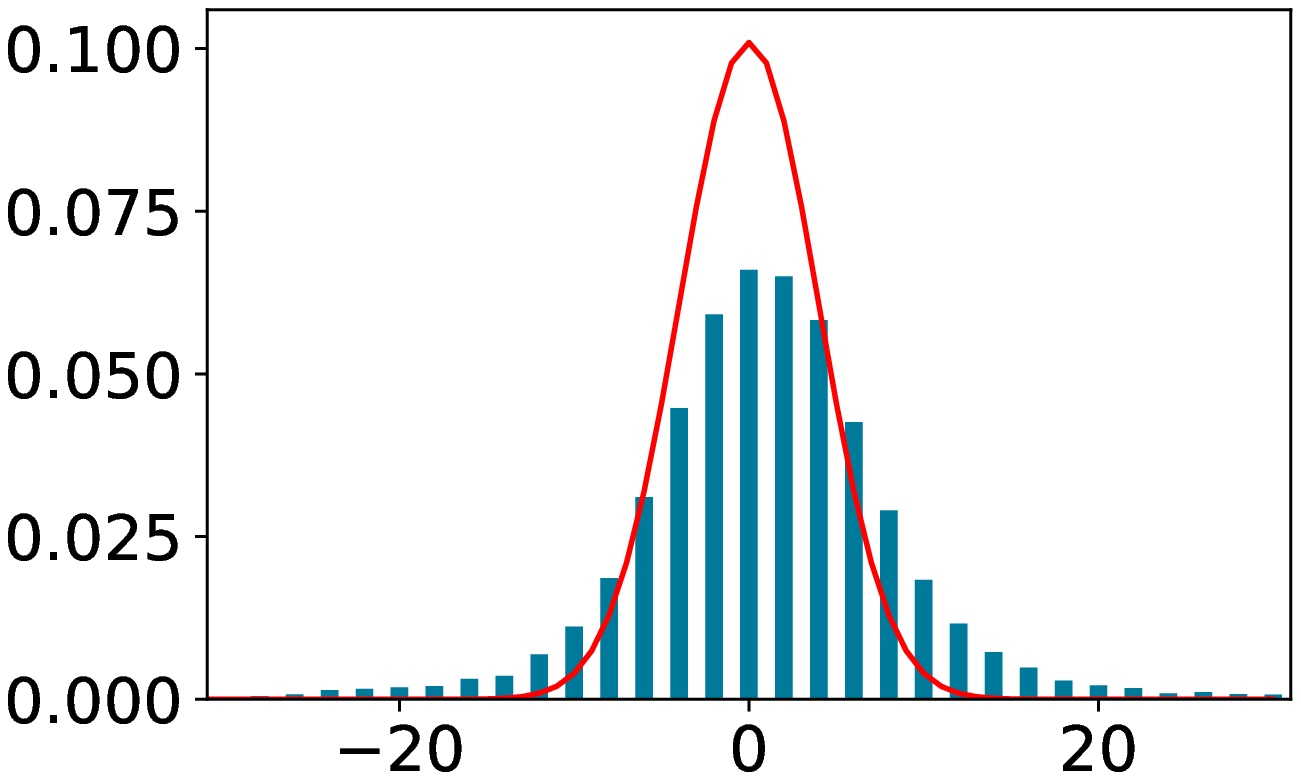}%
		}
		\caption{Distribution of fingerprint random walks for Sch{\"u}rmann and Sigg, Car scenario, $t=10$. Expected binomial distribution in red.}
		\label{fig:res-afp-randomness}
  	\end{minipage}\hfill
	\begin{minipage}[b]{.47\textwidth}
		\centering
		\subcaptionbox{Device A, $t=5$}{%
			\includegraphics[width=0.49\textwidth]{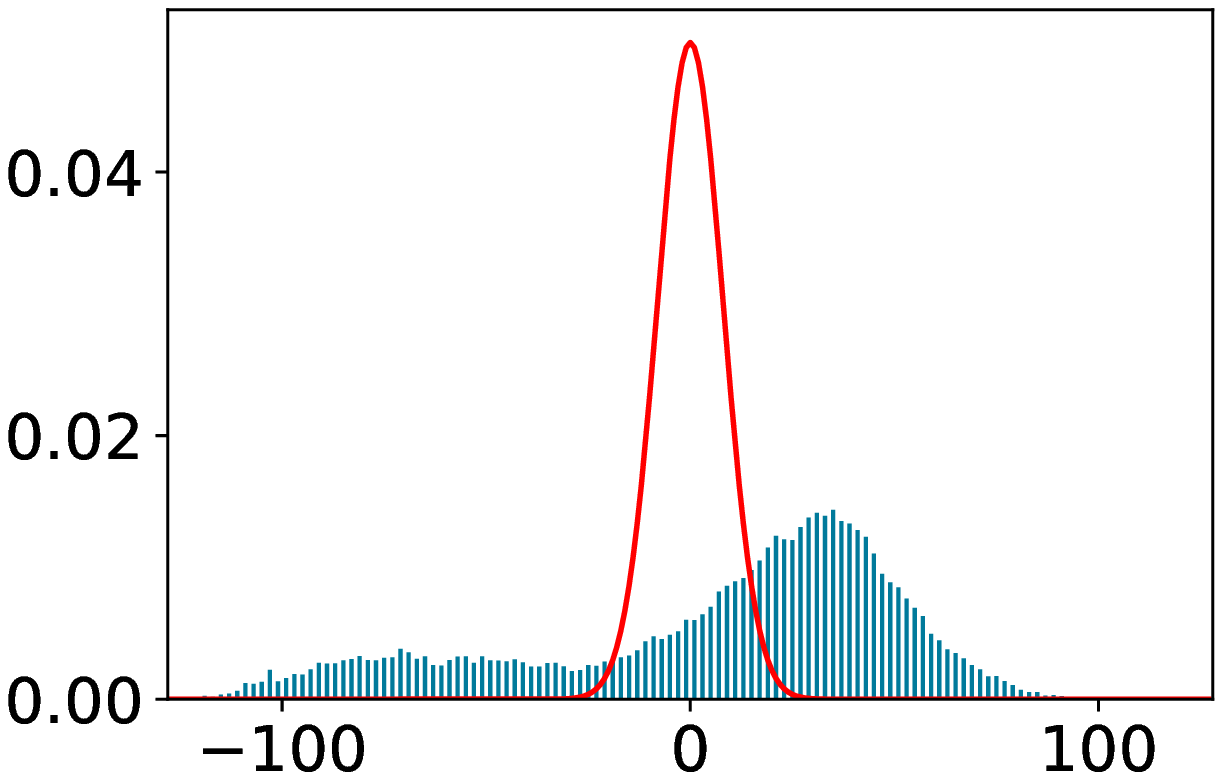}%
		}	
		\subcaptionbox{Device B, $t=5$}{%
			\includegraphics[width=0.49\textwidth]{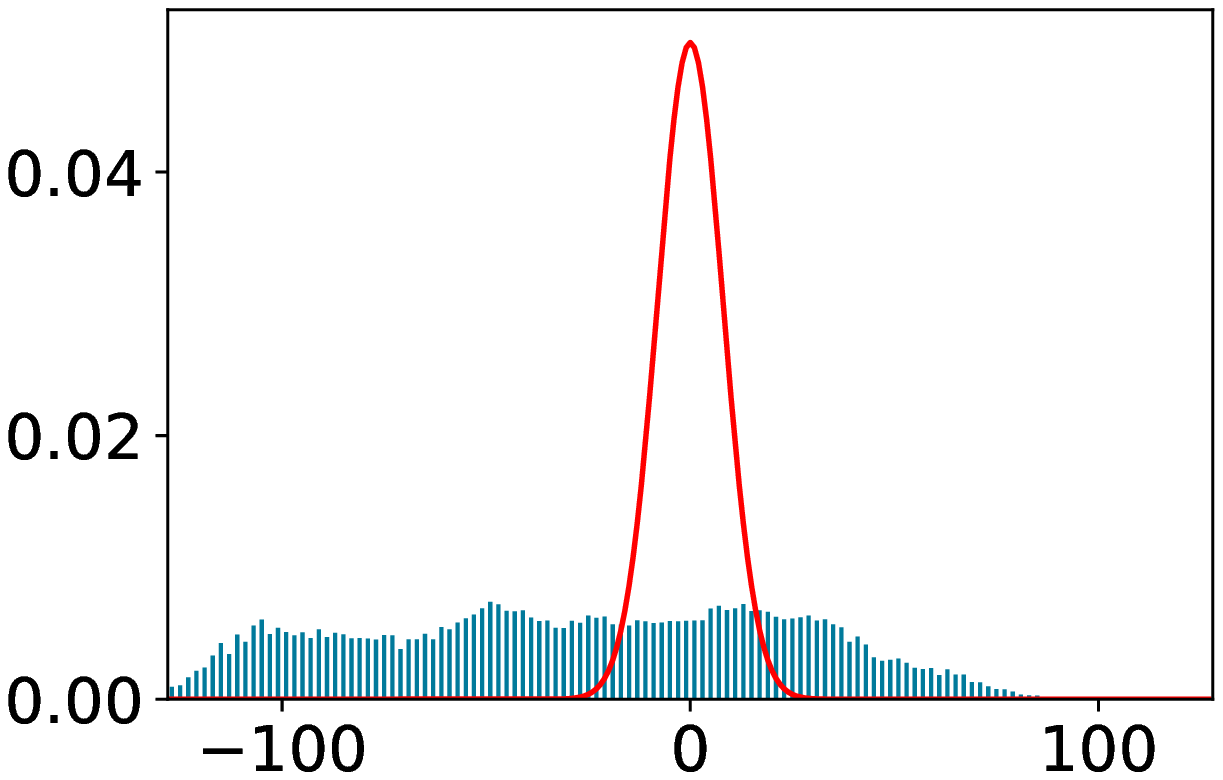}%
		}
	
		\subcaptionbox{Device A, $t=120$}{%
			\includegraphics[width=0.49\textwidth]{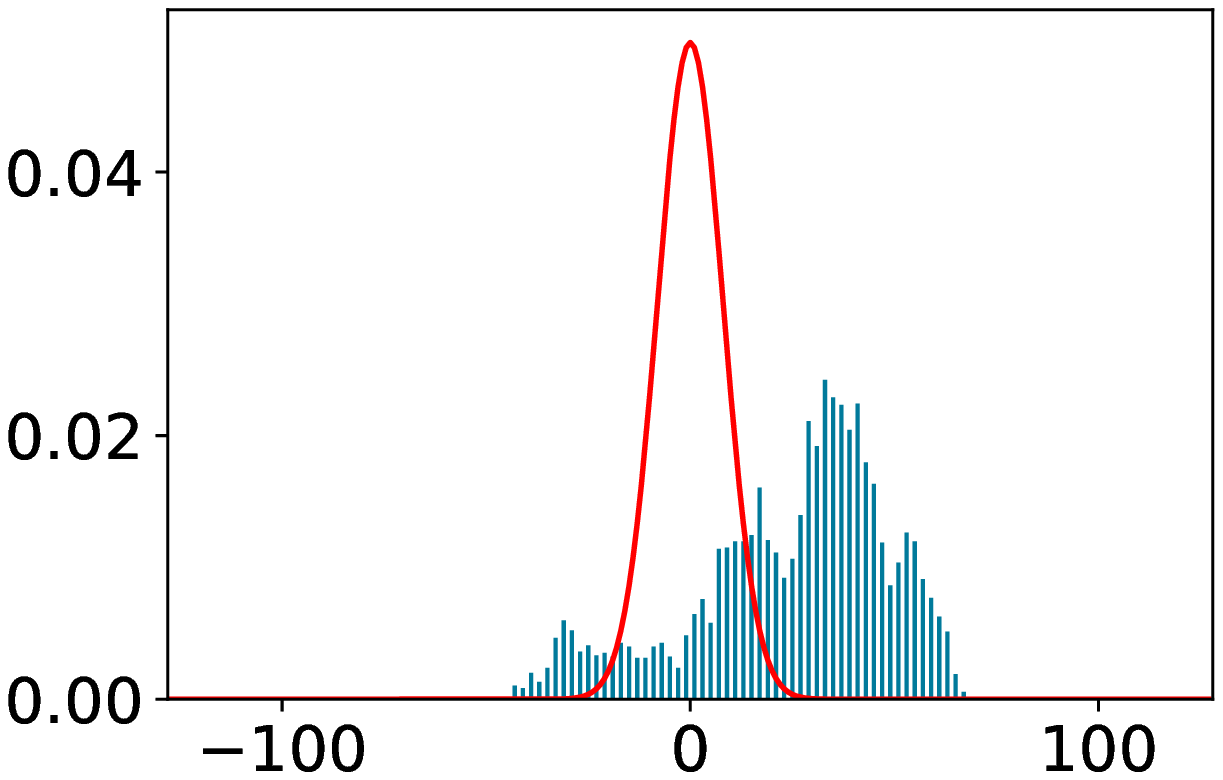}%
		}
		\subcaptionbox{Device B, $t=120$}{%
			\includegraphics[width=0.49\textwidth]{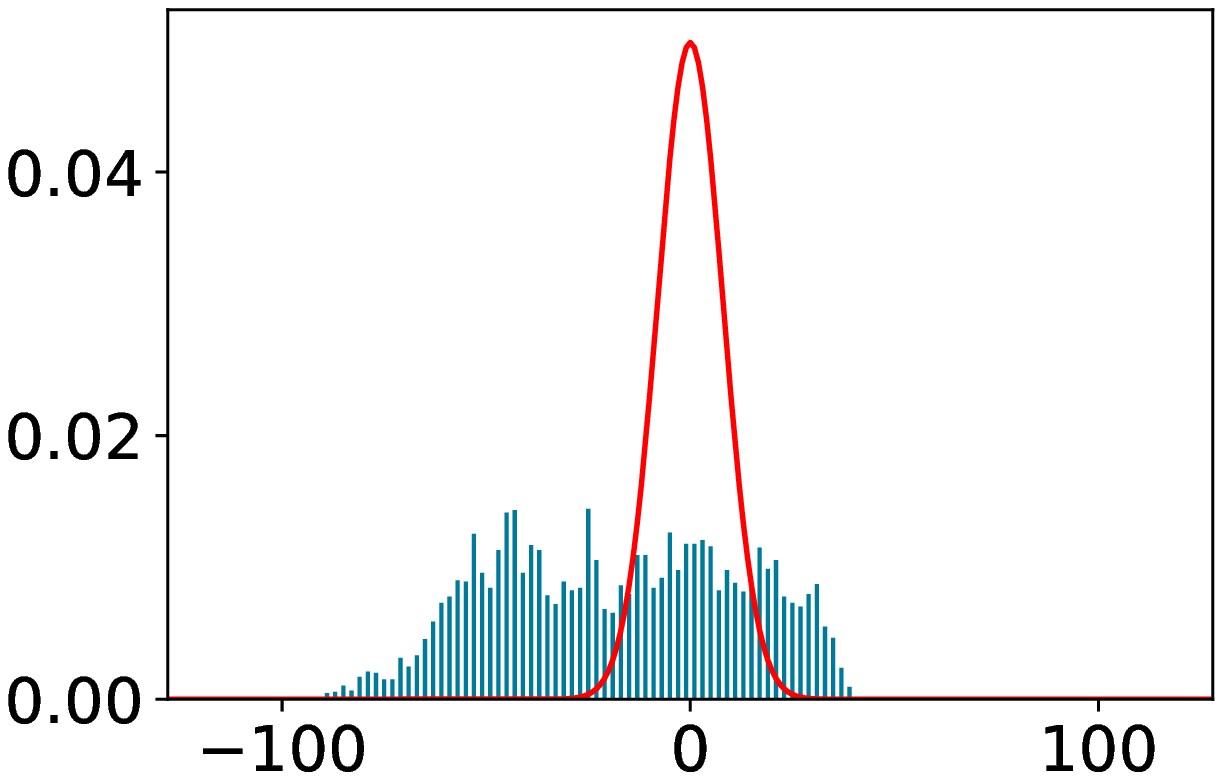}%
		}
		\caption{Distribution of fingerprint random walks of two representative devices for audio feature of Miettinen \textit{et al.}, Office, $b=128$. Expected binomial distribution in red.}
		\label{fig:res-miet-randomness}	
	\end{minipage}
\end{figure}

The fingerprints exhibit good randomness across all devices.
Their Markov property is good, with $P(b=1)\approx0.5$ for all bits.
When interpreting fingerprints as random walks, the resulting distribution of endpoints is close to the expected binomial distribution (cf. \autoref{fig:res-afp-randomness}).
When splitting the 496-bit fingerprints into their constituent 31-bit fingerprints and analyzing them separately, the random walks show a more varied distribution.
Some are close to the expected binomial distribution (cf. \autoref{sf:afp-s11-sub}), while others show a flatter distribution (cf. \autoref{sf:afp-s6-sub}), indicating more fingerprints contain a larger number of 1- or 0-bits than expected.
Investigating these sensors in more detail, we found that their microphones were affixed to surfaces that vibrated more than average.
As fingerprints are derived from variations in signal energy over time, the biased fingerprints may have been caused by periodic variations in the energy induced by the vibrations.

Applying the threshold from the office scenario to this dataset results in an increase in error rates between 3.6 and 11.2 percentage points, with the larger changes occurring for $t=5$ and 120.
The mob/het threshold increases the error rates by 1.7 to 7.1 percentage points, with the largest changes for $t=15$ and 30, while $t=120$ shows the smallest change.
In subscenarios, the most stable results are obtained between city and highway, changing between 4.1 and 9.5 percentage points in both directions. 
The other combinations show significantly larger error rate increases, in some cases up to 25.7 percentage points.
This indicates that the scheme has limited robustness in cases where the environments are similar, but is not robust to larger changes in environmental characteristics.

\subsubsection{Office}
In the office, we observe generally increased \glspl{eer}, ranging from 0.241 to 0.419 and decreasing with increasing interval lengths (cf. \autoref{tab:res-schuermann}).
These error rates are explained by the higher overlaps between colocated and non-colocated classes, which lie between 48\% and 79\%.
In particular, we observe that the computed similarities between some non-colocated devices exceeded the similarities with all of their respective colocated devices, especially using smaller interval sizes $t$.
Investigating these anomalous pairs in more detail revealed that the high similarities occur mostly at night and on the weekend, i.e., at times of very low ambient activity.
However, the question why these particular devices were affected while others behaved normally remains unanswered.

When optimizing for a low \gls{far}, the resulting \glspl{frr} for the full scenario are universally above 0.5 (cf. \autoref{sf:afp-office-all}).
Once again, the system performs best in environments with high audio activity, in this case the weekdays, showing significantly reduced error rates compared to the night and weekend.

The fingerprints again show good randomness, with a strong Markov property and random walks close to the expected distribution for the full fingerprints.
When investigating the sub-fingerprints, we observe a slight bias towards 0 in the lowest three bits of some devices, with $P(b=1)\approx 0.48$.
Most of the affected devices were located in office 2, but there is no discernible pattern in which devices exhibit this behavior and no obvious explanation.

Applying the car threshold to this dataset results in error rate increases of 3.9 to 10.9 percentage points, with the largest changes at $t=5$ and 120.
Conversely, the threshold obtained in the mob/het scenario will increase error rates by 5.6 to 12.6 percentage points, with the largest changes at $t=10$, 15 and 30.
The error rates of the night and weekend subscenarios remain almost completely stable when exchanging their thresholds.
All other combinations show larger changes, often showing swings of more than 10 percentage points.

\subsubsection{Mob/het}
The error rates in the mob/het scenario exhibit a larger spread than in the other scenarios, with \glspl{eer} ranging from 0.140 to 0.363 and decreasing with rising interval lengths (cf. \autoref{tab:res-schuermann}).
They once again correlate with the overlaps in similarity between colocated and non-colocated devices, which ranges from 27\% to 62\%.
When optimizing for low \glspl{far}, the resulting \glspl{frr} range from close to 1.0 to 0.40 (cf. \autoref{sf:afp-mobile-all}).

Although the Markov property is universally good, the randomness of the fingerprints shows significant variation.
While the fixed sensors show decent randomness, the mobile devices (smartphones and smart watches) deviate from the expected distribution, showing similar behavior to the biased sensors in the car scenario.
Part of this deviation can likely be explained by the different characteristics of the microphones (cf. \autoref{sec:res:spf:mobile}).
Devices that were covered (i.e., smartphones in pockets and smart watches worn under long-sleeved clothing) showed the largest deviation from the expected distribution, with strong biases towards sub-fingerprints consisting of mostly 1- or 0-bits.
This is likely related to the movement of cloth over the devices generating wide-band scratching noises, in combination with sound attenuation caused by the clothing.

Applying the car threshold to this dataset results in increases in error rates between 1.5 and 7.4 percentage points, with the largest changes for $t=15$ and 30.
The office threshold increases error rates by 5.3 to 11.8 percentage points, with the highest increases for $t=15$ and 30.

\subsubsection{Conclusion}
The scheme by Sch{\"u}rmann and Sigg is unable to reliably distinguish colocated from non-colocated devices in our scenarios.
We also observed unexplained high similarities for specific non-colocated device pairs in the office scenario.
In particular, the scheme breaks down in environments with low ambient activity (a limitation also noted by the original authors), however, even in high-activity environments like a driving car, the error rates exceed 10\% for almost all parameters.
Still, it may be possible to increase the overall performance of the scheme by excluding low-energy samples with a power threshold (similar to Karapanos \textit{et al.}).

The fingerprints exhibit good randomness in many cases, however, they struggle with noisy inputs, like vibration- or friction-induced sounds, and will in some cases generate fingerprints that consist almost entirely of 1- or 0-bits.
In particular, devices carried in pockets or under long sleeves seem to cause problems.

While the scheme is robust in some pairs of subscenarios, the robustness is very limited.
Interestingly, the intervals behave differently for different combinations of scenarios---while the error rates of $t=120$ are almost unaffected in some pairs, in others, they show very large changes.
The same is true for other intervals like $t=15$.

Sch{\"u}rmann and Sigg do not report error rates in their evaluation, so a direct comparison is impossible.
However, the average separation between colocated and non-colocated fingerprints they report is larger than that observed in our scenarios.
One possible explanation may be a tighter synchronization of audio signals in their experiment, as their samples were recorded by a single device with two microphones, thus avoiding any problems related to recordings not being exactly in sync.
In a practical setting, such a tight synchronization between two devices will be more challenging to achieve (our synchronization method is described in \autoref{sec:appx1}).

\subsection{Miettinen \textit{et al.}}
\label{sec:res:miettinen}
The scheme proposed by Miettinen \textit{et al.} \cite{Miettinen:2014} uses two context features, one based on audio and the other on luminosity.
In both cases, changes over extended timeframes are recorded and encoded into a binary \textit{context fingerprint} of a fixed length $b$.
The similarity of these fingerprints is then used to decide if devices can establish a connection by serving as a shared secret to bootstrap a key exchange using fuzzy commitments.
Due to this usage, the randomness of the fingerprints is once again of interest.
The authors evaluated their scheme in an office, a home scenario, and a mobile scenario simulating wearable devices.
They also propose an optional extension to ensure sufficient fingerprint quality by discarding fingerprints with an insufficient \textit{surprisal}, which measures how unexpected a fingerprint is for the current time of day.
However, they did not evaluate the effect of this proposal.
More details are given in Appendix \ref{subsec:appx1-miettinen}.

\subsubsection{Methodology}
Our methodology is identical to that used for the paper by Sch{\"u}rmann and Sigg (cf. \autoref{sec:res:schurmann}).
As the fingerprints generated by the scheme span long timeframes (up to 34 hours), we omit the subscenario evaluation, as allocating fingerprints to specific subscenario timeframes is impossible.

\begin{table*}[t]
	\caption{\gls{eer}$^*$ results for Miettinen \textit{et al.}}
	\label{tab:res-miettinen}
	\begin{tabular}{cc|cccccc|cccccc}
		\toprule
		&& \multicolumn{6}{c|}{Audio} & \multicolumn{6}{c}{Luminosity} \\
		& & $t=5$ & 10 & 15 & 30 & 60 & 120 & 5 & 10 & 15 & 30 & 60 & 120 \\
		\midrule
		\parbox[t]{2mm}{\multirow{5}{*}{\rotatebox[origin=c]{90}{Car}}}& $b=64$ & 0.377 & 0.389 & 0.396 & 0.384 & 0.382 & 0.263 & 0.506 & 0.505 & 0.504 & 0.505 & 0.501 & 0.517\\
		& 128 & 0.358 & 0.368 & 0.370 & 0.372 & 0.362 &  & 0.507 & 0.506 & \textbf{0.492} & 0.499 & 0.516 & \\
		& 256 & 0.329 & 0.335 & 0.328 & 0.295 &  &  & 0.505 & 0.499 & 0.498 & 0.514 &  & \\
		& 512 & 0.344 & 0.294 & 0.287 &  &  &  & 0.497 & 0.504 & 0.517 &  &  & \\
		& 1024 & 0.297 & \textbf{0.226} &  &  &  &  & 0.498 & 0.522 &  &  &  & \\
		\midrule
		\parbox[t]{2mm}{\multirow{5}{*}{\rotatebox[origin=c]{90}{Office}}}& 64 & 0.249 & 0.228 & 0.218 & 0.204 & 0.202 & 0.193 & 0.495 & 0.491 & 0.486 & 0.468 & 0.444 & 0.425\\
		& 128 & 0.226 & 0.206 & 0.203 & 0.190 & 0.184 & 0.172 & 0.487 & 0.477 & 0.469 & 0.447 & 0.418 & 0.406\\
		& 256 & 0.212 & 0.196 & 0.193 & 0.180 & 0.165 & 0.147 & 0.483 & 0.470 & 0.459 & 0.421 & 0.397 & 0.403\\
		& 512 & 0.203 & 0.188 & 0.185 & 0.166 & 0.136 & 0.131 & 0.471 & 0.454 & 0.440 & 0.397 & 0.362 & 0.400\\
		& 1024 & 0.197 & 0.184 & 0.178 & 0.135 & \textbf{0.120} & 0.126 & 0.454 & 0.437 & 0.426 & \textbf{0.344} & 0.363 & 0.362\\
		\midrule
		\parbox[t]{2mm}{\multirow{5}{*}{\rotatebox[origin=c]{90}{Mob/het$^\dagger$}}}& 64 & 0.377 & 0.368 & 0.364 & 0.383 & 0.349 & 0.314 & \textbf{0.517} & 0.520 & 0.520 & 0.521 & 0.525 & 0.524\\
		& 128 & 0.356 & 0.344 & 0.339 & 0.371 & 0.325 &  & 0.521 & 0.523 & 0.522 & 0.525 & 0.519 & \\
		& 256 & 0.331 & 0.322 & 0.305 & 0.365 &  &  & 0.521 & 0.528 & 0.520 & 0.524 &  & \\
		& 512 & 0.306 & 0.308 & 0.291 &  &  &  & 0.522 & 0.526 & 0.518 &  &  & \\
		& 1024 & \textbf{0.287} &  &  &  &  &  & 0.525 &  &  &  &  & \\
		\bottomrule
	\end{tabular}
	\smallskip\centering
	\center{Empty cell denotes insufficient data to generate fingerprint. Best value in scenario marked in \textbf{bold}.\\$^\dagger$ Computed on subset.}
\end{table*}

\subsubsection{Car}
\begin{figure}
\centering
\begin{subfigure}[b]{0.33\textwidth}
	\centering
   	\includegraphics[width=0.99\textwidth]{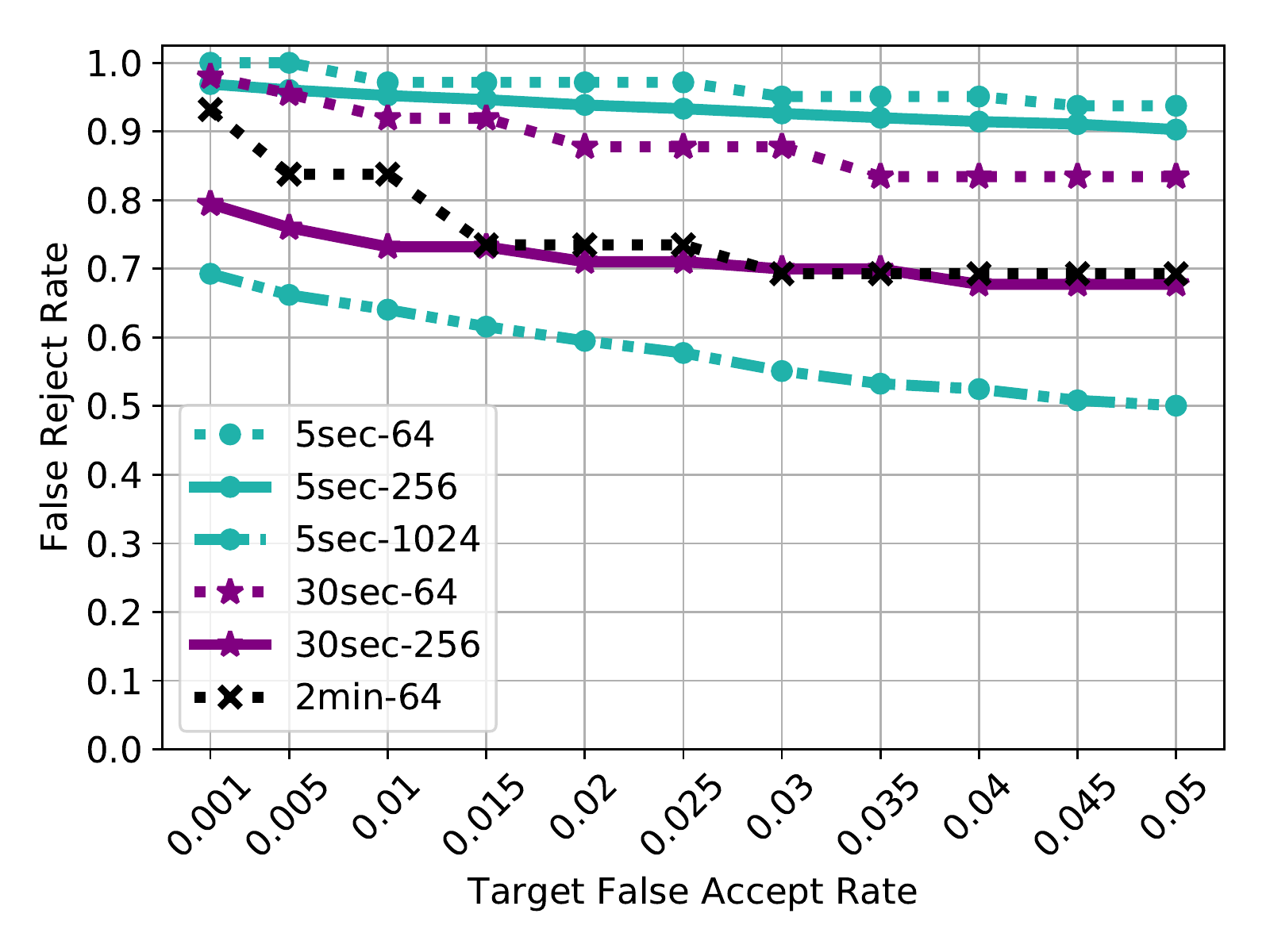}
   	\caption{Audio feature, Car}
   	\label{sf:nfp-car}
\end{subfigure}
\begin{subfigure}[b]{0.33\textwidth}
	\centering
   	\includegraphics[width=0.99\textwidth]{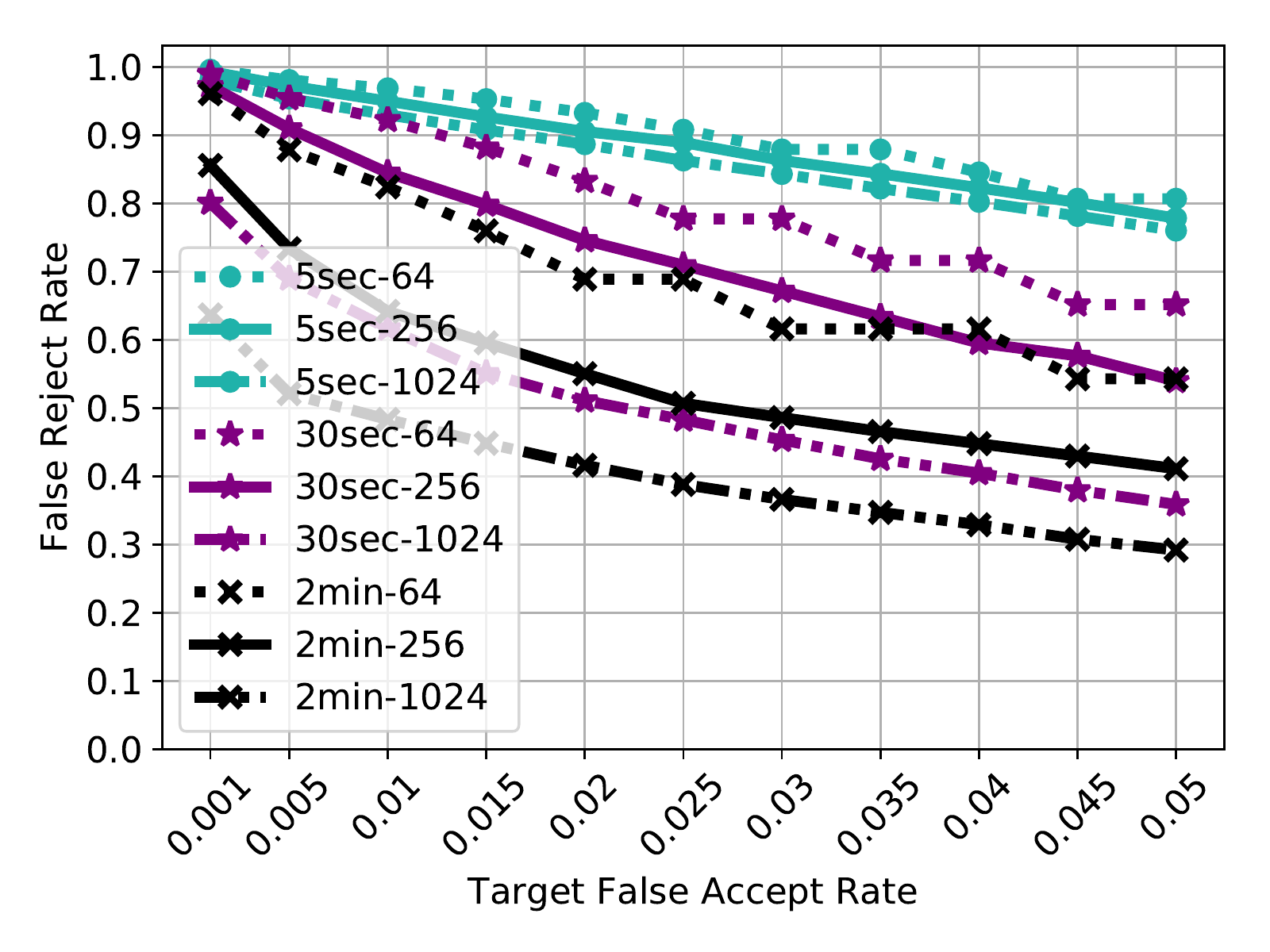}
   	\caption{Audio feature, Office}
    \label{sf:nfp-office}
\end{subfigure}
\begin{subfigure}[b]{0.33\textwidth}
	\centering
   	\includegraphics[width=0.99\textwidth]{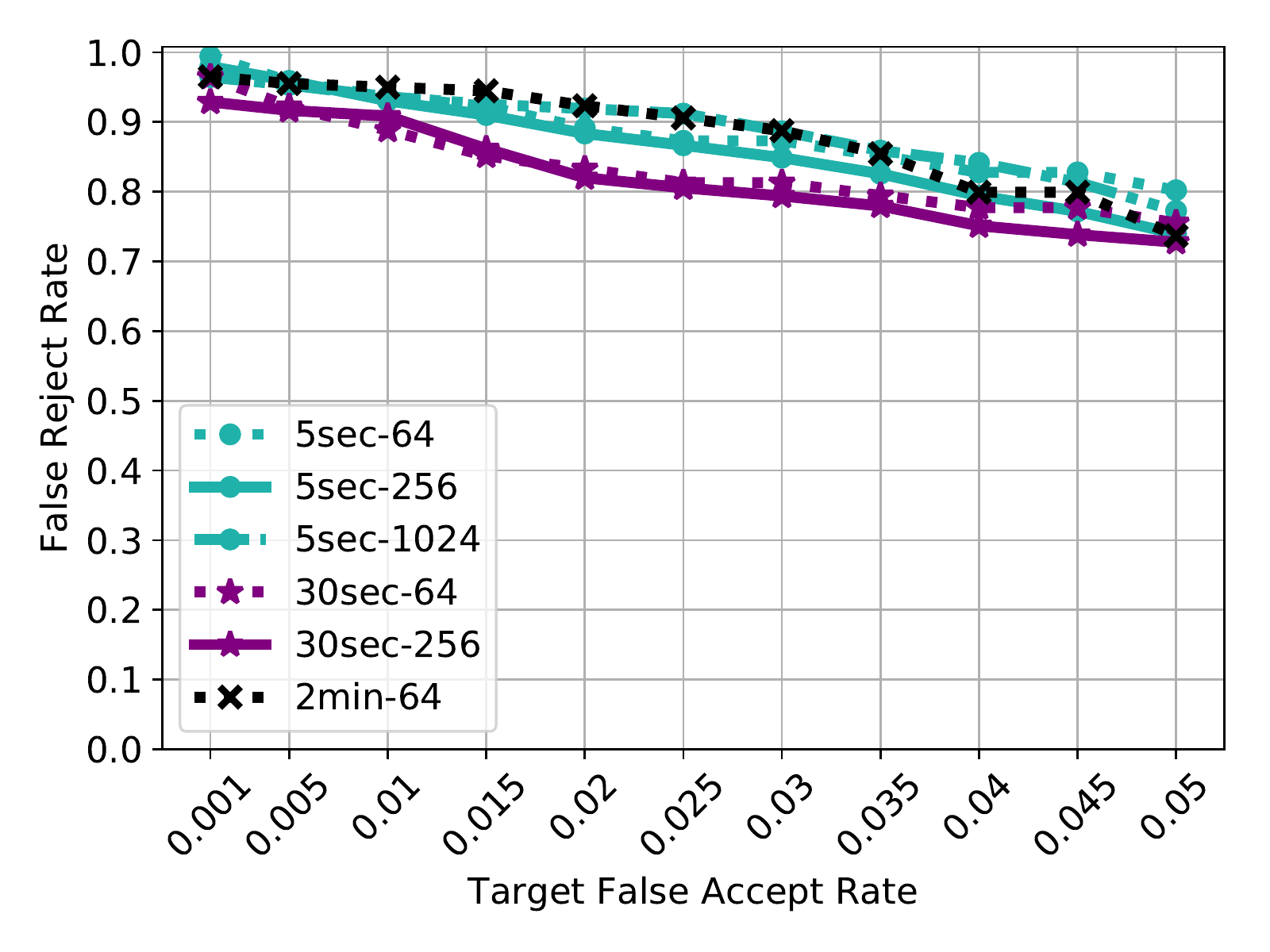}
   	\caption{Audio feature, Mob/het}
   	\label{sf:nfp-mobile}
\end{subfigure}
\caption{FRRs with target FARs for Miettinen \textit{et al.} in the full car, office and mob/het scenarios for a selection of parameters.}
\end{figure}

Both luminosity- and audio-based fingerprints show relatively high error rates, with the lowest observed \gls{eer}$^*$ being 0.492 and 0.226, respectively (cf. \autoref{tab:res-miettinen}).
These high error rates can be explained by the high overlap of similarity percentages between the colocated and non-colocated groups, showing overlaps between 83\% and 96\% for the luminosity fingerprints.
The overlaps are lower, but still significant for the audio fingerprints, with overlaps between 39\% and 79\% being observed.
When aiming for a specific \gls{far}, the resulting \gls{frr} is universally above 0.5 for the audio fingerprint (cf. \autoref{sf:nfp-car}).
For the luminosity feature, the \glspl{frr} are 1.0 for all targeted \glspl{far}, indicating that all samples are rejected, making the scheme usability unacceptable. 

Once again, the security of the scheme does not only depend on the error rates, but also on the randomness of the generated fingerprints.
Here, we observe the luminosity fingerprints to be heavily biased towards zero.
The audio fingerprints contain more 1-bits, but still do not show sufficient randomness.
This limited randomness and high bias also explain the high overlap in the fingerprint similarity distributions.
Rejecting fingerprints with insufficient \textit{surprisal} excluded over 90\% of the luminosity fingerprints even for the smallest specified surprisal value and consistently increased error rates for all attempted thresholds.
For audio fingerprints, we evaluated a series of thresholds for different parameters and found that in many cases, the error rates do not decrease significantly and in some cases will even increase, unless over 95\% of the dataset is excluded.

Applying the threshold from the office scenario increases the error rates for audio fingerprints by varying amounts, in some cases remaining stable, in others increasing by close to 25 percentage points, where higher values of $b$ and $t$ result in higher robustness.
For luminosity fingerprints, increasing $b$ reduces robustness and can lead to all samples being rejected, while smaller values of $b$ with large $t$ sometimes show stable error rates.
With the mob/het threshold, the system rejects all audio fingerprints.
On luminosity fingerprints, it shows unpredictable behavior, being robust for certain parameters and rejecting all samples for others, with no discernible patterns.

\subsubsection{Office}
In the office scenario, we observe lower error rates, with \glspl{eer} between 0.249 and 0.120 for the audio fingerprints (cf. \autoref{tab:res-miettinen}), which can be explained by the decreased overlaps between the fingerprint similarity percentages of colocated and non-colocated devices (between 24\% and 49\%).
For the luminosity fingerprints, the error rates remain high, with the lowest observed \gls{eer} being 0.344, which can be explained by overlaps between 80\% and 99\%.
In many cases, \gls{far} and \gls{frr} only become equal with thresholds close to 100\% similarity, at which point the \gls{far} becomes 0.0 and the \gls{frr} 1.0.
When aiming for a low \gls{far}, the resulting \glspl{frr} remain large for both audio (cf. \autoref{sf:nfp-office}) and luminosity fingerprints (where the error rate is almost universally 1.0).

The luminosity fingerprints consist overwhelmingly of 0-bits, which explains the observed overlaps in similarity percentages.
This can be explained by the low variance in luminosity in offices, which are often lit by electric lighting with only very infrequent changes.
Audio fingerprints show more variance but are usually biased, with the probability of obtaining a 1-bit varying between 0.4 and 0.63.
The distribution within the fingerprint is also unequal, as the fingerprints are almost completely zero at night, leading to further biases (cf. \autoref{fig:res-miet-randomness}).
Rejecting fingerprints with insufficient surprisal once again excluded most of the dataset in the luminosity feature, and only led to improvements of 1-2 percentage points in the audio fingerprints while excluding 10-20\% of the dataset.

Applying the car threshold to the dataset again results in varying increases in the audio fingerprint error rates, following the same trends outlined for this combination in the car section.
The luminosity feature accepts almost all fingerprints, with \glspl{far} between 0.9 and 1.0.
With the threshold from the mob/het scenario, the scheme rejects all audio fingerprints, and either accepts or rejects all luminosity fingerprints, following no discernible pattern.

\subsubsection{Mob/het}
In the mob/het scenario, the colocation of devices changes over time, as they move between offices.
This makes it impossible to perform a comprehensive evaluation of the scheme proposed by Miettinen \textit{et al.}, as the mobile devices often do not stay colocated with any device long enough to establish a pairing.
We thus limit our evaluation to a timeframe of approx. 2.5 hours at the beginning of the recording, during which the colocation of all devices remains static.

The error rates for both luminosity and audio fingerprints are increased compared to the other scenarios, in some cases significantly so (cf. \autoref{tab:res-miettinen}).
The \glspl{eer} of the luminosity fingerprints are above 0.5 for all combinations of parameters, and the best observed \gls{eer} of the audio fingerprint exceeds those of the car and office scenarios by more than four percentage points.
Aiming for a low \gls{far} will result in unacceptably high \glspl{frr} (cf. \autoref{sf:nfp-mobile}).
For audio fingerprints, this decreased performance can be attributed to the varying microphone characteristics leading to sounds being received with different amplitudes, resulting in deviating fingerprints.
The luminosity fingerprints are challenged by the different positions of the mobile devices, which are in some cases carried in pockets and thus do not receive the same luminosity readings as other devices.

Luminosity fingerprints remain heavily biased towards zero, and the audio fingerprints also frequently show strong biases towards 1 or 0, following no discernible dependence on the parameters $t$ and $b$.
Using the surprisal thresholds leads to small improvements (less than 2 percentage points) in the error rates for audio fingerprints, at the cost of excluding 10-20\% of the dataset.
For luminosity fingerprints, even the smallest threshold excludes 96\% of the dataset and does not improve the error rates significantly.

Applying the car threshold to the dataset will result in varying error rates, often rejecting all samples, and never coming close to the original error rates for the audio fingerprint.
The luminosity fingerprints will occasionally reach error rates close to the original, following no particular pattern, but will often reject all fingerprints as well.
The behavior of the office threshold is similar, rejecting close to all samples for fingerprint types.

\subsubsection{Conclusion}
Our evaluation has shown that the scheme is unable to provide good separation between colocated and non-colocated devices, exhibiting large \glspl{far} and \glspl{frr}.
Low \glspl{far} can only be obtained at the cost of large \glspl{frr}.
The best performance is achieved using audio fingerprints in the office scenario, likely because of the homogeneous hardware and low level of background noise.
We also investigate the impact of using the surprisal thresholds proposed by Miettinen \textit{et al.} and find that it will in some cases slightly increase the performance of the scheme but excludes a significant fraction of the dataset in the process, reducing the availability.

The randomness of the generated fingerprints is limited, with devices often showing strong biases towards either 1 or 0, enabling adversaries to break the scheme in a practical deployment by guessing the fingerprint.
This illustrates the importance of using an environmental data source with sufficient variability (unlike fixed electric lighting) and a quantization scheme that ensures a roughly equal proportion of 1- and 0-bits, e.g., \cite{Schurmann:2013}.

Miettinen \textit{et al.} did not compute error rates but observed an average colocated luminosity and audio fingerprint similarity of 95\% and 91.8\%, respectively, using an interval of $t=120$ in their office scenario.
For non-colocated devices, they saw similarities between 68\% and 88\% for luminosity and 62\% to 71\% for audio.
We were unable to achieve this degree of similarity on our dataset.

\subsection{Truong \textit{et al.}}
\label{sec:res:truong}
Truong \textit{et al.} \cite{Truong:2014} propose combining multiple types of context information to increase the reliability and performance of \gls{zia} schemes.
They collect WiFi, Bluetooth, GPS, and audio data and compute a number of context features, aggregated over a time interval $t$.
Features based on the first three modalities are computed based on distances between sets of observed devices and signal strengths, while the audio data is used to calculate the maximum cross-correlation and time-frequency distance between the audio snippets.
Colocation is determined using a machine learning classifier, which has been trained with a labeled dataset of colocated and non-colocated features.
Due to technical limitations of the used hardware, we were unable to capture GPS data.
However, Truong \textit{et al.} found that the GPS feature contained the least amount of discriminative power in their dataset, which was obtained by having volunteers in two cities collect context information and colocation ground-truth data using smartphones and tablets in locations of their choice.
The full details of the scheme are given in \autoref{subsec:appx1-truong}.

\subsubsection{Methodology}
To investigate the performance of machine learning colocation prediction, we use the H2O framework \cite{h2o_Java_software} to train a set of classifiers and pick the best performers.
We evaluated Gradient Boosting Machines (GBMs) \cite{friedman2001greedy} and Random Forests (DRFs) \cite{Breiman2001} as classifiers, and then select the algorithm that gives the best cross-validated performance.
These classifiers perform well in a wide range of datasets \cite{fernandez2014we}, they are fast, and they can handle instances with missing data directly in the model, allowing us to use instances with missing data in our datasets, which would otherwise have to be discarded.
This is desirable, as in the real world, data may be incomplete (e.g., due to missing GPS fixes).
These partial instances still provide information about the generating distribution and therefore are beneficial for the model, as shown by Tang \textit{et al}. \cite{tang2017random}.
When building the cross-validation folds, H2O uses stratified sampling. 
This helps alleviate issues that can arise from class imbalances such as in datasets that contain more non-colocated than colocated instances.

To rank the classifiers, we use 10-fold \gls{cv} and estimate the \gls{auc}, which measures the quality of the predictions irrespective of the selected thresholds. A higher \gls{auc} indicates a more accurately discriminative model. 
Using this measure is valid in this case, as we are interested in lower false accept and false reject errors along the predicting threshold domain.

For the learning, we let H2O split the data into training and validation datasets of 80\% and 20\% respectively. 
H2O will train a set of models independently from each other and automatically perform a parameter search to find optimal parameters for the specific dataset.
Once we have found the top performing models, we get the cross-validated predictions $\hat{y} \in [0,1]$. 
To convert those predictions to actual classes we use a threshold $T$ and classify predictions that satisfy $\hat{y}>T$ as colocated. 
By optimizing the threshold, we balance the values between \gls{far} and \gls{frr} to obtain the \glspl{eer} or our target \glspl{far}.
We also evaluate the impact of the individual features in the process using the normalized relative importance.
The authors evaluated different interval length and came to the conclusion that increasing $t$ above 10 seconds did not significantly increase the performance of the scheme. 
To validate this result, we evaluate two datasets, with $t=10$ and 30.
We present our results in \autoref{sec:res:truong:tab:EERcar} and \ref{sec:res:truong:tab:EERoff}.

\begin{table}
	\centering
	\begin{minipage}[b]{.5\textwidth}
		\centering
		\captionof{table}{Classification results for Truong \textit{et al.}, Car}			\label{sec:res:truong:tab:EERcar}
		\begin{tabular}{lc|cccc}
			\toprule
			Scenario   & $t$ & Model & \gls{eer} & \gls{auc} &  Acc.  \\ \midrule
			Car        & 10  &  GBM  &   0.111   &   0.961   & 88.8\% \\
			-- City    & 10  &  GBM  &   0.038   &   0.993   & 96.1\% \\
			-- Highway & 10  &  GBM  &   0.026   &   0.995   & 97.4\% \\
			-- Parked  & 10  &  GBM  &  0.271*   &   0.813   & 72.9\% \\ \midrule
			Car        & 30  &  GBM  &  0.104*   &   0.967   & 89.6\% \\
			-- City    & 30  &  GBM  &   0.032   &   0.995   & 96.8\% \\
			-- Highway & 30  &  GBM  &   0.022   &   0.997   & 97.7\% \\
			-- Parked  & 30  &  GBM  &  0.282*   &   0.803   & 71.7\% \\ \bottomrule
		\end{tabular}
	\end{minipage}%
	\begin{minipage}[b]{.5\textwidth}
		\centering
		\captionof{table}{Classification results for Truong \textit{et al.}, Office}			\label{sec:res:truong:tab:EERoff}
		\begin{tabular}{lc|cccc}
			\toprule
			Scenario   & $t$ & Model & \gls{eer} & \gls{auc} &  Acc.  \\ \midrule
			Office     & 10  &  GBM  &  0.084*   &   0.974   & 91.5\% \\
			-- Night   & 10  &  GBM  &   0.08*   &   0.976   & 91.9\% \\
			-- Weekday & 10  &  DRF  &   0.087   &   0.973   & 91.3\% \\
			-- Weekend & 10  &  GBM  &   0.071   &   0.981   & 92.9\% \\ \midrule
			Office     & 30  &  GBM  &   0.069   &   0.982   & 93.1\% \\
			-- Night   & 30  &  GBM  &  0.063*   &   0.984   & 93.6\% \\
			-- Weekday & 30  &  GBM  &  0.072*   &   0.981   & 92.8\% \\
			-- Weekend & 30  &  GBM  &   0.053   &   0.989   & 94.6\% \\ \bottomrule
		\end{tabular}
	\end{minipage}
\end{table}

\begin{figure}
\centering
\begin{subfigure}[b]{0.33\textwidth}
	\centering
   	\includegraphics[width=0.99\textwidth]{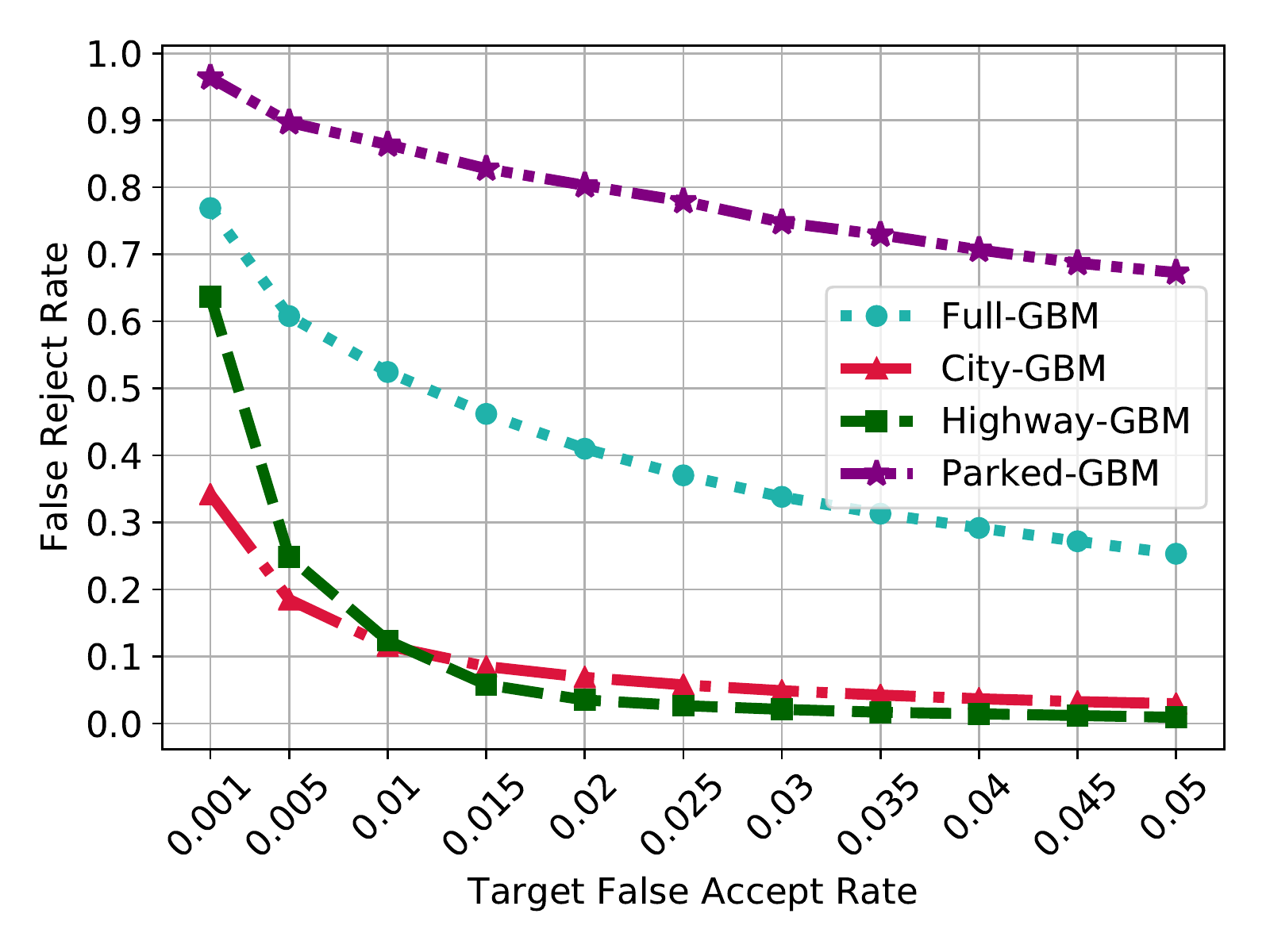}
   	\caption{Car}
   	\label{sf:truong-car}
\end{subfigure}
\begin{subfigure}[b]{0.33\textwidth}
	\centering
   	\includegraphics[width=0.99\textwidth]{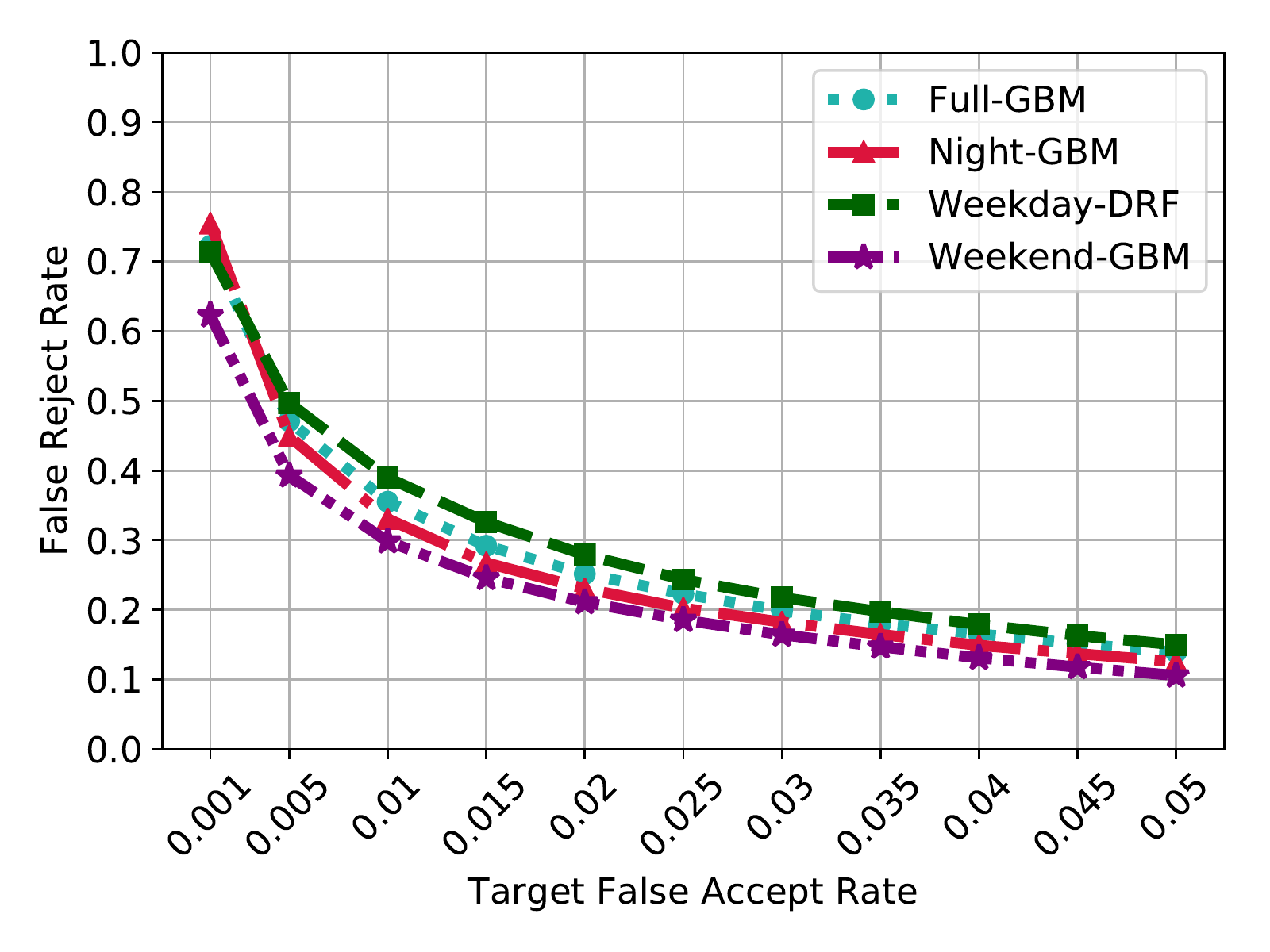}
   	\caption{Office}
    \label{sf:truong-office}
\end{subfigure}
\begin{subfigure}[b]{0.33\textwidth}
	\centering
   	\includegraphics[width=0.99\textwidth]{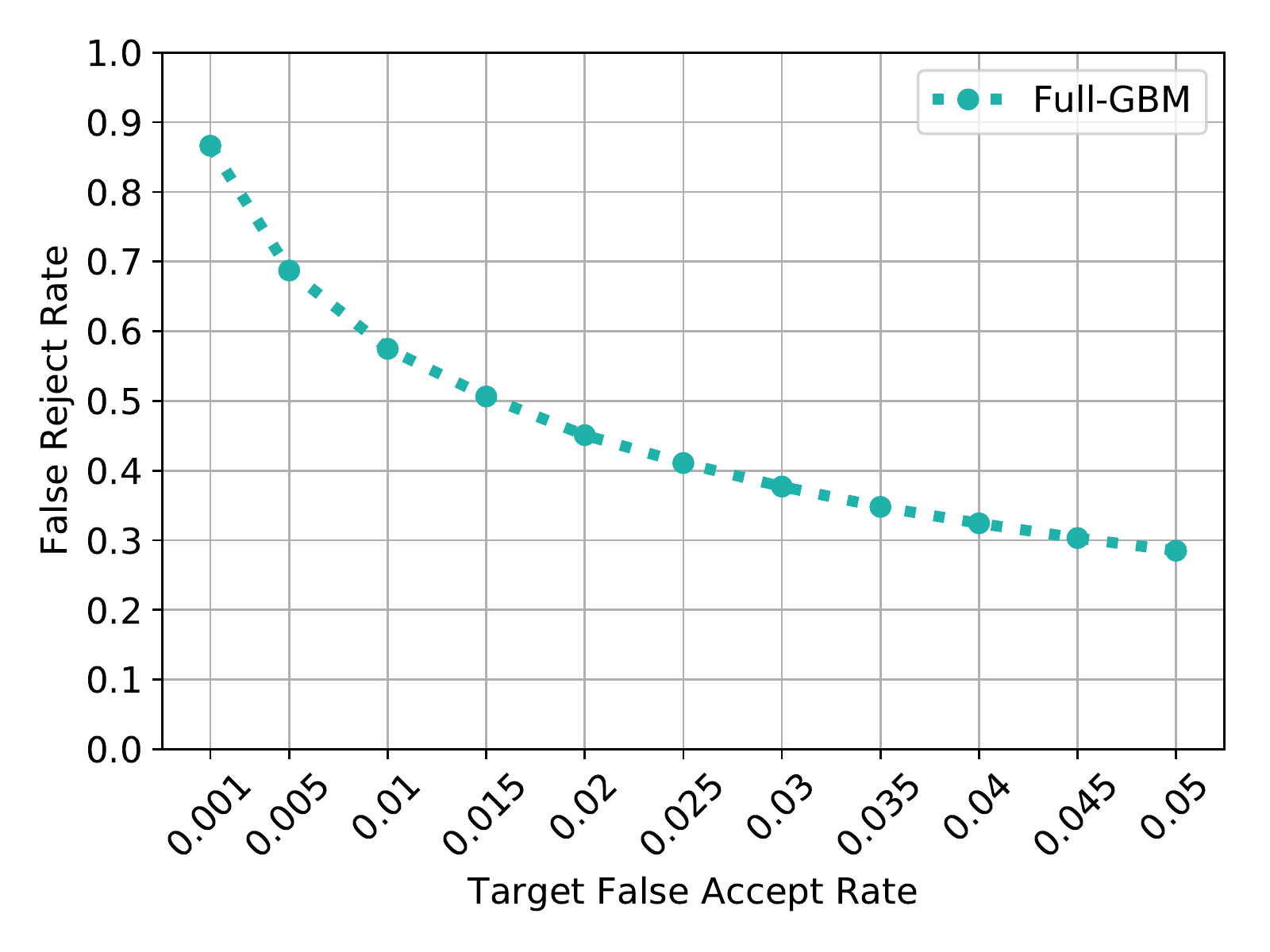}
   	\caption{Mob/het}
   	\label{sf:truong-mobile}
\end{subfigure}
\caption{\glspl{frr} with target \glspl{far} for Truong \textit{et al.} ($t = 10$)}
\end{figure}

\subsubsection{Car}
In this scenario, we obtain an \gls{eer} of 0.111 and 0.104 for $t=10$ and 30, respectively.
Cross-correlation and time-frequency distance of the audio recordings account for 85\% of the relative feature importance.
This is expected, as the route passed through many areas without WiFi \glspl{ap} and \gls{ble} devices.

When investigating subscenarios, we observe that the parked subscenario exhibits a significantly higher \gls{eer} than the other subscenarios.
In this subscenario, the models also show a lower reliance on audio features, with those features making up only 64\% of importance, and a higher precedence being given to WiFi and BLE features.
This can likely be explained by the low audio activity in this subscenario, leading the model to use these less reliable features and thus reducing classification performance.

The \glspl{frr}, given in \autoref{sf:truong-car}, show similar trends: city and highway exhibit the lowest \glspl{frr} for the desired \glspl{far}, with the parked car significantly above them, and the full dataset somewhere in between.
The \glspl{frr} also show a steeper drop in the beginning that tapers off later.

To test the robustness of the model, we use it to obtain predictions on the data from the other scenarios, applying the \gls{eer} threshold determined before.
This results in significantly increased error rates for all combinations of scenarios and intervals, with \glspl{far} larger than 0.44 for the office scenario, and \glspl{frr} in excess of 0.6 for the mob/het scenario, indicating that the model's performance will deteriorate when used on data from a scenario it has not been trained on and thus is not robust to being operated in different environments.

The robustness of models trained on subscenario datasets shows significant variation.
Combinations of the city and highway subscenarios show changes between 0 and 4 percentage points, while combinations involving the parked subscenario show changes between 25 and 82 percentage points.
This indicates that the models are robust to small changes in the environment, but cannot adapt to significant deviations.

\subsubsection{Office}
We observe a slightly improved \gls{eer} of 0.084 ($t=10$) and 0.069 ($t=30$). 
Surprisingly, the WiFi features are not more relevant, despite WiFi being one of the best features reported by Truong \textit{et al.}, and one would expect more stable signals for stationary devices compared to the mobile car scenario.
However, our results show the audio features are even more relevant with a combined relative importance of 91\%.
To investigate if this is caused by the missing WiFi data in the dataset (cf. \autoref{sec:disc}), we repeat the analysis, excluding instances where the WiFi data is missing due to a scan error, and obtain unchanged results.
Thus, even in a dataset that contains WiFi data for all samples, the feature does not become more relevant for the classifier.
The subscenarios show a much more similar behavior than in the previous scenario, with \glspl{eer} between 0.071 (weekend) and 0.087 (weekday).
This trend is also shown in the \gls{frr} evaluation in \autoref{sf:truong-office}, where the curves are all closely matched.

When running the model on the car dataset for $t=10$, we obtain an \gls{far} and \gls{frr} of 0.174 and 0.412, respectively, with $t=30$ increasing the error even further.
Applying it to the mob/het dataset yields error rates of 0.022 and 0.711, respectively, once again increasing further for $t=30$.
This shows that the models are sufficiently different such that generalization is low and therefore robustness of the scheme suffers.
Switching between the different subscenarios results in less pronounced changes, but still in some cases doubles the error rates.
The scheme appears especially challenged when applying the weekend model to the other subscenarios, often doubling the error rates, while the weekday model is fairly robust, with only minor changes to most error rates.
This is likely due to the higher complexity of the weekday dataset, which contains data from a more diverse set of situations.

\subsubsection{Mob/het}
In the mob/het scenario, we obtain \glspl{eer}$^*$ of 0.127 and 0.123, respectively (cf. \autoref{sec:res:truong:tab:EERmobile}).
Once again, the most important features are audio-based, although their importance is less pronounced, making up only 60\% and 56\% of relative feature importance for $t=10$ and 30, respectively.
Optimizing for a low \gls{far} will result in \glspl{frr} between 0.9 and 0.3 (cf. \autoref{sf:truong-mobile}).
This lower overall performance and the reduced prominence of the audio features is likely related to the issue of heterogeneous microphone characteristics, which we previously observed in the scheme proposed by Karapanos \textit{et al.} (cf. \autoref{sec:res:spf:mobile}), as Truong \textit{et al.} use similar audio features.

Using the model to classify the car and office datasets results in significantly increased error rates (FAR $>$ 0.7, FRR $\leq$ 0.22 for all combinations), showing that the model is not robust to different environments.

\subsubsection{Conclusion}
Our evaluation shows that the scheme can achieve a good \gls{eer} in some of our scenarios, although it does not reach the error rates of the original paper, which observed a \gls{far} and \gls{frr} of 0.0198 and 0.0167 for $t=10$.
We also see that models generated in one scenario show a significant loss in accuracy when being used in another scenario, and that the scheme encounters problems when using heterogeneous microphones.
The authors also performed an experiment where pairs of devices were placed in close proximity (which matches our office scenario), obtaining a \gls{far} of 0.0476, but did not report the \gls{frr}, which prevents a direct comparison.

Contrary to the original evaluation, the classification performance increased with larger intervals.
We also saw a much higher importance of the audio feature than the original paper and a correspondingly lower importance of the WiFi feature.
This is likely related to the collection strategy employed by the authors, who collected their dataset in different locations across two cities, which can be easily distinguished by their different WiFi signals.

The subscenario evaluation shows that the system does not work well in environments with little context activity, like cars parked in areas without WiFi and BLE devices.
The differences between the subscenarios were less pronounced in the office scenario, where a larger number of WiFi and BLE devices were visible at all times.

Two factors limit the validity of our results.
First, we did not collect GPS data, used by  the authors.
We assume that the impact would have been low in the office and mob/het scenarios, where devices were located close to each other and mostly static, however, it may have improved performance in the car scenario.
Second, we use a different classifier than the authors, who used a Multiboost classifier \cite{Webb2000}, which is not supported in H2O.
Still, DRFs and GBMs use ensemble methods similar to Multiboost and are unlikely to give significantly worse results.

\subsection{Shrestha \textit{et al.}}
\label{sec:res:shrestha}
Shrestha \textit{et al.} \cite{Shrestha:2014} propose combining readings from temperature, humidity, altitude, and precision gas sensors to decide if two devices are colocated.
They compute the absolute difference between the readings of two devices and use a Multiboost classifier \cite{Webb2000} trained on a labeled dataset to distinguish colocated and non-colocated devices.
As our devices did not feature a precision gas sensor, we omit this feature.
The sensor readings are not averaged over time intervals but used individually.
Their dataset was obtained by collecting data from several locations using a pair of devices.
Any data collected at different locations and times is interpreted as non-colocated.
Additional details of the scheme are given in \autoref{subsec:appx1-shrestha}.

\subsubsection{Methodology}
Although the machine learning methodology is identical to that used for the paper by Truong \textit{et al.} (cf. \autoref{sec:res:truong} for details), the characteristics of the datasets and volume of data demand different treatment.
One assumption made by any classifier in machine learning is that it estimates a surjective function from a vector of features $\hat{x}$ to a particular class $c$, i.e., all unique instances in the dataset map to exactly one class.
However, our datasets do not fulfill this requirement, as several identical instances map from the same feature values to different classes.
As the classifier has no additional data to base its decision on, it is unable to distinguish these ambiguous instances and thus can never reach a performance of 100\%, i.e., the \gls{eer} has a lower bound larger than 0.
This indicates that more features are needed to discriminate the classes properly.
We show the percentage of these ambiguous instances (Amb.) in each dataset in \autoref{sec:res:shrestha:tab:EER}.

At the same time, it also indicates a potential for compression.
Indeed, after analyzing the original office dataset with a size of 81~GB, we observe that many instances are repeated.
Therefore, we introduce a pre-processing step before training, where we group all equal instances and keep a count of how many times they appear. 
These counts are used as weights in the later learning stage, which acts as a lossless compression mechanism.
This way, we reduce the dataset to approximately 600 MB, which allows us to train models much faster and with significantly lower computational resources without sacrificing classification performance.

\begin{table}
	\centering
	\begin{minipage}[b]{.5\textwidth}
		\centering
		\captionof{table}{Classification results for Truong \textit{et al.}, Mob/het}			\label{sec:res:truong:tab:EERmobile}
		\begin{tabular}{lc|cccc}
			\toprule
			Scenario & $t$ & Model & \gls{eer} & \gls{auc} &  Acc.   \\ \midrule
			Mob/het   & 10  &  GBM  &  0.127*   &   0.946   & 0.873\% \\
			Mob/het   & 30  &  GBM  &   0.123   &   0.949   & 0.877\% \\ \bottomrule
		\end{tabular}
	\end{minipage}%
	\begin{minipage}[b]{.5\textwidth}
		\centering
		\captionof{table}{Classification results for Shrestha \textit{et al.} }
		\label{sec:res:shrestha:tab:EER}
		\begin{tabular}{l|ccccc}
			\toprule
			Scenario   & Model & \gls{eer} & \gls{auc} &  Acc.  & Amb.   \\ \midrule
			Car        &  DRF  &   0.115   &   0.960   & 88.5\% & 8.8\%  \\
			-- City    &  DRF  &  0.081*   &   0.977   & 91.9\% & 5.5\%  \\
			-- Highway &  DRF  &   0.08    &   0.979   & 91.9\% & 6.8\%  \\
			-- Parked  &  DRF  &  0.034*   &   0.995   & 96.5\% & 2.2\%  \\ \midrule
			Office     &  DRF  &  0.247*   &   0.834   & 75.2\% & 25.5\% \\
			-- Night   &  DRF  &  0.155*   &   0.911   & 84.4\% & 15.5\% \\
			-- Weekday &  DRF  &  0.271*   &   0.824   & 72.9\% & 25.5\% \\
			-- Weekend &  GBM  &  0.148*   &   0.928   & 85.1\% & 14.1\% \\ \midrule
			Mob/het     &  DRF  &  0.141*   &   0.942   & 85.9\% & 12.2\% \\ \bottomrule
		\end{tabular}
	\end{minipage}
\end{table}

\begin{figure}
\centering
\begin{subfigure}[b]{0.33\textwidth}
	\centering
   	\includegraphics[width=0.99\textwidth]{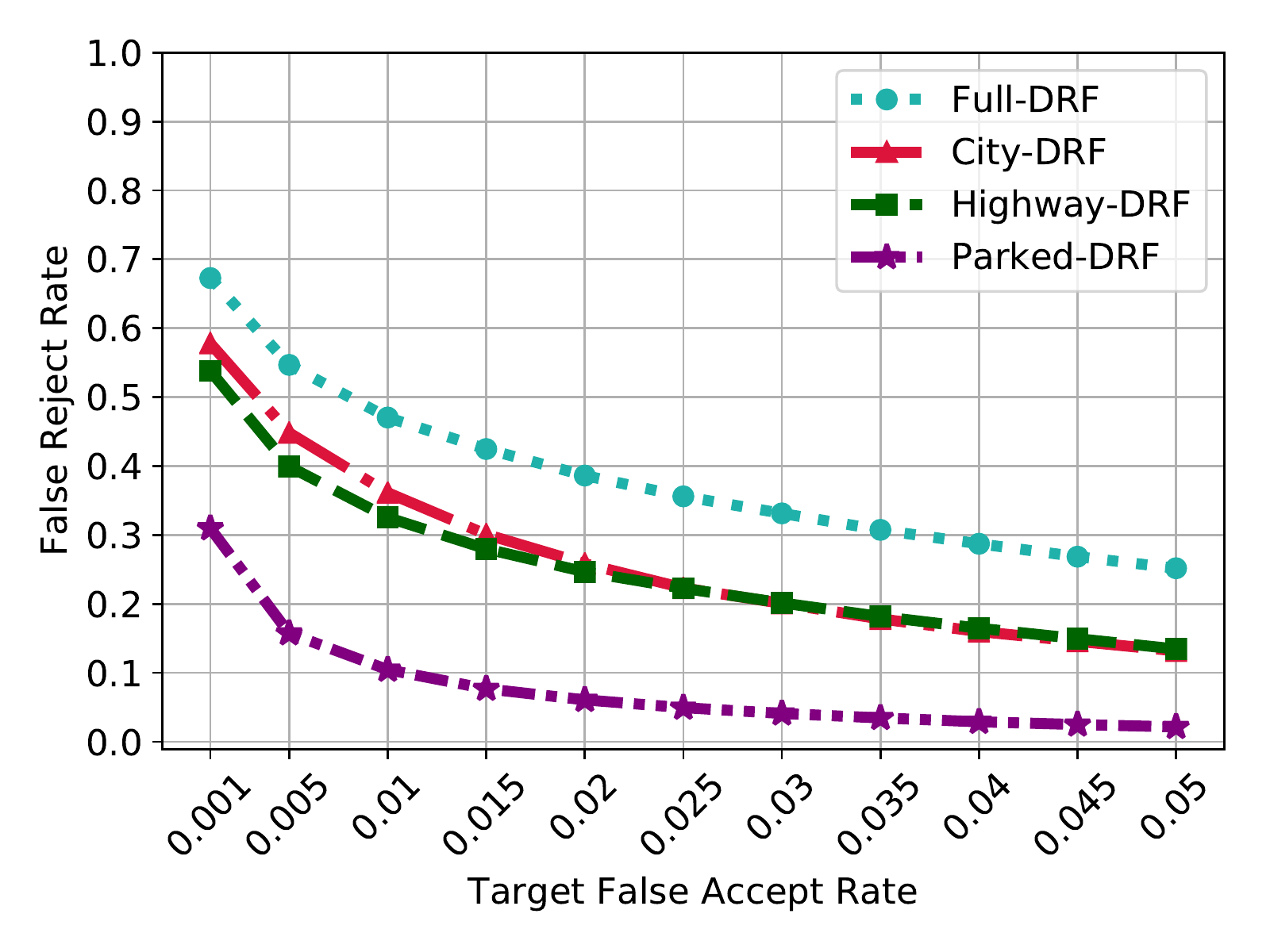}
   	\caption{Car}
   	\label{sf:shrestha-car}
\end{subfigure}
\begin{subfigure}[b]{0.33\textwidth}
	\centering
   	\includegraphics[width=0.99\textwidth]{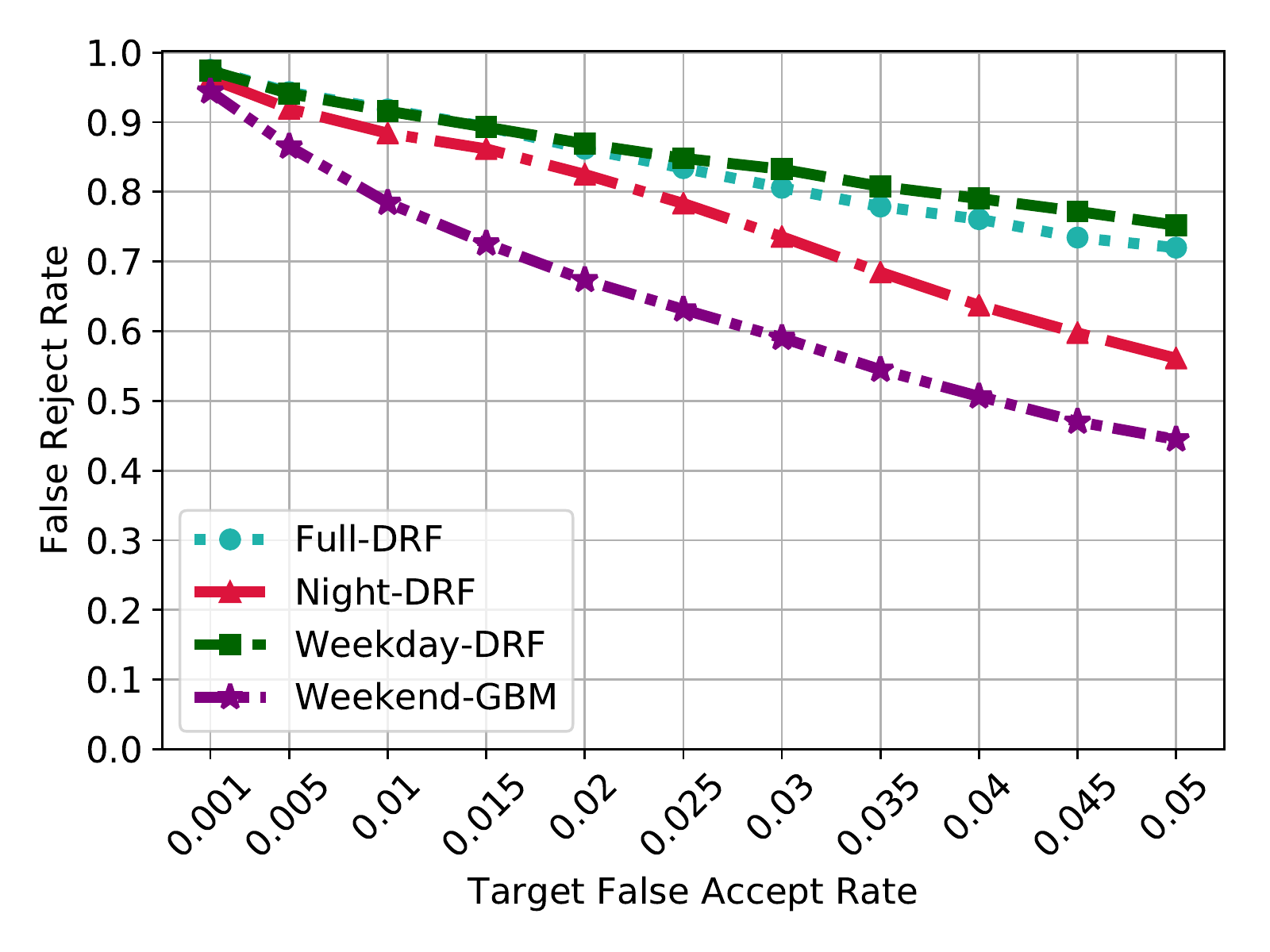}
   	\caption{Office}
    \label{sf:shrestha-office}
\end{subfigure}
\begin{subfigure}[b]{0.33\textwidth}
	\centering
   	\includegraphics[width=0.99\textwidth]{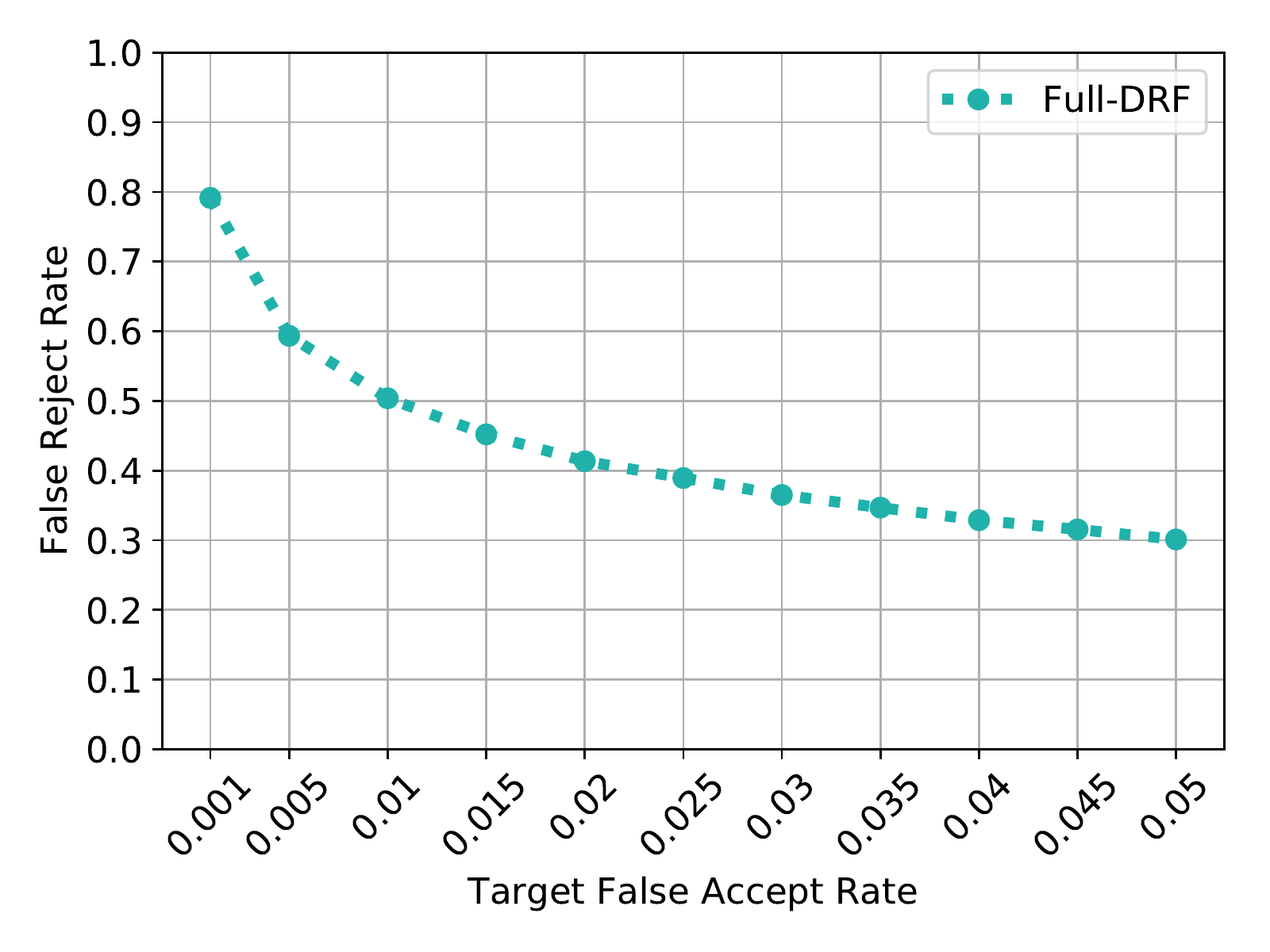}
   	\caption{Mob/het}
   	\label{sf:shrestha-mobile}
\end{subfigure}
\caption{\glspl{frr} with target \glspl{far} for Shrestha \textit{et al.}}
\end{figure}

\subsubsection{Car}
For the car scenario, we obtain an \gls{eer}$^*$ of 0.115, with the classifier relying almost evenly on all the features to make the predictions.
The individual subscenarios achieve even lower \glspl{eer}, showing rates of 0.034 (parked), 0.08 (highway) and 0.081 (city).
This performance correlates with the percentage of ambiguous instances, with subscenarios with more ambiguous instances obtaining higher error rates.
The low error rate of the parked subscenario is likely related to the use of temperature sensors, which captured the different rates of heat dissipation of the cars after they were parked.
When aiming for a specific \gls{far}, the differences between the subscenarios are maintained, with the parked subscenario showing consistently lower \glspl{frr} (cf. \autoref{sf:shrestha-car}).

The model is not very robust, showing significantly increased error rates when applied to the office or mob/het dataset (FAR 0.342, FRR 0.503 for office, 0.276 / 0.717 for mob/het).
Similarly, applying models specific to one subscenario to another increases the error rates by at least 32 percentage points.

\subsubsection{Office}
Here, the classifier reaches an \gls{eer}$^*$ of 0.247, showing a significantly lower performance than in the car scenario.
It relies strongly on the temperature differences to make the predictions.
Such a focus on a feature with a low range of potential values may make the classifier more vulnerable to active attacks and is thus undesirable.
The low performance is mirrored in the subscenarios, with error rates of 0.148 (weekend) to 0.271 (weekday), which is also borne out in the high \glspl{frr} when aiming for a specific \gls{far} (cf. \autoref{sf:shrestha-office}).
Once again, higher percentages of ambiguous instances translate to higher error rates.
We also observe that the percentage of ambiguous instances grows with the size of the dataset.
This is to be expected, as a larger dataset has a higher chance of obtaining these instances, as the range of potential values is limited.

When the office model is used on the other two datasets, the error rates increase significantly (FAR 0.186, FRR 0.693 for car, 0.184 / 0.787 for mob/het), showing that the model is not robust to being used in different environments.
Of the three subscenario models, the weekday model is the most robust, with error rate increases of below 6 percentage points when applied to different subscenario datasets.
However, applying the night model to weekday data increases the error rate by 35 percentage points, and the weekend model increases its error rate by over 43 percentage points with the weekday dataset.
This shows that the robustness is limited.

\subsubsection{Mob/het}
Due to a lack of humidity and temperature sensors, the laptops and robot were not included in the evaluation.
The classifier reaches an \gls{eer}$^*$ of 0.141 (cf. \autoref{sec:res:shrestha:tab:EER}), relying primarily on the altitude readings, with a lower importance given to temperature and humidity.
A closer investigation revealed that the phones' barometric pressure readings deviated from those of the watches and SensorTags, showing an offset of approximately 2 hPa.
The temperature and humidity readings varied widely, with the position of the device (smartphone in pocket, sensor on screen, ...) having a much larger influence than the room they operate in.
The error rate is likely related to this challenging environment, as well as the number of ambiguous instances, which make up 12.2\% of the dataset.
When optimizing for a low \gls{far}, the resulting \gls{frr} are at least 0.3 (cf. \autoref{sf:shrestha-mobile}).

Applying the model to the car dataset results in notably increased error rates (FAR 0.302, FRR 0.672).
The office dataset gives similar error rates (0.306 / 0.606), showing limited robustness of the model in different environments.

\subsubsection{Conclusion}
Overall, the scheme by Shrestha \textit{et al.} cannot reliably separate colocated from non-colocated devices in most scenarios.
This is in stark contrast to the error rates reported by the authors, who obtained an \gls{far} and \gls{frr} of 0.0581 and 0.0296, respectively.
This deviation can partially be explained by the lack of precision gas features in our datasets, reducing the number of dimensions the models can discriminate on.
Another explanation is the more challenging environment our data was collected in---the authors collected their data in widely spaced locations at different times of day, and their non-colocated class consisted of pairings between different locations.
This also explains the high discriminative power of the altitude readings reported by the authors, and indicates that the scheme will likely have a much better performance if only coarse colocation is required.

The high number of ambiguous instances shows that the scheme would benefit from incorporating additional sensors to improve its discriminative power.
Additionally, features tracking the change in values over time may be a more promising approach, as our dataset shows these changes to be more consistent between devices in the same room than the sensor readings themselves.
The results also show that even if high classification performance can be reached, it is still highly specific to the environment it was collected in, and does not transfer well into other environments, i.e., the robustness of the scheme is limited.


\section{Discussion}
\label{sec:disc}
In this section, we discuss the implications of the obtained results, the limitations of our method, and avenues for future work.

\paragraph{Performance Comparison}
Our results, summarized in \autoref{tab:contents}, show that the scheme by Truong \textit{et al.} obtains the best \glspl{eer} in the office (0.069) and mob/het (0.123) scenarios, while the scheme by Karapanos \textit{et al.} achieves a significantly lower error rate (0.006) in the car scenario.
This indicates that depending on the use case, both are solid choices.
We also observed a large variation in performance on different subscenarios, ranging from perfect accuracy (Karapanos \textit{et al.}, $t=120$, highway) to significantly degraded performance compared to using the full dataset (Sch{\"u}rmann \textit{et al.}, $t=120$, parked), illustrating the importance of fine-grained test scenarios.

\paragraph{Adaptiveness}
Some schemes struggle to adapt to times of low ambient activity like the night, where we observed a lower separation between colocated and non-colocated devices.
These times need to be taken into account when designing a scheme intended for continuous operation.
Karapanos \textit{et al.} and Miettinen \textit{et al.} added measures to reduce the impact of these times by dynamically discarding samples with low ambient activity \cite{Karapanos:2015} or high predictability \cite{Miettinen:2014}, trading off availability for security.

\paragraph{Robustness}
Even if they can operate in environments with low activity, most schemes suffer from a lack of robustness, i.e., parameters that are optimal for one scenario do not give good performance on the other.
Some schemes can achieve a certain degree of robustness for specific parameters (e.g., interval sizes), most notably that by Karapanos \textit{et al.}, but no scheme is robust with all parameters.
The same trend holds when exchanging parameters or models between different subscenarios, like day and night, especially if they have significantly different ambient activity levels.
This further illustrates the importance of testing schemes in a wide variety of settings.
We urge researchers to pay special attention to robustness, to facilitate use in the wide variety of different and sometimes unexpected environments the \gls{iot} will be deployed in.

\paragraph{Heterogeneity}
Even if schemes can provide good performance in settings with homogeneous devices (i.e., the same hardware), they may still fail when encountering devices with different characteristics.
These challenges have also been encountered in other research fields such as participatory sensing.
Examples from our dataset include microphones with varying sensitivity and frequency response \cite{Kardous2016,maisonneuve2009noisetube,Lu2009}, which leads to lower correlations, or incorrectly calibrated sensors (e.g., air pressure) measuring with a fixed offset from one another.
In addition, the way a device is carried influences the observed sensor data \cite{Miluzzo2010}.
Schemes need to be able to still provide good results under these conditions if they are intended for use cases where the used hardware is not carefully controlled by a single party and should be tested with heterogeneous devices.

\paragraph{Colocation definition}
Many schemes do not explicitly state their colocation definition.
It is often unclear if they are intended to distinguish personal workspaces inside an office, different rooms, or parts of a city, making it difficult to identify security guarantees schemes provide in any specific situation.
This hinders a fair evaluation and comparison of these schemes, and makes it hard to determine if our results impact its designated use case.
Authors should explicitly define what their scheme considers (non)colocated to allow for fair comparisons.

\paragraph{Limitations}
Technical issues during the recording led to data loss for some features, especially the WiFi captures, which stopped working on some devices.
In the mob/het scenario, three devices stopped the data collection before the eight-hour countdown, resulting in partial loss of audio and sensor data.
This reduces the amount of available data for the evaluation of features based on these modalities.
In the same way, the \textit{SensorTag} platform occasionally delivered incorrect readings for the luminosity, which we detected and excluded.

Our goal was to compare the different \gls{zis} schemes in a fair manner and as specified by their original authors.
While we attempted to stay as close to the published version of the scheme as possible, in some cases, minor changes had to be made to parameters such as interval length or sampling rate.
These deviations are noted in the Appendix, and their influence on the results should be negligible.
We did not attempt to optimize any parameters aside from interval length and the power threshold of Karapanos \textit{et al.} for the mob/het scenario for our dataset, so it is possible that some schemes could perform better when they are instantiated with different parameters.

We also note that our scenarios are challenging, as they include devices in isolated positions (glove compartment, cupboard, pocket), low isolation between two offices, and two cars that traveled next to each other for extended amounts of time.
This is intentional to be able to investigate the performance of the systems in challenging situations, as consumer \gls{iot} deployments rarely follow best practices for a deployment facilitating zero-interaction security.
We also did not include any scenarios in busy areas like shopping malls, which may show different behavior due to the higher environmental variations, as we chose to focus on the likely application domain of zero-interaction technologies, consumer applications.
Additionally, obtaining approval and informed consent for a long-term data collection in a public place would have been infeasible in our jurisdiction.

\paragraph{Future Work}
Our chosen scenarios only cover a subset of interesting \gls{iot} environments.
Other scenarios may pose different challenges for the schemes.
For example, in a \textit{smart building} scenario, an ideal \gls{zis} scheme would have to be able to authenticate all devices within the same building, while excluding adjacent buildings.
In a \textit{caf\'{e}} scenario, schemes would need to be able to distinguish individual tables.
We also did not include any scenarios where devices operated in environments without any humans for extended amounts of time (e.g., automated factory work floors or storage units), which could pose challenges to many schemes due to the potentially low variation in the context information or different noise characteristics.

The collection of additional datasets will assist efforts to create more adaptive and robust schemes and to understand the limitations of existing ones.
Another avenue for future work is the robustness to adversarial settings, where part of the context information can be controlled by an active adversary (e.g., by injecting sound).

\section{Conclusion}
\label{sec:concl}
We reproduced and evaluated five \gls{zip} and \gls{zia} schemes in three realistic scenarios---\textit{smart office}, \textit{connected car} and \textit{smart office with mobile heterogeneous devices}---posing different challenges in aspects like environmental noise, context leakage, and times of low activity.
Our results show that none of the reproduced schemes can perfectly separate devices in all scenarios.
The schemes by Karapanos \textit{et al.} \cite{Karapanos:2015} and Truong \textit{et al.} \cite{Truong:2014} show promising results, but no scheme reliably outperforms all others in all scenarios.
The error rates also indicate that zero-interaction security should not be used as the only access control factor, as even a false accept rate of 1\% would be considered insufficient for some real-world applications.
In fact, Karapanos \textit{et al.} explicitly proposed their \gls{zia} scheme as a more convenient second authentication factor \cite{Karapanos:2015} instead of a stand-alone solution.

We also observed that a good average-case separation of context features aggregated over the whole dataset does not imply a good authentication performance on individual samples.
\gls{zip} and \gls{zia} schemes should thus be evaluated in terms of their error rates in both the average case as well as individual subscenarios to get a realistic impression of their performance.
In the same way, the evaluation should be performed using a set of heterogeneous devices in a realistic, challenging environment, to test the limits of the scheme.

Our evaluation revealed that in many cases, features based on ambient audio performed best.
However, researchers need to take the privacy implications of using audio recordings into account, as this may not be acceptable in some environments like hospitals.
Additionally, the computational costs of processing have to be considered, as expensive audio processing operations may not always be possible on resource-constrained \gls{iot} devices.
Finally, we observed that devices with differing microphone characteristics can significantly degrade the performance of sound-based schemes.
Thus, we encourage researchers to continue to investigate the possibility of using more power-efficient features based on low-power sensors.
Here, we found that instead of using the absolute difference between sensor readings, trends over time may be a more reliable colocation indicator.

We also observed that the robustness and adaptiveness of many schemes varies dramatically for different scenarios.
Schemes should explicitly state which environments they are designed for.
Additionally, they should support robustness and adaptiveness, potentially by automatically adapting their internal parameters to their environment, and should be evaluated on data from different scenarios and devices.

Finally, we release the first extensible open source toolkit \cite{dataset} for researching zero-interaction security, containing reference implementations of the reproduced schemes, the audio recordings of the \textit{mob/het} scenario, and over 1 billion samples of labeled sensor data.
We also release all data generated by our evaluation to facilitate the reproduction of our results and provide a common benchmarking baseline for future schemes.


\section*{Acknowledgments}
\label{sec:acks}
We would like to thank Daniel Wegemer, Jiska Classen, Timm Lippert, Robin Klose, Vanessa Hahn and Santiago Arag\'{o}n for assistance in conducting this research.
Furthermore, we are thankful to Hien Truong, Nikolaos Karapanos, Dominik Sch{\"u}rmann, Markus Miettinen and Babins Shrestha for assistance in reproducing their work.
This work has been co-funded by the DFG within CRC 1119 CROSSING and CRC 1053 MAKI projects, and as part of project C.1 within the RTG 2050 ``Privacy and Trust for Mobile Users''.
Calculations for this research were conducted on the Lichtenberg high performance computer of the TU Darmstadt.

\section*{Appendix}
\appendix

\setcounter{footnote}{0}

\section{Reproduced ZIS schemes}
\label{sec:appx1}
In this appendix, we provide more details about the reproduced \gls{zis} schemes. 
We give a brief overview of each scheme's functionality and use case and describe the implementation of context features utilized by the scheme.

\paragraph{Audio preprocessing:} Before computing audio features we aligned audio recordings from different sensing devices as follows. 
At the beginning of data collection all devices were synchronized using the \gls{ntp}.
First, we performed the coarse-grained alignment using devices' timestamps to synchronize the start of the audio recordings. 
Second, during the feature computation we performed a fine-grained alignment between two input audio recordings 
using the cross-correlation function in Matlab \cite{xcorr:2018}.
Specifically, we considered the first hour of audio recordings to find a lag between them, using the \textit{xcorr} function ($maxlag = 3$ seconds), then we used this lag to align two audio recordings and cut them to the length of the shortest recording. 
These aligned recordings are then split into intervals and used to compute audio features.  

In the mobile scenario, we increased the $maxlag$ to 15 seconds to more precisely find the lag between audio recordings of heterogeneous devices. 
In addition, we found that heterogeneous devices have an inherent audio drift, causing desynchronization of audio recordings. 
We removed this drift by applying a time-stretching effect to audio recordings in the \textit{Audacity} tool (change Tempo).

\subsection{Karapanos \textit{et al.}}
\label{subsec:appx1-karapanos}
\setcounter{equation}{0}

The scheme by Karapanos \textit{et al.} \cite{Karapanos:2015} calculates a similarity score between snippets of ambient audio from two devices to decide if these devices are colocated. 
The similarity score is the average of the maximum cross-correlations between two audio snippets computed on a set of one-third octave bands.
To prevent erroneous authentication when audio activity is low, a power threshold is applied to discard similarity scores from audio snippets with insufficient average power. 
The similarity score is then checked against a fixed similarity threshold to decide if two devices are colocated. 
The scheme is designed to provide colocation evidence between a user's smartphone and a computer with a running browser.
This evidence is utilized as a second authentication factor when a user wants to log-in to an online service such as a bank account. 
In this work, we focus on computing and comparing similarity scores and do not target the specific use case of the second authentication factor.
 
We first provide notations adopted from the original paper in \autoref{tab:karapanos-not}. 
Second, we present parameters of the sound similarity algorithm used in the original and our implementations in  \autoref{tab:karapanos-params}. 
Our goal was to follow the original implementation as close as possible, however, we introduced a few changes, as we did not have tight synchronization between audio snippets. 
Third, we present our implementation of the sound similarity algorithm in Section \ref{subsubsec:spf}.

\begin{table}[!htb]
	\caption{Notations used by Karapanos \textit{et al.}}
	\label{tab:karapanos-not}
	\begin{tabular}{cl}
		\toprule
		Notation & Explanation \\
		\midrule
		$x, y$ & input audio snippets \\
		$L$ & length of input audio snippets in seconds \\
		$l_{max}$ & max cross-correlation lag in seconds \\ 
		$r$ & sampling rate of input audio snippets in kHz \\ 
		$\tau_{dB}$ & average power threshold in dB \\
		$B$ & set of considered one-third octave bands \\ 
		$n$ & number of considered one-third octave bands \\ 
		$S_{x,y}$ & similarity score \\
		\bottomrule
	\end{tabular}
\end{table}

\begin{table}[!htb]
	\captionsetup{justification=centering}
	\caption{Parameters of the sound similarity algorithm used in \cite{Karapanos:2015} (highlighted) and our implementations}
	\label{tab:karapanos-params}
	\begin{tabular}{ccccc}
		\toprule
		$L$, sec&$l_{max}$, sec&$r$, kHz&$\tau_{dB}$, dB&$B \ (n)$\\
		\midrule
		\rowcolor{Gray}
		3 & 0.15 & 44.1 & 40 & 50Hz -- 4kHz (20) \\
		5 & 1 & 16 & 40/38/35 & 50Hz -- 4kHz (20) \\
		10 & 1 & 16 & 40/38/35  & 50Hz -- 4kHz (20) \\
		15 & 1 & 16 & 40/38/35  & 50Hz -- 4kHz (20) \\
		30 & 1 & 16 & 40/38/35  & 50Hz -- 4kHz (20) \\
		60 & 1 & 16 & 40/38/35  & 50Hz -- 4kHz (20) \\
		120 & 1 & 16 & 40/38/35  & 50Hz -- 4kHz (20) \\
		\bottomrule
	\end{tabular}
\end{table}

As shown in Table \ref{tab:karapanos-params}, our implementation differs with respect to these parameters from the implementation by Karapanos \textit{et al.}
First, we increase both the length of input audio snippets $L$ from 3 to 5 seconds and the length of the maximum cross-correlation lag $l_{max}$ from 0.15 to 1 second to achieve a comparable level of authorization to the authors.
We observed that even after the alignment procedure (cf. \textit{Audio preprocessing}) there might be an offset within long audio recordings (24 hours), which can affect synchronization between audio snippets. 
That is why, we set $l_{max} = 1$ to maintain a balance between security ($l_{max}$ thwarts attackers trying to guess the audio environment) and non-tight synchronization, which can happen in a realistic \gls{iot} scenario.
The increase of $l_{max}$ leads to the increase of the audio snippet length $L$ to 5 seconds.
Second, we use a lower sampling rate for the input audio snippets $r$: 16 vs. 44.1 kHz, which does not affect the sound similarity algorithm itself, but can be used to speed up the computations, as a smaller number of samples needs to be processed. 
Despite the lower sampling rate and, thus, narrower audio spectrum (8 kHz) we cover the same set of octave bands as the original implementation. 

As stated in Section \ref{sec:res}, we evaluate the performance of the scheme on a number of intervals from 5 to 120 seconds. 

\subsubsection{Implementation of the sound similarity algorithm}
\label{subsubsec:spf}

\begin{enumerate}
	\setcounter{enumi}{-1}
	
	\item As input, we have two aligned audio snippets $x$ and $y$ of equal length $L$ with a sampling rate $r$.
	
	\item Both $x$ and $y$ are split into $n$ one-third octave bands using a bank of band-pass filters:
	\begin{equation}
	\begin{aligned}
	(x_{B_{1}},\dotsc,x_{B_{n}}) = \text{BP\_filter\_bank}(x) \\
	(y_{B_{1}},\dotsc,y_{B_{n}}) = \text{BP\_filter\_bank}(y)
	\end{aligned}
	\end{equation}  
	
	Table \ref{tab:octave-bands} shows the used one-third octave bands $B$ from 50 Hz to 4 kHz, and each band-pass filter is constructed as a 20th-order Butterworth filter \cite{BP:2018} with cut-off frequencies $[F_{l},F_{h}]$.
	
	\item For each $x_{B_{i}}$ and $y_{B_{i}} \forall i \in [1,n]$ the normalized maximum cross-correlation $\hat{C}_{x,y}(l)$ is computed as the function of the lag $l \in [0, l_{max}]$ (we omit $B_{i}$ indexes for simplicity):
	\begin{equation}
	\label{eq:spf_main}
	\hat{C}_{x,y}(l) = \max_{l}(|C_{x,y}^{\prime}(l)|) = \max_{l}\bigg(\Bigl\lvert{\dfrac{C_{x,y}(l)}{\sqrt{C_{x,x}(0) \cdot C_{y,y}(0)}}}\Bigr\rvert\bigg)
	\end{equation}
	
	In \autoref{eq:spf_main} the term $C_{x,y}(l)$ is a cross-correlation function between two discrete signals $x$ and $y$:
	\begin{equation}
	C_{x,y}(l) = \displaystyle\sum_{i=0}^{N-1} x(i) \cdot y(i-l)
	\end{equation}
	$N$ is the number of samples in the signals\footnote{We assume signals $x$ and $y$ have the same length} and the lag  is bounded within a range $l \in [0,N-1]$. The normalization term $\sqrt{C_{x,x}(0) \cdot C_{y,y}(0)}$ accounts for different amplitudes of signals $x$ and $y$, with $C_{x,x}(0)$ and $C_{y,y}(0)$ being the auto-correlation functions. The resulting maximum cross-correlation is bounded within a range $\hat{C}_{x,y}(l) \in [0,1]$, because we take the absolute value of the normalized cross-correlation $|C_{x,y}^{\prime}(l)|$.
	
	\item The resulting similarity score between two audio snippets $x$ and $y$ is obtained by taking the average of the normalized maximum cross-correlations computed in each one-third octave band:
	\begin{equation}
	S_{x,y} = \dfrac{1}{n}\sum_{i=1}^{n}\hat{C}_{x_{B_{i}},y_{B_{i}}}(l)
	\end{equation}
	
	The similarity score is only used if the input audio snippets have sufficient average power: $\overline{P}_{x}, \overline{P}_{y} > \tau_{dB}$.
	Otherwise, it is discarded and no authentication is attempted.
	
\end{enumerate}

\begin{table}[!htb]
	\caption{Used one-third octave bands}
	\label{tab:octave-bands}
	\begin{tabular}{cccc}
		\toprule
		Band Number &$F_{l}$, Hz&$F_{c}$, Hz&$F_{h}$, Hz\\
		\midrule
		6 & 44.194 & 49.606 (50) & 55.681 \\
		7 & 55.681 & 62.500 (63) & 70.154 \\
		8 & 70.154 & 78.745 (80) & 88.388 \\
		9 & 88.388 & 99.213 (100) & 111.362 \\
		10 & 111.362 & 125.000 (125) & 140.308 \\
		11 & 140.308 & 157.490 (160) & 176.777 \\
		12 & 176.777 & 198.425 (200) & 222.725 \\
		13 & 222.725 & 250.000 (250) & 280.616 \\
		14 & 280.616 & 314.980 (315) & 353.553 \\
		15 & 353.553 & 396.850 (400) & 445.449 \\
		16 & 445.449 & 500.000 (500) & 561.231 \\
		17 & 561.231 & 629.961 (630) & 707.107 \\
		18 & 707.107 & 793.701 (800) & 890.899 \\
		19 & 890.899 & 1000.000 (1000) & 1122.462 \\
		20 & 1122.462 & 1259.921 (1250) & 1414.214 \\
		21 & 1414.214 & 1587.401 (1600) & 1781.797 \\
		22 & 1781.797 & 2000.000 (2000) & 2244.924 \\
		23 & 2244.924 & 2519.842 (2500) & 2828.427 \\
		24 & 2828.427 & 3174.802 (3150) & 3563.595 \\
		25 & 3563.595 & 4000.000 (4000) & 4489.848 \\
		\bottomrule
	\end{tabular}
	\smallskip\centering
	\center{$F_{l}$ - lower band frequency, $Fc$ - calculated center frequency (nominal frequency), $F_{h}$ - upper band frequency}
\end{table}

\subsection{Sch{\"u}rmann and Sigg}
\label{subsec:appx1-schurmann}

\setcounter{equation}{0}

The scheme by Sch{\"u}rmann and Sigg \cite{Schurmann:2013} computes a binary fingerprint from a snippet of ambient audio, based on energy differences in successive frequency bands.
Two devices wishing to establish a pairing compute such fingerprints from their ambient environments.
These fingerprints are used in a fuzzy commitment scheme to obtain a shared secret. 
One device uses its fingerprint to hide a randomly chosen secret and sends this commitment to the other device, which can only retrieve the random secret from the commitment if it has a sufficiently similar fingerprint.  
In this work, we focus on deriving and comparing binary fingerprints and we do not target a specific use case of establishing a shared secret key. 

We first provide notations adopted from the original paper in \autoref{tab:schurmann-not}. 
Second, we present parameters of the audio fingerprinting algorithm used in the original and our implementations in \autoref{tab:schurmann-params}, where we introduce a few changes, as our audio snippets have a lower sampling rate. 
Third, we present our implementation of the audio fingerprinting algorithm in Section \ref{subsubsec:apf}.

\begin{table}[!htb]
  \caption{Notations used by Sch{\"u}rmann and Sigg}
  \label{tab:schurmann-not}
  \begin{tabular}{cl}
    \toprule
    Notation & Explanation \\
    \midrule
    $S$ & input audio snippet \\
    $l$ & length of the input audio snippet in seconds \\ 
    $r$ & sampling rate of the input audio snippet in kHz \\ 
    $n$ & number of frames to split the input audio snippet \\ 
    $m$ & number of frequency bands to split each frame \\
    $d$ & length of each frame in seconds (duration) \\
    $b$ & width of each frequency band in Hz \\
    $f$ & binary fingerprint of length $(n-1)\cdot(m-1)$ in bits \\ 
  \bottomrule
\end{tabular}
\end{table}

\begin{table}[!htb]
  \captionsetup{justification=centering}
  \caption{Parameters of the audio fingerprinting algorithm used in \cite{Schurmann:2013} (highlighted) and our implementations}
  \label{tab:schurmann-params}
  \begin{tabular}{ccccccc}
    \toprule
    $f$, bits&$l$, sec&$r$, kHz&$n$, frames&$m$, bands&$d$, sec&$b$, Hz\\
    \midrule
    \rowcolor{Gray}
    512 & 6.375 & 44.1 & 17 & 33 & 0.375 & 250 \\
    496 & 5 & 16 & 17 & 32 & $\sim$0.29 & 250 \\
    496 & 10 & 16 & 17 & 32 & $\sim$0.59 & 250 \\
    496 & 15 & 16 & 17 & 32 & $\sim$0.88 & 250 \\
    496 & 30 & 16 & 17 & 32 & $\sim$1.76 & 250 \\
    496 & 60 & 16 & 17 & 32 & $\sim$3.53 & 250 \\
    496 & 120 & 16 & 17 & 32 & $\sim$7.06 & 250 \\
  \bottomrule
\end{tabular}
\end{table}

As shown in \autoref{tab:schurmann-params}, our implementation differs with respect to some parameters from the implementation by Sch{\"u}rmann and Sigg. 
First, we use a lower sampling rate $r$ of 16 kHz instead of the original 44.1 kHz, which affects the number of frequency bands $m$ we can split our frames $n$ into.
With a 16 kHz sampling rate our audio spectrum is only 8 kHz, thus we can only obtain 32 non-overlapping frequency bands, each of width 250 Hz $b$. 
Having 32 frequency bands instead of 33 as in the original implementation results in shorter binary fingerprints $f$ of 496 instead of 512 bits.  
Second, we vary the lengths $l$ from 5 to 120 seconds, which also affects the length of a single frame $d$, which varies between 0.29 and 7.06 seconds.
We note that shorter audio frames (e.g. $d = 0.29$) are more susceptible to synchronization issues between input audio snippets, thus reducing the similarity of binary fingerprints generated from these snippets. 
However, starting from $l = 10$ our frame length $d$ is bigger than in the original implementation, which makes our results comparable and allows us to access the performance of the scheme (i.e. distinguishing between colocated and non-colocated devices) on longer audio snippets.

\subsubsection{Implementation of the audio fingerprinting algorithm}
\label{subsubsec:apf}

\begin{enumerate}
  \setcounter{enumi}{-1}
  
  \item As input, we have an audio snippet $S$ of length $l$ with a sampling rate $r$ (audio snippets from different devices are aligned). The number of frames $n$ and the number of frequency bands $m$ are selected to obtain the binary fingerprint of the desired length:
\begin{equation}
L_{f} =(n-1)\cdot(m-1)
\end{equation}

The width of a frequency band depends not only on the number of bands but also on the available audio spectrum which is limited by the Nyquist frequency ($f_{N} = \frac{r}{2}$):
\begin{equation}
b=\dfrac{\text{maxfreq}(S)-\text{minfreq}(S)}{m}
\end{equation}

  \item The audio snippet \textit{S} is split into \textit{n} successive frames $F_{1},\dotsc,F_{n}$ of equal length $d=r\cdot\frac{l}{n}$ in samples ($F_{i}$ is a $d$x1 vector). 
  
  \item Each frame $F_{1},\dotsc,F_{n}$ is split into $m$ non-overlapping frequency bands of width $b$ using a bank of band-pass filters:
\begin{equation}
(F_{B_{1}},\dotsc,F_{B_{m}})_{i} = \text{BP\_filter\_bank}(F_{i}),\quad\forall i \in [1,n]
\end{equation}

In our implementation the available audio spectrum is 8 kHz, thus we split it into the following 32 bands of width 250 Hz: $B_{1} = [1,250], B_{2} = [251,500],\dotsc, B_{m} = [7751,7999]$, using a 20th-order Butterworth filter \cite{BP:2018} for each band. 

  \item For each frame $F_{1},\dotsc,F_{n}$ the energy of each frequency band $B_{1},\dotsc,B_{m}$ is computed as (superscript $T$ denotes transpose):
\begin{equation}
(E_{B_{j}})_{i} = (F_{B_{j}}^T \cdot F_{B_{j}})_{i},\quad\forall i \in [1,n]; \forall j \in [1,m]
\end{equation}

  \item The results of energy computation are stored in the energy matrix ($\forall i \in [1,n]; \forall j \in [1,m]$):
\begin{equation}
E_{i,j} = 
\begin{pmatrix}
  E_{F_{1},B_{1}} & E_{F_{1},B_{2}} & \cdots & E_{F_{1},B_{m}} \\
  E_{F_{2},B_{1}} & E_{F_{2},B_{2}} & \cdots & E_{F_{2},B_{m}} \\
  \vdots  & \vdots  & \ddots & \vdots  \\
  E_{F_{n},B_{1}} & E_{F_{n},B_{2}} & \cdots & E_{F_{n},B_{m}} 
\end{pmatrix}
\end{equation}

   \item The binary fingerprint $f$ is obtained by iterating over consecutive frames $\forall i \in [1,n-1]$ and frequency bands $\forall j \in [1,m-1]$. Each bit of the fingerprint is generated by checking the energy difference between successive frequency bands of two consecutive frames ($\forall k \in [1,L_{f}]$):
\begin{equation}
f_{k} =
  \begin{cases}
    1, &  (E_{i+1,j}-E_{i+1,j+1})-(E_{i,j}-E_{i,j+1}) > 0 \\
    0, &  \text{otherwise}
  \end{cases}
\end{equation}

\end{enumerate} 

\subsection{Miettinen \textit{et al.}}
\label{subsec:appx1-miettinen}
The scheme by Miettinen \textit{et al.} \cite{Miettinen:2014} is inspired by the audio fingerprinting scheme proposed by Sch{\"u}rmann and Sigg (cf. \ref{subsec:appx1-schurmann}) but works on longer timescales.
It uses noise level and luminosity measurements to derive long-term binary fingerprints, which can defend against adversaries that are colocated for short timeframes.
The scheme utilizes such fingerprints in a fuzzy commitment scheme (as described in \ref{subsec:appx1-schurmann}) to gradually evolve a shared secret key to achieve pairing between two devices that are colocated for a sustained period of time. 
In this work, we focus on deriving and comparing long-term binary fingerprints and we do not target a specific use case of establishing a shared secret key. 
 
We first provide notations adopted from the original paper in \autoref{tab:miettinen-not}.
Second, we present parameters of the context fingerprinting algorithm used in the original and our implementations in  \autoref{tab:miettinen-params}. 
Our goal was to follow the original implementation as close as possible, however, we introduced a few changes, as we use audio with a higher sampling rate to generate noise levels.  
We discuss the effect of those changes on the parameters of the context fingerprinting algorithm. 
Third, we present our implementation of the context fingerprinting algorithm in Section \ref{subsubsec:nfp}.

\setcounter{equation}{0}

\begin{table}[!htb]
	\caption{Notations used by Miettinen \textit{et al.}}
	\label{tab:miettinen-not}
	\begin{tabular}{cl}
		\toprule
		Notation & Explanation \\
		\midrule
		$w$ & length of the context snapshot in seconds \\
		$f$ & a new snapshot is recorded every $f$ seconds \\ 
		$r$ & sampling rate of recorded audio in kHz \\ 
		$m_{w}$ & measurement window in seconds \\ 
		$\Delta_{rel}$ & relative threshold for fingerprint generation \\ 
		$\Delta_{abs}$ & absolute threshold for fingerprint generation \\
		\bottomrule
	\end{tabular}
\end{table}

\begin{table}[!htb]
	\caption{Parameters of the context fingerprinting algorithm used in \cite{Miettinen:2014} (highlighted) and our implementations}
	\label{tab:miettinen-params}
	\begin{tabular}{cccccc}
		\toprule
		$w$, sec&$f$, sec&$r$, kHz&$m_{w}$, sec&$\Delta_{rel}$ & $\Delta_{abs}$\\
		\midrule
		\rowcolor{Gray}
		120 & 120 & 8 & 0.1/1 & 0.1 & 10 \\
		5 & 5 & 16 & 1 & 0.1 & 10 \\
		10 & 10 & 16 & 1 & 0.1 & 10 \\
		15 & 15 & 16 & 1 & 0.1 & 10 \\
		30 & 30 & 16 & 1 & 0.1 & 10 \\
		60 & 60 & 16 & 1 & 0.1 & 10 \\
		120 & 120 & 16 & 1 & 0.1 & 10 \\
		\bottomrule
	\end{tabular}
\end{table}

As shown in \autoref{tab:miettinen-params}, our implementation differs with respect to these parameters from the implementation by Miettinen \textit{et al.} \cite{Miettinen:2014}. 
First, we use audio with a higher sampling rate $r$: 16 vs. 8 kHz to generate noise levels. 
The noise levels are generated by averaging absolute amplitudes of audio samples over $m_{w}$ seconds, given by the 
measurement window.
Thus, for $m_{w} = 1$, we obtain one noise level from 16000 audio samples, whereas the original implementation computes one noise level from only 8000 audio samples, which makes our noise levels more fine-grained.
The original implementation uses two different measurement windows $m_{w}$: 0.1 and 1 sec. 
The shorter measurement window speeds up the fingerprint generation but may be susceptible to synchronization 
issues, thus we opt for a longer measurement window. 
For luminosity measurements we do not use the measurement window. We collect luminosity readings at 10 samples per second and use all samples generated during context snapshot length $w$ to obtain the fingerprint.  
Second, we evaluate the context fingerprinting algorithm on the context snapshots of different lengths $w$ from 5 to 120 seconds. 
Thus, we can assess the performance of the scheme (i.e. distinguishing between colocated and non-colocted devices) on shorter context snapshots.  

\subsubsection{Implementation of the context fingerprinting algorithm}
\label{subsubsec:nfp}

\begin{enumerate}
  \setcounter{enumi}{-1}
  
  \item As input we have sets of noise level $S_{nl}$ and luminosity $S_{lux}$  measurements generated from context information collected in our scenarios (i.e. car and office) as stated above.
The number of bits $b$ in the resulting context fingerprints is given by $\frac{|S_{nl}|}{f}$ and $\frac{|S_{lux}|}{f}$, where $|\cdot|$ denotes the set cardinality. 

  \item The \textit{context snapshot} $c_w$ for a timeslot $t$ consists of all measurements $m$ taken in the timeslot of $w$ seconds, $c_{w}(t) = (m_{i},m_{i+1},\dotsc,m_{i+n})$. For each context fingerprint the average value $\overline{c}(t)$ is computed as: 
\begin{equation}
\overline{c}(t) = \dfrac{\displaystyle\sum_{m_{i} \in c(t)}m_{i}}{|\{m_{i} \in c(t)\}|}
\end{equation}

  \item Each set of measurements ($S_{nl}$ or $S_{lux}$) can be represented as a sequence of context snapshots $C(t,t+nf) = (c(t),c(t+f),\dotsc,c(t+nf))$. Then the fingerprint bit $b(t_{i})$ which corresponds to each snapshot $c(t_{i})$ is generated as: 
\begin{equation}
b(t_{i}) =
\begin{cases}
1, &  \Bigl\lvert{\dfrac{\overline{c}(t_{i})}{\overline{c}(t_{i}-f)}-1}\Bigr\rvert > \Delta_{rel} \ \land \ \Bigl\lvert{\overline{c}(t_{i})-\overline{c}(t_{i}-f)}\Bigr\rvert > \Delta_{abs} \\
0, &  \text{otherwise}
\end{cases}
\end{equation}

We note that the values for $\Delta_{rel}$ and $\Delta_{abs}$ (cf. \autoref{tab:miettinen-params}) are not given in the original paper but were provided by the authors in private communication.

  \item The resulting fingerprint for the set of measurements ($S_{nl}$ or $S_{lux}$) is obtained as: 
\begin{equation}
\phi(C(t,t+nf)) = (b(t),b(t+f),\dotsc,b(t+nf))
\end{equation}

\end{enumerate}

To avoid using fingerprints that are exclusively zero in times of low ambient noise and light, Miettinen \textit{et al.} proposed an extension to their system: they propose to compute the \textit{surprisal} of a fingerprint before using it.
The surprisal of a single bit $b$ of the fingerprint is defined as its self-information $I$, measured in bits:
\begin{equation}
\sigma(b) = I(b) = -log_2(P(B=b))
\end{equation}
The surprisal of the whole fingerprint $F$ is the sum of the surprisal of its individual bits:
\begin{equation}
\sigma(F) = \sum_{b \in F}\sigma(b)
\end{equation}
Calculating this surprisal requires knowledge about how often bits occur at specific positions of the fingerprint during specific times of the day, indicated as $P(B=b)$ in the formula.
Miettinen \textit{et al.} do not state the time resolution, but it is implied that the probabilities are tracked on a per-hour basis.
For the office scenario, which covers multiple days, we track the probabilities independently for the individual days, i.e., fingerprints generated on weekdays do not influence the probabilities and thus surprisals for the weekend.

Miettinen \textit{et al.} propose to set a surprisal threshold $\sigma_{thr}$ that the surprisal of a fingerprint has to exceed in order to be considered valid for pairing.
This avoids the problem of attacks by adversaries guessing the low-entropy fingerprints generated at night.
The threshold is defined as
\begin{equation}
\sigma_{thr} = t + \sigma_{marg}
\end{equation}
$t$ denotes the number of incorrect bits the fuzzy commitment will tolerate and $\sigma_{marg}$ denotes an extra security margin.
However, the authors do not state how this margin should be chosen.
Our margin choice is described in \autoref{sec:res:miettinen}

\subsection{Truong \textit{et al.}}
\label{subsec:appx1-truong}

The scheme by Truong \textit{et al.} \cite{Truong:2014} uses WiFi, Bluetooth, GPS and ambient audio collected by two devices to compute a number of context features, which are then fed into a machine learning classifier that outputs a prediction if these devices are colocated. 
This scheme is designed to provide colocation evidence to thwart relay attacks on wireless channels between a user's device and terminal, which employ \gls{zia} (e.g. unlock a computer if a user's smartphone is nearby). 
In this work, we focus on computing context features and obtaining classification results from the machine learning algorithms and we do not target the specific use case of thwarting relay attacks.

We first provide notations adopted from the original paper in \autoref{tab:truong-not}. 
Second, we describe how different context features are computed. 
Third, we provide details of our machine learning methodology, where we discuss our datasets, the parameters of machine learning algorithms that we use and the evaluation procedure. 

Due to a lack of GPS support in the used hardware, we were unable to collect GPS information. However, since our office scenario is static and the car scenario mostly considers geographically close cars, the information value of the GPS features would have been low. In addition, the original authors report that the GPS feature contains the least amount of discriminative power in their dataset.

\setcounter{equation}{0}

\subsubsection{Non-audio features}
\begin{table}[!htb]
	\caption{Notations used by Truong \textit{et al.}}
	\label{tab:truong-not}
	\begin{tabular}{cl}
		\toprule
		Notation & Explanation \\
		\midrule
		$m_i^{(a)}$ & identifier of the $i$th beacon observed by device $a$ \\
		$s_i^{(a)}$ & signal strength of $i$th beacon observed by device $b$ \\
		$\theta$ & value substituted for missing signal strengths \\
		$S_a$ & set of records sensed by device $a$ \\
		$n_a$ & number of different beacons observed by device $a$ \\
		$S_\cap$ & beacons seen by $a$ and $b$ \\
		$S_\cup$ & beacons seen by $a$ or $b$, $\theta$ substituted for missing $s$\\
		$x, y$ & input audio snippets \\
		$L$ & length of input audio snippets in seconds \\
		$r$ & sampling rate of input audio snippets in kHz \\
		\bottomrule
	\end{tabular}
\end{table}

The features for WiFi, Bluetooth and GPS are defined over a number of sets. 
Individual samples for each context information are defined as a tuple $(m,s)$, where $m$ denotes the identifier of the observed beacon (i.e., BLE MAC address, WiFi BSSID) and $s$ denotes the received signal strength.
The set of records observed by devices $a$ and $b$ is denoted as $S_a$ and $S_b$, respectively, while $n_a$ and $n_b$ denote the number of unique beacons observed by the devices.
The notation is also given in \autoref{tab:truong-not}.
Given these preconditions, the following sets are defined:

\begin{align*}
	S_{a} = & \{(m_{i}^{(a)}, s_{i}^{(a)}) \ | \ i \in \mathbb{Z}_{n_{a}-1}\} \\
	S_{b} = & \{(m_{i}^{(b)}, s_{i}^{(b)}) \ | \ i \in \mathbb{Z}_{n_{b}-1}\} \\
	S_{a}^{(m)} = &\{m \ \forall (m,s) \in S_{a}\} \\
	S_{b}^{(m)} = &\{m \ \forall (m,s) \in S_{b}\} \\
	S_{\cap} = &\{(m,s^{(a)},s^{(b)}) \ \forall m \ | \ (m,s^{(a)}) \in S_{a}, \ (m,s^{(b)}) \in S_{b}\} \\
	S_{\cup} = &S_{\cap} \ \cup \ \{(m,s^{(a)},\theta) \ \forall m \ | \ (m,s^{(a)}) \in S_{a}, \ m \notin S_{b}^{(m)}\} \\ &\cup \{(m,\theta,s^{(b)}) \ \forall m \ | \ (m,s^{(b)}) \in S_{b}, \ m \notin S_{a}^{(m)}\} \\
	S_{\cap}^{(m)} = &\{m \ \forall m \ | \ (m,s^{(a)},s^{(b)}) \in S_{\cap}\} \\
	S_{\cup}^{(m)} = &\{m \ \forall m \ | \ (m,s^{(a)},s^{(b)}) \in S_{\cup}\} \\
	L_{a}^{(s)} = &\{s^{a} \ | \ (m,s^{(a)},s^{(b)}) \in S_{\cap}\} \\
	L_{b}^{(s)} = &\{s^{b} \ | \ (m,s^{(a)},s^{(b)}) \in S_{\cap}\} \\
\end{align*}

Truong \textit{et al.} uses these sets to define a total of six features, five of which we implement: the Jaccard Distance $J_{a,b}$, mean Hamming distance $\overline{H}_{a,b}$, Euclidean distance $E_{a,b}$, mean exponential of difference $\overline{Exp}_{a,b}$, the sum of squared ranks $S_{a,b}^{(sr)}$.
The sixth feature, subset count, is only used for the GPS data and thus omitted.
The features are given by the following formulas, where $\theta$ is specific to certain context information. 

\begin{equation}
J_{a,b} = 1 - \dfrac{|S_{\cap}^{(m)}|}{|S_{\cup}^{(m)}|}
\end{equation}

\begin{equation}
\overline{H}_{a,b} = \dfrac{\displaystyle\sum_{k=1}^{|S_{\cup}|}\bigl\lvert{s_{k}^{(a)}-s_{k}^{(b)}}\bigr\rvert}{|S_{\cup}|}
\end{equation}

\begin{equation}
E_{a,b} = \sqrt{\displaystyle\sum_{k=1}^{|S_{\cup}|}(s_{k}^{(a)}-s_{k}^{(b)})^2}
\end{equation}

\begin{equation}
\overline{Exp}_{a,b} = \dfrac{\displaystyle\sum_{k=1}^{|S_{\cup}|}\exp{\bigl\lvert{s_{k}^{(a)}-s_{k}^{(b)}}\bigr\rvert}}{|S_{\cup}|} 
\end{equation}

\begin{equation}
S_{a,b}^{(sr)} = \displaystyle\sum_{k=1}^{|S_{\cap}|}(r_{k}^{(a)}-r_{k}^{(b)})^2
\end{equation}
$|\cdot|$ denotes the set cardinality; $r_{k}^{(a)}$ ($r_{k}^{(b)}$) is the rank of $s_{k}^{(a)}$ ($s_{k}^{(b)}$) in the set $L_{a}$ ($L_{b}$) sorted in ascending order.

For WiFi, all features are used. The signal strength $s$ for each observed identifier is set to the average observed signal strength for that identifier over all included scans. $\theta$, which is substituted as signal strength for devices that have been observed by one but not the other device, is set to -100.
For \gls{ble}, features 1 and 3 are used, once again using the average observed signal strength for each identifier as $s$ and $\theta = -100$.

In case both sensors observe no beacons, the distances are not defined, and the original paper does not specify a behavior for this case.
In private communication, the authors recommended choosing either zero (if the system should be biased towards accepting) or a very high number (if it should be biased towards rejecting).
In our case, we chose to replace undefined values with the distance $10\,000$ to bias the system towards rejecting when in doubt.

\subsubsection{Audio features}
Truong \textit{et al.} use two audio features: the maximum cross-correlation and time-frequency distance computed on snippets of ambient audio of length $L = 10$ seconds. 
The authors do not provide the sampling rate $r$ of their audio snippets; in our implementation $r = 16$ kHz.
We compute these context features on audio snippets of different lengths $L$ from 5 to 120 seconds. 
In the end, we create and evaluate two different datasets for machine learning, one using $L=10$, the other $L=30$.

In the following, we explain how the maximum cross-correlation and time-frequency distance are computed. We note that this information is not available in the original paper and was obtained via private communication with the authors. 

\begin{enumerate}
  \setcounter{enumi}{-1}
  
  \item As input we have two aligned audio snippets $x$ and $y$ of equal length $L$ with a sampling rate $r$.
  
  \item $x$ and $y$ are normalized as (superscript $T$ denotes transpose): 
  \begin{equation}
  x^{\prime} = \dfrac{x}{\sqrt{x^T \cdot x}} \qquad y^{\prime} = \dfrac{y}{\sqrt{y^T \cdot y}}
  \end{equation} 
  Here, the denominator represents a square root of the signal's energy.
  
  \item The \textit{maximum cross-correlation} between the normalized audio snippets $x^{\prime}$ and $y^{\prime}$ is computed as (we omit prime superscripts in $\hat{C}_{x,y}(l)$ for simplicity):
  \begin{equation}
\hat{C}_{x,y}(l) = \max(|C_{x,y}(l)|) =  \max\bigg(\Bigl\lvert\displaystyle\sum_{i=0}^{N-1} x^{\prime}(i) \cdot y^{\prime}(i-l)\Bigr\rvert\bigg)
  \end{equation} 
  $|\cdot|$ denotes the absolute value, $N$ is the number of samples in audio snippets, and the lag $l$ is set to the default value $2N - 1$ \cite{xcorr:2018}. The resulting maximum cross-correlation is bounded within a range $\hat{C}_{x,y}(l) \in [0,1]$, because we take the absolute value $|C_{x,y}(l)|$.
  
  \item To compute the frequency distance between audio snippets $x$ and $y$ a \gls{fft} weighted by a Hamming window is applied: 
  \begin{equation}
	X = FFT(HW(x)) \qquad Y = FFT(HW(y))
 \end{equation}
 
 \item Since the \gls{fft} is symmetric, only a half of the \gls{fft} values is taken to construct frequency vectors for $x$ and $y$:
 \begin{equation}
X_{h} = \Bigl\lvert{X[1,\frac{L_{X}}{2}]}\Bigr\rvert \qquad Y_{h} = \Bigl\lvert{Y[1,\frac{L_{Y}}{2}]}\Bigr\rvert
\end{equation}
$|\cdot|$ denotes the absolute value, $L_{X}$ and $L_{Y}$ are lengths of \gls{fft} vectors $X$ and $Y$.

 \item Frequency vectors $X_{h}$ and $Y_{h}$ are normalized similarly to step (1):
 \begin{equation}
X_{h}^{\prime} = \dfrac{X_{h}}{\sqrt{X_{h}^T \cdot X_{h}}} \qquad Y_{h}^{\prime} = \dfrac{Y_{h}}{\sqrt{Y_{h}^T \cdot Y_{h}}}
\end{equation}

 \item The frequency distance between audio snippets $x$ and $y$ is given by:
\begin{equation}
D_{f, xy} = \sqrt{\displaystyle\sum\left((X_{h}^{\prime}-Y_{h}^{\prime})*(X_{h}^{\prime}-Y_{h}^{\prime})\right)}
\end{equation}
$*$ denotes element-wise multiplication.

 \item The time distance between audio snippets $x$ and $y$ is given by:
 \begin{equation}
D_{t, xy} = 1-\hat{C}_{x,y}(l)
\end{equation}

  \item The \textit{time-frequency distance} between audio snippets $x$ and $y$ is given by:
  \begin{equation}
D_{tf, xy} = \sqrt{D_{t, xy}^2 + D_{f, xy}^2}
\end{equation}

\end{enumerate}

\subsubsection{Machine Learning}

After calculating these features over their dataset, Truong \textit{et al.} used the machine learning suite \textit{Weka} \cite{Hall2009} using Multiboost \cite{Webb2000} with grafted C4.5 decision trees \cite{Webb1999} as weak learners in their evaluation.
As Weka does not support large datasets, we chose to use the H2O framework \cite{h2o_Java_software} instead.
For the training of the classifiers, we set the seed to 1619 and the early stopping to 5 rounds.
This means that the training is repeatable when using the same seed and dataset, and the system will consider learning complete once no improvements have been made for five iterations.
We let H2O train a set of independent models and perform a hyperparameter search to optimize the parameters (e.g., number of trees in the random forest) for the dataset, maximizing the cross-validated \gls{auc}.
Afterwards, we select the top performing model and determine its \gls{eer} as described in \autoref{sec:res:truong}.

\subsection{Shrestha \textit{et al.}}
\label{subsec:appx1-shrestha}
\setcounter{equation}{0}

The scheme by Shrestha \textit{et al.} \cite{Shrestha:2014} utilizes ambient temperature, humidity, pressure, and precision gas collected by two devices to compute a number of context features, which are then fed into a machine learning classifier that outputs a prediction if these devices are colocated.
Similarly to Truong \textit{et al.}, this scheme addresses relay attacks by providing colocation evidence between two devices involved in \gls{zia}. 
In this work, we focus on computing context features and obtaining classification results from the machine learning algorithms and we do not target a specific use case of thwarting relay attacks.

We first provide notations adopted from the original paper in \autoref{tab:shrestha-not}. 
Second, we describe how different context features are computed. 
Third, we provide details of our machine learning methodology, where we discuss our datasets, the parameters of machine learning algorithms that we use and the evaluation procedure.

Due to a lack of hardware support, we were unable to collect precision gas and thus omit this context feature. 

\begin{table}[!htb]
	\caption{Notations used by Shrestha \textit{et al.}}
	\label{tab:shrestha-not}
	\begin{tabular}{cl}
		\toprule
		Notation & Explanation \\
		\midrule
		$s_{a}^{(k)}$ & sample of context information $k$ by device $a$ \\
		$D_{a,b}^{(k)}$ & distance between samples of devices $a$ and $b$  \\  
		\bottomrule
	\end{tabular}
\end{table}

\subsubsection{Context features}
The authors convert ambient pressure $P$ in millibars to altitude in meters using the following formula before computing context features.
\begin{equation}
h_{altitude} = \left(1-\left(\frac{P_{station}}{1013.25}\right)^{0.190284}\right)*145366.45*0.3048
\end{equation}
For each of the considered context information (ambient temperature, humidity, and altitude), the context feature is given by the absolute difference between two samples of context information collected devices $a$ and $b$ at time $t$:
\begin{equation}
D_{a,b}^{(k)} = |s_{a}^{(k)} - s_{b}^{(k)}|
\end{equation}

\subsubsection{Machine learning}
The resulting distances are passed to a Multiboost classifier \cite{Webb2000}, with random forests \cite{Breiman2001} as a weak learner, using Weka \cite{Hall2009}.
The process for machine learning is identical to that described in the previous section.

\section{Study design}

\autoref{tab:sensing-devices} presents hardware used to collect context information in car, office and mob/het scenarios. 

\begin{table}[!hb]
	\caption{Sensing device used for data collection}
	\label{tab:sensing-devices}
	\begin{tabular}{l|cccc}
		\toprule
		Sensor type &\multicolumn{4}{c}{Sensing device (sampling rate)} \\
		& {\makecell{TI SensorTag CC2650 + \\ Raspberry Pi 3 + Samson Go}} & {Samsung Galaxy S6} & {Samsung Gear S3} & {RuuviTag+} \\
		\midrule
		Audio & 16 kHz & 16 kHz & 16 kHz & \raisebox{-0.75ex}{\APLminus} \\
		Barometric pressure & 10 Hz & 5 Hz & 10 Hz & 10 Hz \\
		Humidity & 10 Hz & \raisebox{-0.75ex}{\APLminus} & \raisebox{-0.75ex}{\APLminus} & 10 Hz \\
		Luminosity & 10 Hz & 5 Hz & 10 Hz & \raisebox{-0.75ex}{\APLminus} \\
		Temperature & 10 Hz & \raisebox{-0.75ex}{\APLminus} & \raisebox{-0.75ex}{\APLminus} & 10 Hz \\
		\gls{ble} beacons & 0.1 Hz & 0.1 Hz & 0.1 Hz & \raisebox{-0.75ex}{\APLminus} \\
		WiFi beacons & 0.1 Hz & 0.1 Hz & 0.1 Hz & \raisebox{-0.75ex}{\APLminus} \\
		Accelerometer & 10 Hz & 50 Hz & 50 Hz & \raisebox{-0.75ex}{\APLminus} \\
		Gyroscope & 10 Hz & 50 Hz & 50 Hz & \raisebox{-0.75ex}{\APLminus} \\
		Magnetometer & 10 Hz & 50 Hz & 50 Hz & \raisebox{-0.75ex}{\APLminus} \\
		\bottomrule
	\end{tabular}
	\smallskip\centering
	\center{\raisebox{-0.75ex}{\APLminus{}} = sensor not available}
\end{table}

\begin{figure}[H]
	\centering
	\begin{minipage}[t]{.5\textwidth}
	  \vspace{0pt}
	  \centering
	  \captionof{table}{Device location mapping in the car scenario}
	  \label{tab:car-mapping}
	  \begin{tabular}{ccc}
	    \toprule
		Car 1 Device & Device location & Car 2 Device \\
		\midrule
		01 & Dashboard & 07 \\
		02 & Glove compartment & 08 \\
		03 & Between front seats & 09 \\
		04 & Right back handhold & 10 \\
		05 & Left back handhold & 11 \\
		06 & Trunk & 12 \\
		\bottomrule
	  \end{tabular}
	\end{minipage}%
	\begin{minipage}[t]{.5\textwidth}
	  \centering
	  \vspace{0pt}
	  \includegraphics[width=.63\linewidth]{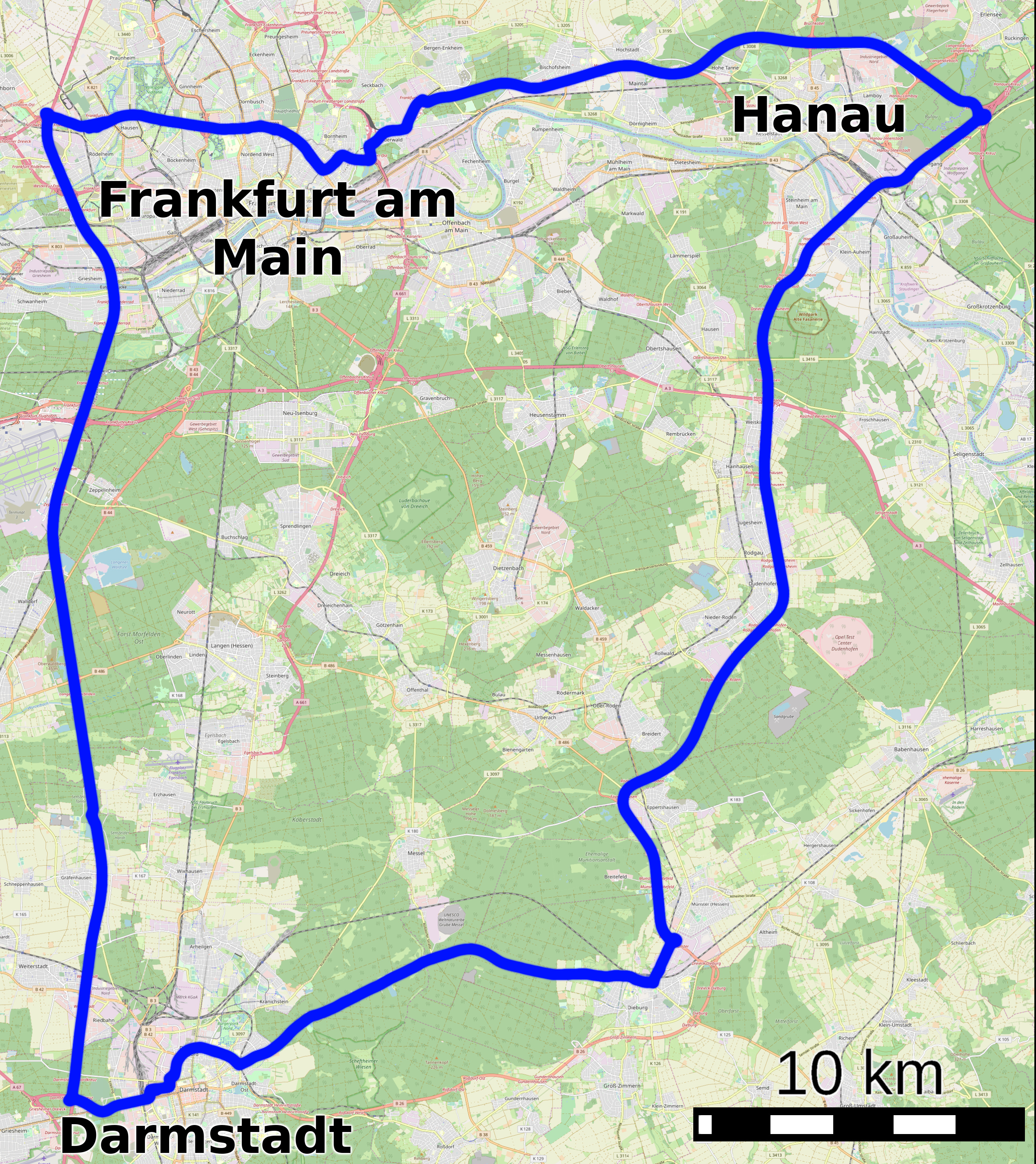}
	  \captionof{figure}{Route driven in the car scenario}
	  \label{fig:car-route}
	\end{minipage}
\end{figure}

\autoref{tab:car-mapping} contains a description of the device deployment in the car scenario, while \autoref{tab:office-mapping} contains the mapping for the office scenario, and \autoref{tab:mob-het} shows device locations in the mob/het scenario. 

\autoref{fig:car-route} shows the route the cars took during the car scenario (cf. \autoref{sec:design:car}).
The route covers city traffic, country roads and highways between the cities of Darmstadt and Frankfurt in the state of Hesse in Germany (the actual GPS traces can be found in \cite{dataset}). 

\begin{table}[H]
	\caption{Device location mapping in the office scenario}
	\label{tab:office-mapping}
	\begin{tabular}{ll|ll|ll}
		\toprule
		\multicolumn{2}{c|}{Office 1} & \multicolumn{2}{c|}{Office 2} &	\multicolumn{2}{c}{Office 3} \\
		{Dev.} & {Location} & {Dev.} & {Location} & {Dev.} & {Location} \\
		\midrule
		01 & Near WiFi access point (\textit{h}) & 09 & Screen of User 2 (\textit{m}) & 17 & Wall behind Users 2 and 3 (\textit{h}) \\
		02 & Window sill (\textit{m}) & 10 & Window sill (\textit{m}) & 18 & Window sill (\textit{m}) \\
		03 & Above door to Office 2 (\textit{h}) & 11 &  Above door to Office 1 (\textit{h}) & 19 & Lamp above User 1 (\textit{h}) \\
		04 & Lamp above User 1 (\textit{h}) & 12 &  Lamp above User 1 (\textit{h}) & 20 & Screen of User 1 (\textit{m}) \\
		05 & Screen of User 1 (\textit{m}) & 13 &  Right screen of User 1 (\textit{m}) & 21 & Screen of User 2 (\textit{m}) \\      
		06 & Screen of User 2 (\textit{m}) & 14 & Left screen of User 1 (\textit{m}) & 22 & Screen of User 3 (\textit{m}) \\
		07 & In the cupboard (\textit{h}) & 15 & In the cupboard (\textit{l}) & 23 & Shelf next to the door (\textit{m}) \\
		08 & Wall next to the door (\textit{h}) & 16 & Shelf left of the door (\textit{m}) & 24 & In the cupboard (\textit{h}) \\
		\bottomrule
	\end{tabular}
	\smallskip\centering
	\center{\textit{h}\textcolor{white}{-}= high position; \textit{m}\textcolor{white}{-}= medium position; \textit{l}\textcolor{white}{-}= low position}
\end{table}

\begin{table}[H]
	\caption{Device location mapping in the mob/het scenario, initial configuration}
	\label{tab:mob-het}
	\begin{tabular}{ll|ll|ll}
		\toprule
		\multicolumn{2}{c|}{Office 1} & \multicolumn{2}{c|}{Office 2} &	\multicolumn{2}{c}{Office 3} \\
		{Dev.} & {Location} & {Dev.} & {Location} & {Dev.} & {Location} \\
		\midrule
		01 & Screen of User 1 (\textit{m}) & 11 & Screen left from User 3 (\textit{m}) & 18 & Screen in front of User 4 (\textit{m}) \\
		02 & Screen of User 2 (\textit{m}) & 12 & Screen of User 3 (\textit{m}) & 19 & Screen of User 4 (\textit{m}) \\
		03 & Near a power plug (\textit{l}) & 13 & Near a power plug (\textit{l}) & 20 & Near a power plug (\textit{l}) \\
		04 & On top of robot station (\textit{l}) & 14 & Near a fan (\textit{l}) & 21 & On top of coffee machine (\textit{m}) \\
		05 & Smartphone of User 1$^\dagger$ & 15 & Smartphone of User 3$^\dagger$ & 22 & Smartphone of User 4$^\dagger$ \\
		06 & Smartwatch of User 1$^\dagger$ & 16 & Smartwatch of User 3$^\dagger$ & 23 & Smartwatch of User 4$^\dagger$ \\
		07 & Laptop of User 1$^\dagger$ & 17 & Laptop of User 3$^\dagger$ & 24 & Laptop of User 4$^\dagger$ \\
		08 & Smartphone of User 2$^\dagger$ & &  & & \\
		09 & Smartwatch of User 2$^\dagger$ & &  & & \\
		10 & Laptop of User 2$^\dagger$ & &  & & \\
		25 & Smartphone on top of robot$^\dagger$ & &  & & \\
		\bottomrule
	\end{tabular}
	\smallskip\centering
	\center{\textit{h}\textcolor{white}{-}= high position; \textit{m}\textcolor{white}{-}= medium position; \textit{l}\textcolor{white}{-}= low position; $^\dagger$\textcolor{white}{-}= mobile device}
\end{table}

\bibliographystyle{ACM-Reference-Format}
\bibliography{bibliography}


\begin{thebibliography}{39}


\ifx \showCODEN    \undefined \def \showCODEN     #1{\unskip}     \fi
\ifx \showDOI      \undefined \def \showDOI       #1{#1}\fi
\ifx \showISBNx    \undefined \def \showISBNx     #1{\unskip}     \fi
\ifx \showISBNxiii \undefined \def \showISBNxiii  #1{\unskip}     \fi
\ifx \showISSN     \undefined \def \showISSN      #1{\unskip}     \fi
\ifx \showLCCN     \undefined \def \showLCCN      #1{\unskip}     \fi
\ifx \shownote     \undefined \def \shownote      #1{#1}          \fi
\ifx \showarticletitle \undefined \def \showarticletitle #1{#1}   \fi
\ifx \showURL      \undefined \def \showURL       {\relax}        \fi
\providecommand\bibfield[2]{#2}
\providecommand\bibinfo[2]{#2}
\providecommand\natexlab[1]{#1}
\providecommand\showeprint[2][]{arXiv:#2}

\bibitem[\protect\citeauthoryear{ANSI/ASA S1.11}{ANSI/ASA S1.11}{2004}]%
        {ANSI:2004}
ANSI/ASA S1.11 \bibinfo{year}{2004}\natexlab{}.
\newblock \bibinfo{booktitle}{\emph{{Specification for Octave-Band and
  Fractional-Octave-Band Analog and Digital Filters}}}.
\newblock \bibinfo{type}{Standard}. \bibinfo{institution}{American National
  Standards Institute}.
\newblock


\bibitem[\protect\citeauthoryear{Benureau and Rougier}{Benureau and
  Rougier}{2018}]%
        {Benureau:2017}
\bibfield{author}{\bibinfo{person}{Fabien C.~Y. Benureau} {and}
  \bibinfo{person}{Nicolas~P. Rougier}.} \bibinfo{year}{2018}\natexlab{}.
\newblock \showarticletitle{{Re-run, Repeat, Reproduce, Reuse, Replicate:
  Transforming Code into Scientific Contributions}}.
\newblock \bibinfo{journal}{\emph{Frontiers in Neuroinformatics}}
  \bibinfo{volume}{11} (\bibinfo{year}{2018}), \bibinfo{pages}{69}.
\newblock
\showISSN{1662-5196}


\bibitem[\protect\citeauthoryear{Breiman}{Breiman}{2001}]%
        {Breiman2001}
\bibfield{author}{\bibinfo{person}{Leo Breiman}.}
  \bibinfo{year}{2001}\natexlab{}.
\newblock \showarticletitle{{Random Forests}}.
\newblock \bibinfo{journal}{\emph{Machine Learning}} \bibinfo{volume}{45},
  \bibinfo{number}{1} (\bibinfo{year}{2001}), \bibinfo{pages}{5--32}.
\newblock
\showISBNx{0885-6125}
\showISSN{08856125}


\bibitem[\protect\citeauthoryear{Br{\"u}sch, Nguyen, Sch{\"u}rmann, Sigg, and
  Wolf}{Br{\"u}sch et~al\mbox{.}}{2018}]%
        {Brusch:2018}
\bibfield{author}{\bibinfo{person}{Arne Br{\"u}sch}, \bibinfo{person}{Ngu
  Nguyen}, \bibinfo{person}{Dominik Sch{\"u}rmann}, \bibinfo{person}{Stephan
  Sigg}, {and} \bibinfo{person}{Lars Wolf}.} \bibinfo{year}{2018}\natexlab{}.
\newblock \showarticletitle{{On the Secrecy of Publicly Observable Biometric
  Features: Security Properties of Gait for Mobile Device Pairing}}.
\newblock \bibinfo{journal}{\emph{CoRR}}  \bibinfo{volume}{abs/1804.03997}
  (\bibinfo{year}{2018}).
\newblock


\bibitem[\protect\citeauthoryear{Elkhodr, Shahrestani, and Cheung}{Elkhodr
  et~al\mbox{.}}{2016}]%
        {Elkhodr:2016}
\bibfield{author}{\bibinfo{person}{Mahmoud Elkhodr}, \bibinfo{person}{Seyed
  Shahrestani}, {and} \bibinfo{person}{Hon Cheung}.}
  \bibinfo{year}{2016}\natexlab{}.
\newblock \showarticletitle{{The Internet of Things: New Interoperability,
  Management and Security Challenges}}.
\newblock \bibinfo{journal}{\emph{International Journal of Network Security and
  its Applications}} \bibinfo{volume}{8}, \bibinfo{number}{2}
  (\bibinfo{year}{2016}), \bibinfo{pages}{85--102}.
\newblock


\bibitem[\protect\citeauthoryear{Fern{\'a}ndez-Delgado, Cernadas, Barro, and
  Amorim}{Fern{\'a}ndez-Delgado et~al\mbox{.}}{2014}]%
        {fernandez2014we}
\bibfield{author}{\bibinfo{person}{Manuel Fern{\'a}ndez-Delgado},
  \bibinfo{person}{Eva Cernadas}, \bibinfo{person}{Sen{\'e}n Barro}, {and}
  \bibinfo{person}{Dinani Amorim}.} \bibinfo{year}{2014}\natexlab{}.
\newblock \showarticletitle{{Do We Need Hundreds of Classifiers to Solve Real
  World Classification Problems}}.
\newblock \bibinfo{journal}{\emph{J. Mach. Learn. Res}} \bibinfo{volume}{15},
  \bibinfo{number}{1} (\bibinfo{year}{2014}), \bibinfo{pages}{3133--3181}.
\newblock


\bibitem[\protect\citeauthoryear{Fomichev, \'{A}lvarez, Steinmetzer,
  Gardner-Stephen, and Hollick}{Fomichev et~al\mbox{.}}{2018}]%
        {Fomichev:2018}
\bibfield{author}{\bibinfo{person}{Mikhail Fomichev}, \bibinfo{person}{Flor
  \'{A}lvarez}, \bibinfo{person}{Daniel Steinmetzer}, \bibinfo{person}{Paul
  Gardner-Stephen}, {and} \bibinfo{person}{Matthias Hollick}.}
  \bibinfo{year}{2018}\natexlab{}.
\newblock \showarticletitle{{Survey and Systematization of Secure Device
  Pairing}}.
\newblock \bibinfo{journal}{\emph{IEEE Communications Surveys Tutorials}}
  \bibinfo{volume}{20}, \bibinfo{number}{1} (\bibinfo{year}{2018}),
  \bibinfo{pages}{517--550}.
\newblock


\bibitem[\protect\citeauthoryear{Fomichev, Maass, Almon, Molina, and
  Hollick}{Fomichev et~al\mbox{.}}{2019a}]%
        {audio_dataset}
\bibfield{author}{\bibinfo{person}{Mikhail Fomichev}, \bibinfo{person}{Max
  Maass}, \bibinfo{person}{Lars Almon}, \bibinfo{person}{Alejandro Molina},
  {and} \bibinfo{person}{Matthias Hollick}.} \bibinfo{year}{2019}\natexlab{a}.
\newblock \bibinfo{title}{Audio Data from Mobile Scenario from "Perils of
  Zero-Interaction Security in the Internet of Things"}.
\newblock
\newblock
\urldef\tempurl%
\url{https://doi.org/10.5281/zenodo.2537984}
\showDOI{\tempurl}


\bibitem[\protect\citeauthoryear{Fomichev, Maass, Almon, Molina, and
  Hollick}{Fomichev et~al\mbox{.}}{2019b}]%
        {dataset}
\bibfield{author}{\bibinfo{person}{Mikhail Fomichev}, \bibinfo{person}{Max
  Maass}, \bibinfo{person}{Lars Almon}, \bibinfo{person}{Alejandro Molina},
  {and} \bibinfo{person}{Matthias Hollick}.} \bibinfo{year}{2019}\natexlab{b}.
\newblock \bibinfo{title}{Index of Supplementary Files from "Perils of
  Zero-Interaction Security in the Internet of Things"}.
\newblock
\newblock
\urldef\tempurl%
\url{https://doi.org/10.5281/zenodo.2537721}
\showDOI{\tempurl}


\bibitem[\protect\citeauthoryear{Friedman}{Friedman}{2001}]%
        {friedman2001greedy}
\bibfield{author}{\bibinfo{person}{Jerome~H Friedman}.}
  \bibinfo{year}{2001}\natexlab{}.
\newblock \showarticletitle{{Greedy Function Approximation: a Gradient Boosting
  Machine}}.
\newblock \bibinfo{journal}{\emph{Annals of statistics}}
  (\bibinfo{year}{2001}), \bibinfo{pages}{1189--1232}.
\newblock


\bibitem[\protect\citeauthoryear{{Futurae Technologies AG}}{{Futurae
  Technologies AG}}{2017}]%
        {Futurae:2017}
\bibfield{author}{\bibinfo{person}{{Futurae Technologies AG}}.}
  \bibinfo{year}{2017}\natexlab{}.
\newblock \bibinfo{title}{{Futurae Authentication Suite}}.
\newblock
\newblock
\urldef\tempurl%
\url{https://www.futurae.com/product/strongauth/}
\showURL{%
\tempurl}
\newblock
\shownote{[Online, Accessed 2018-04-25].}


\bibitem[\protect\citeauthoryear{Hall, Frank, Holmes, Pfahringer, Reutemann,
  and Witten}{Hall et~al\mbox{.}}{2009}]%
        {Hall2009}
\bibfield{author}{\bibinfo{person}{Mark~A Hall}, \bibinfo{person}{Eibe Frank},
  \bibinfo{person}{Geoffrey Holmes}, \bibinfo{person}{Bernhard Pfahringer},
  \bibinfo{person}{Peter Reutemann}, {and} \bibinfo{person}{Ian~H Witten}.}
  \bibinfo{year}{2009}\natexlab{}.
\newblock \showarticletitle{{The WEKA Data Mining Software: an Update}}.
\newblock \bibinfo{journal}{\emph{SIGKDD Explorations}} \bibinfo{volume}{11},
  \bibinfo{number}{1} (\bibinfo{year}{2009}), \bibinfo{pages}{10--18}.
\newblock
\showISBNx{1931-0145}
\showISSN{19310145}


\bibitem[\protect\citeauthoryear{Han, Chung, Sinha, Harishankar, Pan, Noh,
  Zhang, and Tague}{Han et~al\mbox{.}}{2018}]%
        {Han:2018}
\bibfield{author}{\bibinfo{person}{Jun Han}, \bibinfo{person}{Albert~Jin
  Chung}, \bibinfo{person}{Manal~Kumar Sinha}, \bibinfo{person}{Madhumitha
  Harishankar}, \bibinfo{person}{Shijia Pan}, \bibinfo{person}{Hae~Young Noh},
  \bibinfo{person}{Pei Zhang}, {and} \bibinfo{person}{Patrick Tague}.}
  \bibinfo{year}{2018}\natexlab{}.
\newblock \showarticletitle{{Do You Feel What I Hear? Enabling Autonomous IoT
  Device Pairing Using Different Sensor Types}}. In
  \bibinfo{booktitle}{\emph{2018 IEEE Symposium on Security and Privacy (SP)}}.
  IEEE, \bibinfo{pages}{836--852}.
\newblock


\bibitem[\protect\citeauthoryear{Karapanos, Marforio, Soriente, and
  Capkun}{Karapanos et~al\mbox{.}}{2015}]%
        {Karapanos:2015}
\bibfield{author}{\bibinfo{person}{Nikolaos Karapanos},
  \bibinfo{person}{Claudio Marforio}, \bibinfo{person}{Claudio Soriente}, {and}
  \bibinfo{person}{Srdjan Capkun}.} \bibinfo{year}{2015}\natexlab{}.
\newblock \showarticletitle{{Sound-Proof: Usable Two-Factor Authentication
  Based on Ambient Sound}}. In \bibinfo{booktitle}{\emph{USENIX Security
  Symposium}}. \bibinfo{pages}{483--498}.
\newblock


\bibitem[\protect\citeauthoryear{Kardous and Shaw}{Kardous and Shaw}{2014}]%
        {Kardous2016}
\bibfield{author}{\bibinfo{person}{Chucri~A. Kardous} {and}
  \bibinfo{person}{Peter~B. Shaw}.} \bibinfo{year}{2014}\natexlab{}.
\newblock \showarticletitle{{Evaluation of Smartphone Sound Measurement
  Applications}}.
\newblock \bibinfo{journal}{\emph{The Journal of the Acoustical Society of
  America}} \bibinfo{volume}{135}, \bibinfo{number}{4} (\bibinfo{date}{apr}
  \bibinfo{year}{2014}), \bibinfo{pages}{EL186--EL192}.
\newblock
\showISBNx{4}
\showISSN{0001-4966}


\bibitem[\protect\citeauthoryear{Kolias, Kambourakis, Stavrou, and Voas}{Kolias
  et~al\mbox{.}}{2017}]%
        {mirai:2017}
\bibfield{author}{\bibinfo{person}{C. Kolias}, \bibinfo{person}{G.
  Kambourakis}, \bibinfo{person}{A. Stavrou}, {and} \bibinfo{person}{J. Voas}.}
  \bibinfo{year}{2017}\natexlab{}.
\newblock \showarticletitle{{DDoS in the IoT: Mirai and Other Botnets}}.
\newblock \bibinfo{journal}{\emph{Computer}} \bibinfo{volume}{50},
  \bibinfo{number}{7} (\bibinfo{year}{2017}), \bibinfo{pages}{80--84}.
\newblock
\showISSN{0018-9162}
\urldef\tempurl%
\url{https://doi.org/10.1109/MC.2017.201}
\showDOI{\tempurl}


\bibitem[\protect\citeauthoryear{Lu, Pan, Lane, Choudhury, and Campbell}{Lu
  et~al\mbox{.}}{2009}]%
        {Lu2009}
\bibfield{author}{\bibinfo{person}{Hong Lu}, \bibinfo{person}{Wei Pan},
  \bibinfo{person}{Nicholas~D. Lane}, \bibinfo{person}{Tanzeem Choudhury},
  {and} \bibinfo{person}{Andrew~T. Campbell}.} \bibinfo{year}{2009}\natexlab{}.
\newblock \showarticletitle{{SoundSense: Scalable Sound Sensing for
  People-Centric Applications on Mobile Phones}}. In
  \bibinfo{booktitle}{\emph{Proceedings of the 7th international conference on
  Mobile systems, applications, and services - Mobisys '09}}.
  \bibinfo{publisher}{ACM Press}, \bibinfo{address}{New York, New York, USA},
  \bibinfo{pages}{165}.
\newblock
\showISBNx{9781605585666}
\showISSN{0006-3002}


\bibitem[\protect\citeauthoryear{Maisonneuve, Stevens, Niessen, and
  Steels}{Maisonneuve et~al\mbox{.}}{2009}]%
        {maisonneuve2009noisetube}
\bibfield{author}{\bibinfo{person}{Nicolas Maisonneuve},
  \bibinfo{person}{Matthias Stevens}, \bibinfo{person}{Maria~E. Niessen}, {and}
  \bibinfo{person}{Luc Steels}.} \bibinfo{year}{2009}\natexlab{}.
\newblock \showarticletitle{{NoiseTube: Measuring and Mapping Noise Pollution
  with Mobile Phones}}.
\newblock In \bibinfo{booktitle}{\emph{Information technologies in
  environmental engineering}}. \bibinfo{publisher}{Springer},
  \bibinfo{pages}{215--228}.
\newblock
\showISBNx{9783540883500}
\showISSN{18635520}


\bibitem[\protect\citeauthoryear{Mare, Markham, Cornelius, Peterson, and
  Kotz}{Mare et~al\mbox{.}}{2014}]%
        {Mare:2014}
\bibfield{author}{\bibinfo{person}{Shrirang Mare},
  \bibinfo{person}{Andr{\'e}s~Molina Markham}, \bibinfo{person}{Cory
  Cornelius}, \bibinfo{person}{Ronald Peterson}, {and} \bibinfo{person}{David
  Kotz}.} \bibinfo{year}{2014}\natexlab{}.
\newblock \showarticletitle{{Zebra: Zero-effort Bilateral Recurring
  Authentication}}. In \bibinfo{booktitle}{\emph{Security and Privacy (SP),
  2014 IEEE Symposium on}}. IEEE, \bibinfo{pages}{705--720}.
\newblock


\bibitem[\protect\citeauthoryear{Miettinen, Asokan, Nguyen, Sadeghi, and
  Sobhani}{Miettinen et~al\mbox{.}}{2014}]%
        {Miettinen:2014}
\bibfield{author}{\bibinfo{person}{Markus Miettinen}, \bibinfo{person}{N
  Asokan}, \bibinfo{person}{Thien~Duc Nguyen}, \bibinfo{person}{Ahmad-Reza
  Sadeghi}, {and} \bibinfo{person}{Majid Sobhani}.}
  \bibinfo{year}{2014}\natexlab{}.
\newblock \showarticletitle{{Context-based Zero-Interaction Pairing and Key
  Evolution for Advanced Personal Devices}}. In \bibinfo{booktitle}{\emph{ACM
  Conference on Computer and Communications Security (CCS)}}. ACM,
  \bibinfo{pages}{880--891}.
\newblock


\bibitem[\protect\citeauthoryear{Miluzzo, Papandrea, Lane, Lu, and
  Campbell}{Miluzzo et~al\mbox{.}}{2010}]%
        {Miluzzo2010}
\bibfield{author}{\bibinfo{person}{Emiliano Miluzzo}, \bibinfo{person}{Michela
  Papandrea}, \bibinfo{person}{Nicholas~D Lane}, \bibinfo{person}{Hong Lu},
  {and} \bibinfo{person}{Andrew~T Campbell}.} \bibinfo{year}{2010}\natexlab{}.
\newblock \showarticletitle{{Pocket, Bag, Hand, etc. - Automatically Detecting
  Phone Context through Discovery}}.
\newblock \bibinfo{journal}{\emph{PhoneSense 2010: International Workshop on
  Sensing for App Phones (November 2, 2010), held at ACM SenSys '10 (Zurich,
  Switzerland, November 2-5, 2010)}} (\bibinfo{year}{2010}),
  \bibinfo{pages}{21--25}.
\newblock
\showISSN{1872-7573}


\bibitem[\protect\citeauthoryear{Perera, Zaslavsky, Christen, and
  Georgakopoulos}{Perera et~al\mbox{.}}{2014}]%
        {Perera:2014}
\bibfield{author}{\bibinfo{person}{Charith Perera}, \bibinfo{person}{Arkady
  Zaslavsky}, \bibinfo{person}{Peter Christen}, {and}
  \bibinfo{person}{Dimitrios Georgakopoulos}.} \bibinfo{year}{2014}\natexlab{}.
\newblock \showarticletitle{{Context Aware Computing for the Internet of
  Things: A Survey}}.
\newblock \bibinfo{journal}{\emph{IEEE communications surveys \& tutorials}}
  \bibinfo{volume}{16}, \bibinfo{number}{1} (\bibinfo{year}{2014}),
  \bibinfo{pages}{414--454}.
\newblock


\bibitem[\protect\citeauthoryear{Sch{\"u}rmann, Br{\"u}sch, Sigg, and
  Wolf}{Sch{\"u}rmann et~al\mbox{.}}{2017}]%
        {Schurmann:2017}
\bibfield{author}{\bibinfo{person}{Dominik Sch{\"u}rmann},
  \bibinfo{person}{Arne Br{\"u}sch}, \bibinfo{person}{Stephan Sigg}, {and}
  \bibinfo{person}{Lars Wolf}.} \bibinfo{year}{2017}\natexlab{}.
\newblock \showarticletitle{{BANDANA - Body Area Network Device-to-device
  Authentication Using Natural gAit}}. In \bibinfo{booktitle}{\emph{IEEE
  International Conference on Pervasive Computing and Communications
  (PerCom)}}. IEEE, \bibinfo{pages}{190--196}.
\newblock


\bibitem[\protect\citeauthoryear{Sch{\"u}rmann and Sigg}{Sch{\"u}rmann and
  Sigg}{2013}]%
        {Schurmann:2013}
\bibfield{author}{\bibinfo{person}{Dominik Sch{\"u}rmann} {and}
  \bibinfo{person}{Stephan Sigg}.} \bibinfo{year}{2013}\natexlab{}.
\newblock \showarticletitle{{Secure Communication Based on Ambient Audio}}.
\newblock \bibinfo{journal}{\emph{IEEE Transactions on mobile computing}}
  \bibinfo{volume}{12} (\bibinfo{year}{2013}), \bibinfo{pages}{358--370}.
\newblock


\bibitem[\protect\citeauthoryear{Shepherd, Gurulian, Frank, Markantonakis,
  Akram, Panaousis, and Mayes}{Shepherd et~al\mbox{.}}{2017}]%
        {Shepherd:2017}
\bibfield{author}{\bibinfo{person}{Carlton Shepherd}, \bibinfo{person}{Iakovos
  Gurulian}, \bibinfo{person}{Eibe Frank}, \bibinfo{person}{Konstantinos
  Markantonakis}, \bibinfo{person}{Raja~Naeem Akram},
  \bibinfo{person}{Emmanouil Panaousis}, {and} \bibinfo{person}{Keith Mayes}.}
  \bibinfo{year}{2017}\natexlab{}.
\newblock \showarticletitle{{The Applicability of Ambient Sensors as Proximity
  Evidence for NFC Transactions}}. In \bibinfo{booktitle}{\emph{2017 IEEE
  Security and Privacy Workshops (SPW)}}. \bibinfo{publisher}{IEEE},
  \bibinfo{pages}{179--188}.
\newblock


\bibitem[\protect\citeauthoryear{Shrestha, Mohamed, and Saxena}{Shrestha
  et~al\mbox{.}}{2016a}]%
        {Shrestha:2016}
\bibfield{author}{\bibinfo{person}{Babins Shrestha}, \bibinfo{person}{Manar
  Mohamed}, {and} \bibinfo{person}{Nitesh Saxena}.}
  \bibinfo{year}{2016}\natexlab{a}.
\newblock \showarticletitle{{Walk-Unlock: Zero-Interaction Authentication
  Protected with Multi-Modal Gait Biometrics}}.
\newblock \bibinfo{journal}{\emph{CoRR}}  \bibinfo{volume}{abs/1605.00766}
  (\bibinfo{year}{2016}).
\newblock


\bibitem[\protect\citeauthoryear{Shrestha, Mohamed, Tamrakar, and
  Saxena}{Shrestha et~al\mbox{.}}{2016b}]%
        {Shrestha:2016theft}
\bibfield{author}{\bibinfo{person}{Babins Shrestha}, \bibinfo{person}{Manar
  Mohamed}, \bibinfo{person}{Sandeep Tamrakar}, {and} \bibinfo{person}{Nitesh
  Saxena}.} \bibinfo{year}{2016}\natexlab{b}.
\newblock \showarticletitle{{Theft-Resilient Mobile Wallets: Transparently
  Authenticating NFC Users with Tapping Gesture Biometrics}}. In
  \bibinfo{booktitle}{\emph{Proceedings of the 32nd Annual Conference on
  Computer Security Applications}}. ACM, \bibinfo{pages}{265--276}.
\newblock


\bibitem[\protect\citeauthoryear{Shrestha, Saxena, Truong, and Asokan}{Shrestha
  et~al\mbox{.}}{2014}]%
        {Shrestha:2014}
\bibfield{author}{\bibinfo{person}{Babins Shrestha}, \bibinfo{person}{Nitesh
  Saxena}, \bibinfo{person}{Hien Thi~Thu Truong}, {and} \bibinfo{person}{N
  Asokan}.} \bibinfo{year}{2014}\natexlab{}.
\newblock \showarticletitle{{Drone to the Rescue: Relay-resilient
  Authentication Using Ambient Multi-Sensing}}. In
  \bibinfo{booktitle}{\emph{International Conference on Financial Cryptography
  and Data Security (FC)}}. Springer, \bibinfo{pages}{349--364}.
\newblock


\bibitem[\protect\citeauthoryear{Shrestha, Saxena, Truong, and Asokan}{Shrestha
  et~al\mbox{.}}{2018}]%
        {Shrestha:2015}
\bibfield{author}{\bibinfo{person}{Babins Shrestha}, \bibinfo{person}{Nitesh
  Saxena}, \bibinfo{person}{Hien Thi~Thu Truong}, {and} \bibinfo{person}{N
  Asokan}.} \bibinfo{year}{2018}\natexlab{}.
\newblock \showarticletitle{{Sensor-based Proximity Detection in the Face of
  Active Adversaries}}.
\newblock \bibinfo{journal}{\emph{IEEE Transactions on Mobile Computing}}
  (\bibinfo{year}{2018}).
\newblock


\bibitem[\protect\citeauthoryear{Shrestha, Shirvanian, Shrestha, and
  Saxena}{Shrestha et~al\mbox{.}}{2016c}]%
        {Shrestha:2016spf}
\bibfield{author}{\bibinfo{person}{Babins Shrestha}, \bibinfo{person}{Maliheh
  Shirvanian}, \bibinfo{person}{Prakash Shrestha}, {and}
  \bibinfo{person}{Nitesh Saxena}.} \bibinfo{year}{2016}\natexlab{c}.
\newblock \showarticletitle{{The Sounds of the Phones: Dangers of Zero-Effort
  Second Factor Login based on Ambient Audio}}. In
  \bibinfo{booktitle}{\emph{Proceedings of the 2016 ACM SIGSAC Conference on
  Computer and Communications Security}}. ACM, \bibinfo{pages}{908--919}.
\newblock


\bibitem[\protect\citeauthoryear{Sigg}{Sigg}{2011}]%
        {Sigg:2011}
\bibfield{author}{\bibinfo{person}{Stephan Sigg}.}
  \bibinfo{year}{2011}\natexlab{}.
\newblock \showarticletitle{{Context-based Security: State of the Art, Open
  Research Topics and a Case Study}}. In \bibinfo{booktitle}{\emph{Proceedings
  of the 5th ACM International Workshop on Context-Awareness for Self-Managing
  Systems}}. ACM, \bibinfo{pages}{17--23}.
\newblock


\bibitem[\protect\citeauthoryear{Tang and Ishwaran}{Tang and Ishwaran}{2017}]%
        {tang2017random}
\bibfield{author}{\bibinfo{person}{Fei Tang} {and} \bibinfo{person}{Hemant
  Ishwaran}.} \bibinfo{year}{2017}\natexlab{}.
\newblock \showarticletitle{{Random Forest Missing Data Algorithms}}.
\newblock \bibinfo{journal}{\emph{Statistical Analysis and Data Mining: The ASA
  Data Science Journal}} \bibinfo{volume}{10}, \bibinfo{number}{6}
  (\bibinfo{year}{2017}), \bibinfo{pages}{363--377}.
\newblock


\bibitem[\protect\citeauthoryear{team}{team}{2015}]%
        {h2o_Java_software}
\bibfield{author}{\bibinfo{person}{The~H2O.ai team}.}
  \bibinfo{year}{2015}\natexlab{}.
\newblock \bibinfo{title}{{H2O: Scalable Machine Learning}}.
\newblock
\newblock
\urldef\tempurl%
\url{http://www.h2o.ai}
\showURL{%
\tempurl}
\newblock
\shownote{version 3.1.0.99999 [Online, Accessed 2018-04-25].}


\bibitem[\protect\citeauthoryear{{The MathWorks, Inc.}}{{The MathWorks,
  Inc.}}{2018a}]%
        {BP:2018}
\bibfield{author}{\bibinfo{person}{{The MathWorks, Inc.}}}
  \bibinfo{year}{2018}\natexlab{a}.
\newblock \bibinfo{title}{{Bandpass IIR Filter}}.
\newblock
\newblock
\urldef\tempurl%
\url{https://mathworks.com/help/signal/ref/designfilt.html}
\showURL{%
\tempurl}
\newblock
\shownote{[Online, Accessed 2018-04-25].}


\bibitem[\protect\citeauthoryear{{The MathWorks, Inc.}}{{The MathWorks,
  Inc.}}{2018b}]%
        {xcorr:2018}
\bibfield{author}{\bibinfo{person}{{The MathWorks, Inc.}}}
  \bibinfo{year}{2018}\natexlab{b}.
\newblock \bibinfo{title}{{Cross-correlation}}.
\newblock
\newblock
\urldef\tempurl%
\url{https://mathworks.com/help/signal/ref/xcorr.html#bual1fd-maxlag}
\showURL{%
\tempurl}
\newblock
\shownote{[Online, Accessed 2018-04-25].}


\bibitem[\protect\citeauthoryear{Truong, Gao, Shrestha, Saxena, Asokan, and
  Nurmi}{Truong et~al\mbox{.}}{2014}]%
        {Truong:2014}
\bibfield{author}{\bibinfo{person}{Hien Thi~Thu Truong}, \bibinfo{person}{Xiang
  Gao}, \bibinfo{person}{Babins Shrestha}, \bibinfo{person}{Nitesh Saxena},
  \bibinfo{person}{N Asokan}, {and} \bibinfo{person}{Petteri Nurmi}.}
  \bibinfo{year}{2014}\natexlab{}.
\newblock \showarticletitle{{Comparing and Fusing Different Sensor Modalities
  for Relay Attack Resistance in Zero-Interaction Authentication}}. In
  \bibinfo{booktitle}{\emph{IEEE International Conference on Pervasive
  Computing and Communications (PerCom)}}. IEEE, \bibinfo{pages}{163--171}.
\newblock


\bibitem[\protect\citeauthoryear{Webb}{Webb}{1999}]%
        {Webb1999}
\bibfield{author}{\bibinfo{person}{Geoffrey~I. Webb}.}
  \bibinfo{year}{1999}\natexlab{}.
\newblock \showarticletitle{{Decision Tree Grafting from the All-tests-but-one
  Partition}}. In \bibinfo{booktitle}{\emph{Proceedings of the 16th
  International Joint Conference on Artificial Intelligence - Volume 2}}
  \emph{(\bibinfo{series}{IJCAI'99})}. \bibinfo{publisher}{Morgan Kaufmann
  Publishers Inc.}, \bibinfo{address}{San Francisco, CA, USA},
  \bibinfo{pages}{702--707}.
\newblock


\bibitem[\protect\citeauthoryear{Webb}{Webb}{2000}]%
        {Webb2000}
\bibfield{author}{\bibinfo{person}{Geoffrey~I. Webb}.}
  \bibinfo{year}{2000}\natexlab{}.
\newblock \showarticletitle{{MultiBoosting: A Technique for Combining Boosting
  and Wagging}}.
\newblock \bibinfo{journal}{\emph{Machine Learning}} \bibinfo{volume}{40},
  \bibinfo{number}{2} (\bibinfo{year}{2000}), \bibinfo{pages}{159--196}.
\newblock
\showISBNx{0885-6125}
\showISSN{08856125}


\bibitem[\protect\citeauthoryear{Xi, Qian, Han, Zhao, Zhong, Li, and Zhao}{Xi
  et~al\mbox{.}}{2016}]%
        {Xi:2016}
\bibfield{author}{\bibinfo{person}{Wei Xi}, \bibinfo{person}{Chen Qian},
  \bibinfo{person}{Jinsong Han}, \bibinfo{person}{Kun Zhao},
  \bibinfo{person}{Sheng Zhong}, \bibinfo{person}{Xiang-Yang Li}, {and}
  \bibinfo{person}{Jizhong Zhao}.} \bibinfo{year}{2016}\natexlab{}.
\newblock \showarticletitle{{Instant and Robust Authentication and Key
  Agreement among Mobile Devices}}. In \bibinfo{booktitle}{\emph{Proceedings of
  the 2016 ACM SIGSAC Conference on Computer and Communications Security}}.
  ACM, \bibinfo{pages}{616--627}.
\newblock


\end{thebibliography}

\end{document}